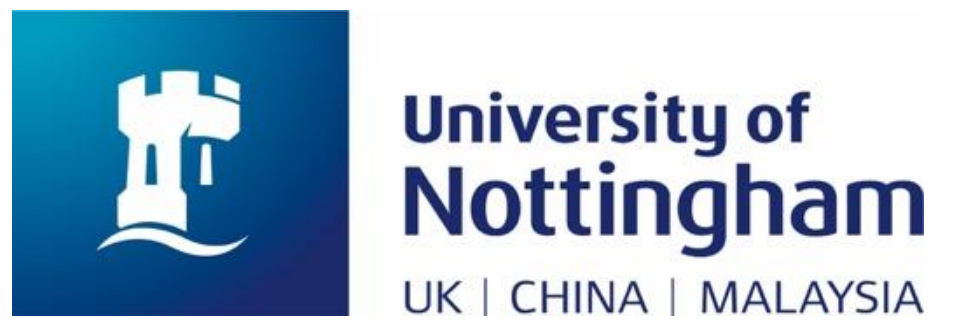

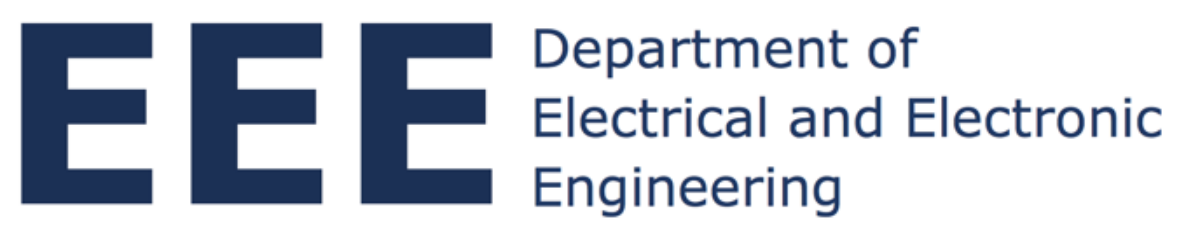

EEEE4008
Final Year Individual Project Thesis

# Investigating reservoir computing for branch prediction in pipelined processors using emerging CMOS memristor devices

| | |
|---|---|
| **AUTHOR:** | **Mr Harvey Samuel George Johnson** |
| **ID NUMBER:** | **20373187** |
| **SUPERVISOR:** | **Dr. Sendy Phang** |
| **MODERATOR:** | **Dr. Paul Blunt** |
| **DATE:** | **19.05.2026** |

This fourth year project Thesis is submitted in part fulfilment of the requirements of the degree of Master of Engineering.

# Contents

# 1 Introduction

## 1.1 Overview

This project aimed to develop a novel reservoir compute (RC) implementation framework targeting high-speed operation and integration with CMOS digital logic. With the target workload of branch prediction (BP) for multistage pipelined central processing unit (CPU) cores. For this, a novel memristor based RC design framework was developed, Section 3.2.1, within the context of the workload requirements. This was then implemented in simulation using industry standard modelling languages of System Verilog (SV) and Verilog-AMS (VAMS).

The developed RC design framework was subsequently verified using a basic sequence detection task before further benchmarking for its effectiveness at BP. The developed RC framework was tested using the Dhrystone performance benchmark, while targeting the RISC-V RV64GC instruction set architecture (ISA). Conducted testing demonstrates that RC shows great promise for application to BP and is capable of achieving impressive overall prediction accuracy. However, testing also shows that further refinement of the developed RC design framework is necessary to address shortfalls in the adaptability of the proposed RC system. As comparison against the state of the art TAGE predictor showed the proposed RC design framework to be 15x slower to adapt to changes in branching behaviour.

## 1.2 Project Motivations

Current CPU core designs are conceptually based on a pipelined Von-Neuman architecture where instructions are fetched, decoded, and executed sequentially. Branches occur when an instruction that conditionally alters this flow is encountered. When a branching instruction is encountered, the CPU core instruction fetch logic cannot determine the next instruction to fetch and has to stall until the branch is resolved. This is known as a control hazard and results in the formation of pipeline bubbles where downstream pipeline stages sit idle waiting for work. As current CPU core designs such as Intel's redwood cove are capable of executing up to 12 instructions at once [1], pipeline bubbles result in a significant reduction in CPU performance. With increasing execution pipeline lengths worsening the performance impact [2].

BP is the technique relied upon to mitigate the pipeline control hazards arising from branching instructions. BP commonly functions by gathering processor contextual information to make a prediction on if a branch will be taken or not taken. This leads to speculative execution, where instructions are fetched, decoded, and executed based on the speculation (prediction) of the branch predictor. This keeps the execution pipeline busy when the correct prediction is made and results in execution pipeline flushing when a misprediction occurs. With the penalty for misprediction being significant in terms of wasted time and energy [3].

BP methods can be divided into static and dynamic, with dynamic methods being the most accurate and of interest. Existing dynamic methods rely on building a statistical model representing branching behaviour. Such as TAGE, considered state of the art in 2011, which uses a series of hash indexed correlation tables increasing in length geometrically that track branching behaviour and identify patterns [4]. Or perceptron based neural methods which both encode branching behaviours in the perceptron's weights and use a global history table to track recent branch outcomes [5].

The common themes of large physical data structures to capture history and complex cascading logic such as hashing functions makes scaling of existing methods costly [2]. Often necessitating pipelining of prediction, resulting in significant latencies and the necessity of hybrid prediction schemes with near and far predictors [6]. Which further complicates branch prediction unit (BPU) design and contributes to significant chip area usage.

At the same time, RC has seen significant strides in its development. With memristor based RC demonstrated to operate at up to 20 million predictions per second at digital logic friendly voltages [7]. And advances in manufacturing delivering high density area efficient reservoir structures with increasing uniformity and endurance [8, 9]. Device engineering of the constituent memristors is also delivering significant advances in switching speed and operating voltage via device shrinkage [9, 10, 11]. Achieving full switching operation at as low as 0.5V [11] and as fast as 15ns [10], making the application of RC to BP a realistic and enticing area of research to explore.

RC works by using the "black box" non-linear behaviour of complex physical systems to map non-linearly separable problems into a higher dimensional space. Allowing them to be solved by a simple linear regression output layer such as single perceptron neuron. It is comprised of three main parts, Fig.1, the input layer which performs data preprocessing and encoding, the reservoir which projects the inputs into higher dimensional space, and the output layer which makes the prediction.

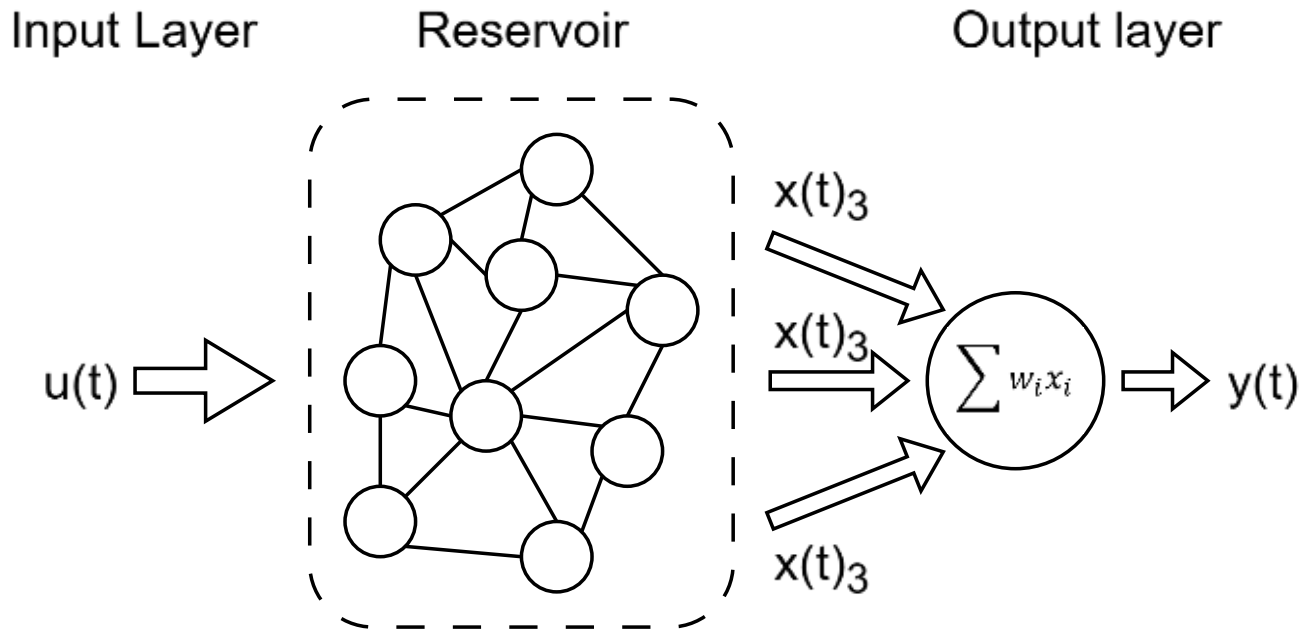

*Figure 1. general RC architecture.*

This project seeks to explore the use of RC by leveraging its memory and non-linear operation [12] to perform branch prediction. Done so by developing a novel reservoir architecture in simulation to target high speed low latency operation. Alongside development of a highly parameterizable RC BPU design framework to explore application of RC to BP workload. With the possibility of successful application to address the increasing need for energy efficient compute for datacentre and AI workloads. With datacentre power usage set to increase 6x in the decade to 2034 [13] and the cost of semiconductor manufacturing surging 300% since 2017 [14] being driving factors for this research.

## 1.3 Research Goals

This project will develop a framework for the implementation of memristor based RC for the application of BP. With the framework subsequently applied to numerically evaluate the suitability and capability of memristor RC for BP workload. Using industry standard behavioural modelling tools, the project aims to develop a RC BPU targeting:

- $P_1$ Prediction latency of 2ns
- $P_2$ 1 billion predictions per second
- $P_3$ Prediction accuracy > 70%
- $P_4$ Integrable with CPU digital logic

The overall project targets have been refined through literature review and represent a compromise between the initial proposal requirements and achievable work. The requirement of prediction latency was changed from 1 ns to 2 ns, and the original requirement is clarified as throughput of 1 billion predictions per second. The proposed target of developing physical design and floor planning has been revised to a stretch goal due to time constraints.

The project is comprised of five deliverables (**D**), Fig.2, being:

- $\boldsymbol{D_1}$ Memristor behavioural model
- $\boldsymbol{D_2}$ Reservoir architecture and design
- $\boldsymbol{D_3}$ Trainable readout layer
- $\boldsymbol{D_4}$ Full system integration
- $\boldsymbol{D_5}$ Workload benchmarking

Each deliverable represents a significant milestone in framework development and project completion. Each deliverable is defined by a set of baseline requirements that must be fulfilled in addition to stretch goals where appropriate.

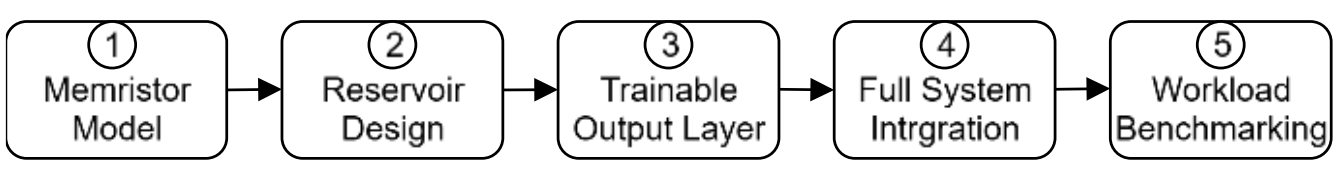

*Figure 2. research project work structure.*

### 1.3.1 Memristor Behavioural Model

The first deliverable is the memristor behavioural model. This deliverable is comprised of the work done to develop a Verilog-AMS behavioural model which implements SiOx memristor device behaviours from literature [10, 15, 16, 11, 17, 18, 19, 9]. This work is foundational to the project as the memristors behaviour, operating conditions, and limitations inform reservoir design and achievable system performance $\boldsymbol{P_1}$. Proposed requirements are:

- $\boldsymbol{M_1}$ Reproduce device behaviour as shown in literature
- $\boldsymbol{M_2}$ Be computationally efficient
- $\boldsymbol{M_3}$ Be numerically stable
- $\boldsymbol{M_4}$ Have configurable behaviour
- $\boldsymbol{M_5}$ Explicit modelling of device geometry
- $\boldsymbol{M_6}$ Model temperature effects on device behaviours

$\boldsymbol{M_1}$ to $\boldsymbol{M_4}$ are the base requirements, ensuring the model replicates correct behaviour from literature adequately enough to allow reservoir design. This can be achieved by adopting the model described in [16] and validating against [10, 9, 20, 15, 21, 18, 19] using simulations. $\boldsymbol{M_5}$ and $\boldsymbol{M_6}$ are stretch requirements over the baseline aimed at enhancing model configurability and better replicating physical devices [10, 15]. Contributing to $\boldsymbol{P_4}$ by the exploration and consideration of device size and operating temperature impact on suitability for CPU integration. Which is conducted after behavioural model enhancement and evaluated referencing literature.

### 1.3.2 Reservoir Architecture and Design

The second deliverable is the reservoir design framework. Encompassing the work to develop the design framework and implement it as a behavioural model in Verilog-AMS. This involves developing the reservoir itself using the memristor behavioural model developed in $\boldsymbol{D_1}$ and the development of reservoir control logic such as analogue to digital interfaces. The Proposed requirements are:

- $\boldsymbol{R_1}$ Demonstrate measurable state change with <1ns latency
- $\boldsymbol{R_2}$ Demonstrate operation at or below 3.3V VDD
- $\boldsymbol{R_3}$ Demonstrate reproducibility and fading memory
- $\boldsymbol{R_4}$ Investigate measurable state change with <0.5ns latency
- $\boldsymbol{R_5}$ Investigate tuneable memory capacity

$\boldsymbol{R_1}$ to $\boldsymbol{R_3}$ are the base requirements for the reservoir design, directly contributing to project goals $\boldsymbol{P_1}$, $\boldsymbol{P_2}$, and $\boldsymbol{P_4}$. With $\boldsymbol{R_2}$ being realistically attainable [9, 11, 10] and allowing easy integration with CPUs as 3.3V power infrastructure already exists on chip for external signalling. $\boldsymbol{R_1}$ and $\boldsymbol{R_3}$ requirements feed into $\boldsymbol{P_1}$, $\boldsymbol{P_2}$, and $\boldsymbol{P_3}$ as reservoir latency is expected to dominate prediction latency. With fading memory and reproducibility being key requirements for effective reservoir compute [12].

$\boldsymbol{R_4}$ and $\boldsymbol{R_5}$ are stretch goals aimed at enhancing the base reservoir design if time allows. Lower prediction latency, $\boldsymbol{P_1}$, can be achieved with a lower reservoir latency, $\boldsymbol{R_4}$. Additionally, tuneable memory capacity, $\boldsymbol{R_5}$, allows variable frequency operation which is a common feature of CPUs.

### 1.3.3 Trainable Output Layer

The third deliverable is the development of a trainable output layer design in System Verilog. This layer provides the taken or not taken predictions based on measured reservoir state. The proposed requirements are:

- $\boldsymbol{O_1}$ In-situ training capability
- $\boldsymbol{O_2}$ Critical path < 1ns
- $\boldsymbol{O_3}$ Full configurability of learning rate and input count
- $\boldsymbol{O_4}$ Critical path < 0.5ns

$\boldsymbol{O_1}$ to $\boldsymbol{O_3}$ are the base requirements for the output layer. $\boldsymbol{O_1}$ allows the final RC BPU to adapt to changing branching patters, a requirement for successful application to BP workload. $\boldsymbol{O_2}$ is derived from $\boldsymbol{P_1}$, with the output layer allocated 50% of the timing budget needed to reach target. $\boldsymbol{O_2}$ can be achieved by using fixed point arithmetic or low bit depth integer arithmetic for operation. $\boldsymbol{O_3}$ contributes to the projects aim to develop an optimal design through iteration. With $\boldsymbol{O_3}$ allowing easy testing of different reservoir configurations and sizes for final BP workload by reducing design modification work required. $\boldsymbol{O_4}$ is a stretch goal over $\boldsymbol{O_2}$ to enhance timing of the output layer, contributing to enhancing project requirements $\boldsymbol{P_1}$ and $\boldsymbol{P_2}$. Making the developed RC BPU more suitable for BP workload.

### 1.3.4 Full System Integration

The fourth deliverable is to develop a fully integrated simulation model for the RC BPU design, involving the successful integration of all subsystem components into a standalone design. This involves the integration of the System Verilog behavioural models with the Verilog-AMS behavioural models using Siemens Symphony AMS simulator. With additional work performed for whole system verification and basic functionality demonstration. The proposed requirements for this deliverable are:

- $\boldsymbol{F_1}$ Achieve >70% accuracy in time series pattern recognition
- $\boldsymbol{F_2}$ Investigate impact of reservoir configuration on $\boldsymbol{F_1}$
- $\boldsymbol{F_3}$ Investigate impact of reservoir memory capacity on $\boldsymbol{F_1}$
- $\boldsymbol{F_4}$ Prediction latency ≤ 2ns
- $\boldsymbol{F_5}$ Modular design for design space exploration

The requirements $\boldsymbol{F_1}$ to $\boldsymbol{F_5}$ represent the goals of full system integration. $\boldsymbol{F_1}$ is important to demonstrate the basic functionality of the developed RC system. Sequence detection serves as an intuitive workload that is a proxy for BP, requiring both non-linear dynamics for dimensional transformation and fading memory for temporal information processing. Achieving $\boldsymbol{F_1}$ thus represents the baseline performance indicative of a functional RC system. $\boldsymbol{F_2}$ and $\boldsymbol{F_3}$ contribute to the understanding of the developed RC BPU, providing a baseline understanding on the design parameters that are relevant to full system performance. Allowing a fuller focus on BP workload for BP workload benchmarking and final system testing. $\boldsymbol{F_4}$ is directly derived from the project aim $\boldsymbol{P_1}$. $\boldsymbol{F_5}$ derives from the projects focus on assessing the possible implementation of RC for BP workload. This necessitates an adaptable design framework to allow design space exploration and investigation of different system configurations within the time constraints of this project.

### 1.3.5 Workload Benchmarking

The fifth and final deliverable is BP workload benchmarking on the complete RC BPU design. This involves the creation of test programs and their conversion into instruction streams using a CPU core functional model which are then used to benchmark the RC BPU design. The requirements for this deliverable are mostly derived from the project aims and are:

$\boldsymbol{W_1}$ Achieve 2ns prediction latency for $\boldsymbol{W_3}$

$\boldsymbol{W_2}$ Demonstrate operation at or below 3.3V $V_{DD}$ for $\boldsymbol{W_3}$

$\boldsymbol{W_3}$ Demonstrate > 70% accuracy

$\boldsymbol{W_4}$ Demonstrate > 90% accuracy

Requirements $\boldsymbol{W_1}$ to $\boldsymbol{W_3}$ represent the minimum viable performance to consider RC successfully applied to BP workload. $\boldsymbol{W_4}$ represents the ideal performance on BP workload, which if achieved would put RC within close proximity to existing state of the art BP methods.

## 1.4 Project Methodology

The project development methodology comprised four fundamental stages. The first stage was a literature review to source inspiration from existing research and inform subsequent project development stages. The second was the development of numerical simulation frameworks and modelling for each required component of an RC implementation. The third stage was the implementation of the proposed RC BPU design by design framework standardisation for each subcomponent. The fourth stage was benchmarking of the developed RC BPU design on BP workload, investigating its effectiveness in various configurations.

The project development follows an iterative agile methodology where whenever possible each deliverable can be worked on in parallel. Providing the advantage of flexible time allocation to reprioritise in the event of unexpected challenges or setbacks.

### 1.4.1 Research and Literature Review

The first stage of project development was to undertake a literature review and conduct background research on existing BP methods and RC implementations. This work, detailed in Section 2.0, informed the initial RC BPU design proposal in addition to building a library of literature to draw from for subsequent development stages.

The tools utilised for this stage were IEEExplore and google scholar. Which provided the ability to search nearly all the worlds scientific and engineering literature to build a foundation of understanding. However, one limitation is the relative lack of literature regarding current CPU and BPU implementation which results from their extreme commercial sensitivity. Previous experience related to CPU architecture was drawn from to guide the literature review to focus on high quality and meaningful sources to mitigate this.

### 1.4.2 Design Framework and Modelling

The second stage of the project's development was to develop a design framework for each of the RC subcomponents. This is undertaken as an iterative process following agile methodology where all deliverables are developed in parallel. With the only critical dependence being $\boldsymbol{D_2}$ necessitating the prior development of $\boldsymbol{D_1}$.

Behavioural modelling was performed using Verilog-AMS and System Verilog for analogue and digital system modelling respectively. The Verilog family of languages are powerful high level behavioural description languages with wide ranging industry support. Their human readability and flexibility provided a critical advantage over the alternatives of SPICE or MATLAB Simulink modelling. Their standardisation and ubiquity for behavioural modelling also provided a significant advantage as detailed language reference manuals and documentation exists to inform their use. Additionally, their standardisation provides a wide variety of compatible simulators from many vendors such as Synopsys, Cadence, and Siemens. Siemens Symphony AMS platform is used to perform behavioural simulations. It allows deep integration of VAMS and SV models to simulate mixed signal designs efficiently. Siemens Symphony is chosen as licenses were already held by the university, so its use incurred no additional cost.

Validation of the developed behavioural models for each deliverable $\boldsymbol{D_1}$, $\boldsymbol{D_2}$, and $\boldsymbol{D_3}$ was conducted using automated verification flows. Which use the python scripting language to control the Symphony simulation environment. Python scripts orchestrate the verification flow generally as shown in Fig.3. Providing the advantage of faster turnaround time through code generation for parallel testing and simulator output data aggregation. MATLAB is then used as the final stage of the flow to visualise the simulator data outputs as plots for analysis and presentation.

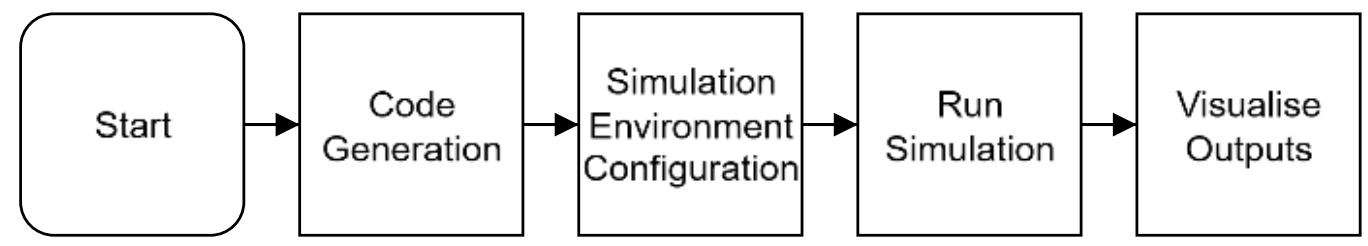

*Figure 3. Python script driven automated simulation verification flow.*

### 1.4.3 Subsystem Integration

The third stage of the project's development was the integration of the RC BPU subcomponents into a single complete design as part of deliverable $\boldsymbol{D_4}$. This involved the standardisation of module interfaces for deliverables $\boldsymbol{D_1}$, $\boldsymbol{D_2}$, and $\boldsymbol{D_3}$. necessitating the completion of design framework development and modelling for each of them before subsystem integration could begin.

Verification of full system integration utilised the Siemens Symphony AMS platform for the same reasons as outlined in Section 1.4.2. Python orchestrated simulations, outlined in Section 1.4.2, were used to test the fully integrated system. This enabled continual iteration on subcomponent design and configuration, as part of the agile methodology, to address design weaknesses.

The Siemens Symphony AMS simulator is used as it provides tight integration between the digital and analogue domain as discussed in Section 1.4.2. Allowing it to simulate the digital logic readout layer, $\boldsymbol{D_3}$, using SV and the analogue logic using VAMS in a single unified simulation environment. The same python driven verification flow methodology from Section 1.4.2 is used to drive automated testing for assessment of relevant metrics. Allowing the autonomous testing of the design's overall accuracy and adaptability.

### 1.4.4 BP Workload Benchmarking

The fourth stage of the project's development is the BP workload benchmarking and testing to reach the final RC BPU design configuration. This is comprised entirely of $\boldsymbol{D_5}$ and involved the initial application of the developed configuration from $\boldsymbol{D_4}$ to BP workload. With subsequent design parameter refinement to achieve an optimal design in terms of feasibility and performance.

Workload benchmarking utilised the Condor fork of the Spike RISC-V functional model [22] and the Dhrystone benchmark [23] for instruction trace generation. After which, the instruction trace was then processed into a set of RC BPU inputs and classification labels. Transient simulations were then run with different RC BPU configurations to iterate on the design towards an optimal RC BPU configuration. Maximising prediction accuracy and adaptability and minimising reservoir size. Siemens Symphony AMS platform is used for simulations, alongside MATLAB for data visualisation.

# 2 Background

This section establishes the background of the project by first examining RC to establish its core principles of operation. Next, the execution pipeline of conventional CPU designs is investigated with a focus on both the performance implications of branching and existing methods of BP.

## 2.1 Reservoir Compute Overview

Reservoir computing is a computational framework derived from liquid state machines and recurrent neural network theory [12, 24, 25]. It is particularly well suited to classification tasks that require some form of memory [25] or have non-linear decision boundaries such as sequence classification.

The three components of the general RC architecture are visualised in Fig.4. The first is the input layer which encodes data, u(t), for input into the reservoir by an appropriate form of preprocessing. The next component is the reservoir which processes the input data through its dynamics, represented abstractly as a cyclic graph in Fig.4. The reservoir is typically a "black box" system with fixed and complex non-linear dynamics. The third and final component of a RC system is the output layer which performs classification on the reservoir state to generate the system output. The output layer is typically a simple linear regression or single perceptron neuron [24, 25]. The key feature of RC is that only the output layer requires training as reservoir dynamics are fixed.

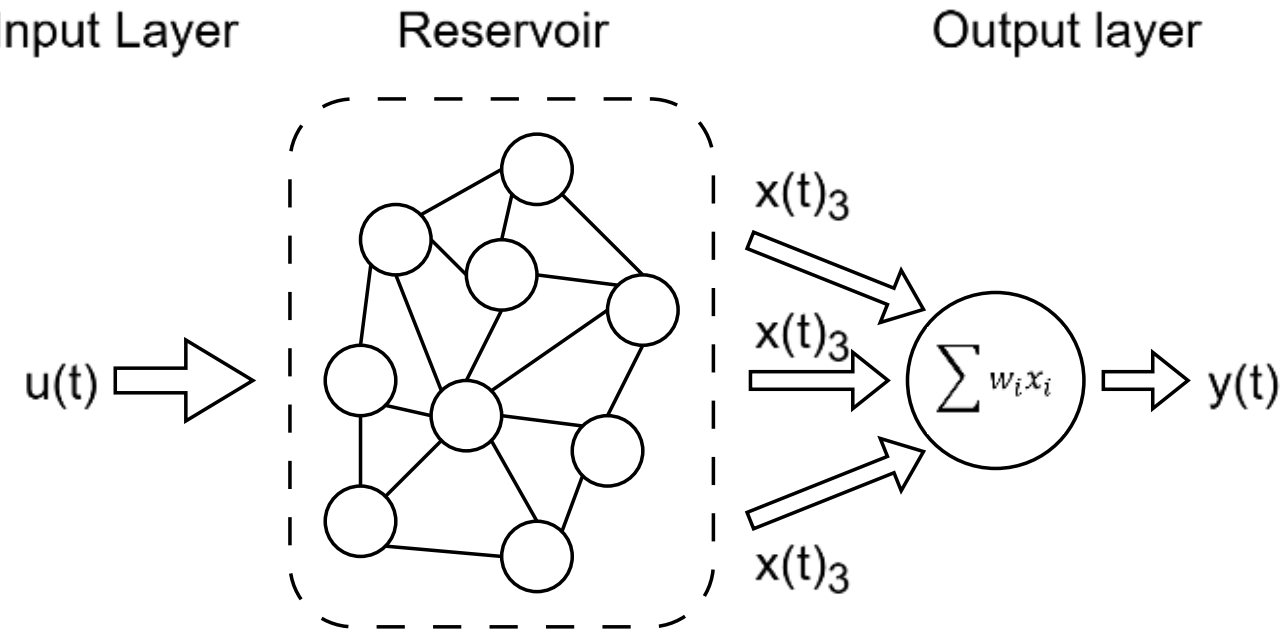

*Figure 4. General RC system architecture.*

The reservoir can be thought of as a state machine with a state $\vec{X}_k$ at time point k with its next state is determined by (1); where $\vec{u}_k$ is the input to the reservoir (excitation / stimuli) at time point k.

$$\vec{X}_{k+1} = \vec{f}(\vec{X}_k, \vec{u}_k) \quad (1)$$

The reservoir possesses a form of memory through the relationship in (1) between the current state and inputs and the next state. This means that the reservoir dynamics act to project a series of inputs, $\vec{U}_N$, into the higher dimensional state space of the reservoir, $\vec{X}^M$ [12]. The specific transform is dependent on the reservoir implementation but is typically irrelevant given that it is sufficiently complex, non-linear, and fixed [12, 25].

The property of fixed reservoir dynamics with only the output layer needing training provides a significant advantage to RC compared to existing machine learning methods. Many physical systems can be described generally in the form of (1) meaning that physical systems can be leveraged to significantly reduce computational requirement. This allows RC to solve complex non-linearly separable problems with a fraction of the compute of existing deep neural network (DNN) and recurrent neural network (RNN) methods [25, 12].

The ability of RC to leverage the powerful computational capability of physical systems has been extensively demonstrated in literature. For example. The RC implementation proposed in [26], Fig.5c, which uses a reservoir comprised of 88 memristor devices in a crossbar was able to achieve 91.1% accuracy on handwritten digit recognition [26]. This utilised simple preprocessing to convert rows of pixels into time series voltage pulses which were input to the reservoir, Fig.5d. Demonstrating the effectiveness of RC to process complex time series data for pattern recognition.

Memristor based RC has also been applied to the task of gesture recognition, achieving 90% accuracy [7]. 9 channel IMU data containing acceleration and orientation information was pre-processed into a 1-2 bit quantisation and input to a series of virtual reservoirs implemented using memristors. Demonstrating the ability of memristor RC to function with low bit quantisation of inputs for high-speed and efficient operation at just 0.78pJ/input [7].

Another example of memristor RC is [27] where an array of memristors was utilised to classify ECG signals and predict chaotic time series. Achieving an accuracy of 99.375% in ECG signal classification and a RMS error of as low as 0.014594 after 40 time steps for time series forecasting.

These examples demonstrate the versatility of memristor RC to many complex tasks with a common theme of suitability to time-series based classification and prediction. Making memristor RC appealing as a potential advancement in BPU design and implementation.

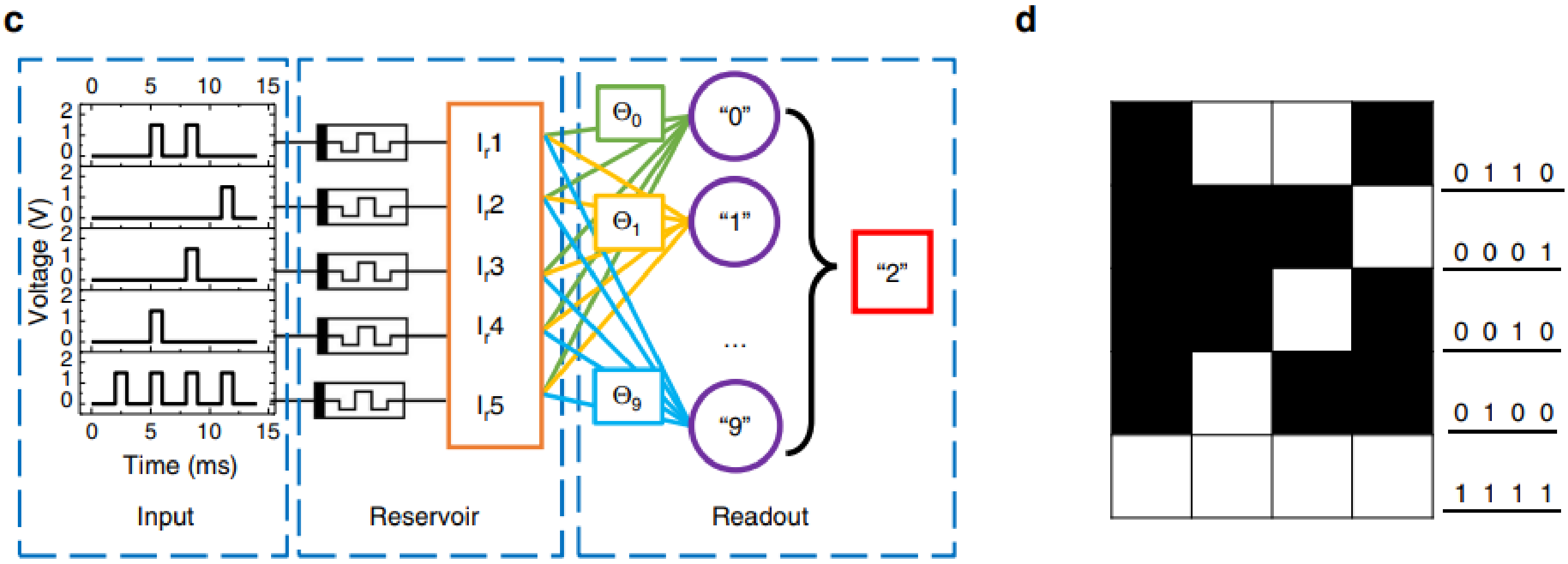

*Figure 5. schematic of RC system with pulse streams as inputs, the memristor reservoir and the readout network for digital classification task of 5x4 images (c). And the example digit of 2 used in the simple digit analysis (d) [26].*

## 2.2 Branch Prediction Overview

Branch prediction is a critical technology in pipelined CPU architectures, responsible for mitigating control hazards which arise from conditional jump instructions [2].

Current CPU core designs implement the Von-Neuman architecture of fetch, decode, execute, and writeback shown in Fig.6. In this architecture, the CPU core processes multiple instructions at different stages in the life cycle at once in a single execution pipeline. Advanced CPU cores often have 10+ stages in the execution pipeline such as Intel's Core microarchitecture from 2006 which has 14 execution pipeline stages [28]. Long pipelines are used to support features that enhance performance such as out of order execution (OOO) and super scalar execution. These allow the processing of multiple instructions per stage to extract instruction level parallelism (ILP) to increase the speed of program execution [1, 3]. This can be seen in designs such as Intel's Golden Cove which allows for parallel execution of up to 12 instructions and parallel decode of 6 instructions at once [1]. Deep and wide pipelines of cores such as Intel's Golden Cove achieve industry leading performance and efficiency when at high utilisation. However, high utilisation is not always practically attainable due to pipeline hazards.

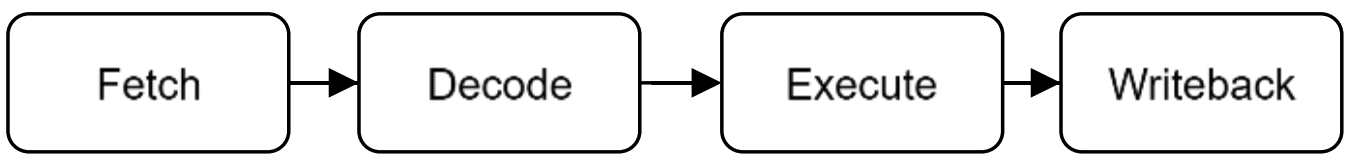

*Figure 6. abstract CPU Von-Neuman architecture.*

Pipelined execution results in hazards that arise from the sequential model of execution that all software assumes. The sequential execution model of software assumes that instructions execute in order one after another and individually. This results in control hazards in pipelined CPU cores when an instruction that changes this sequential nature, such as a branching instruction, is encountered. Branching instructions conditionally change the flow of execution by modifying the program counter (PC). As the PC controls the location to fetch the next instruction from, instruction fetch logic is unable to fetch the next instruction until the outcome of the branch is known. As the outcome of the branch is only known once it is finished executing at the end of the pipeline, the pipeline has to stall. This results in the formation of pipeline bubbles where no further work is issued, and the pipeline stages begin to sit idle. Seen in Fig.7 as the series of wasted cycles after the branching instruction is decoded.

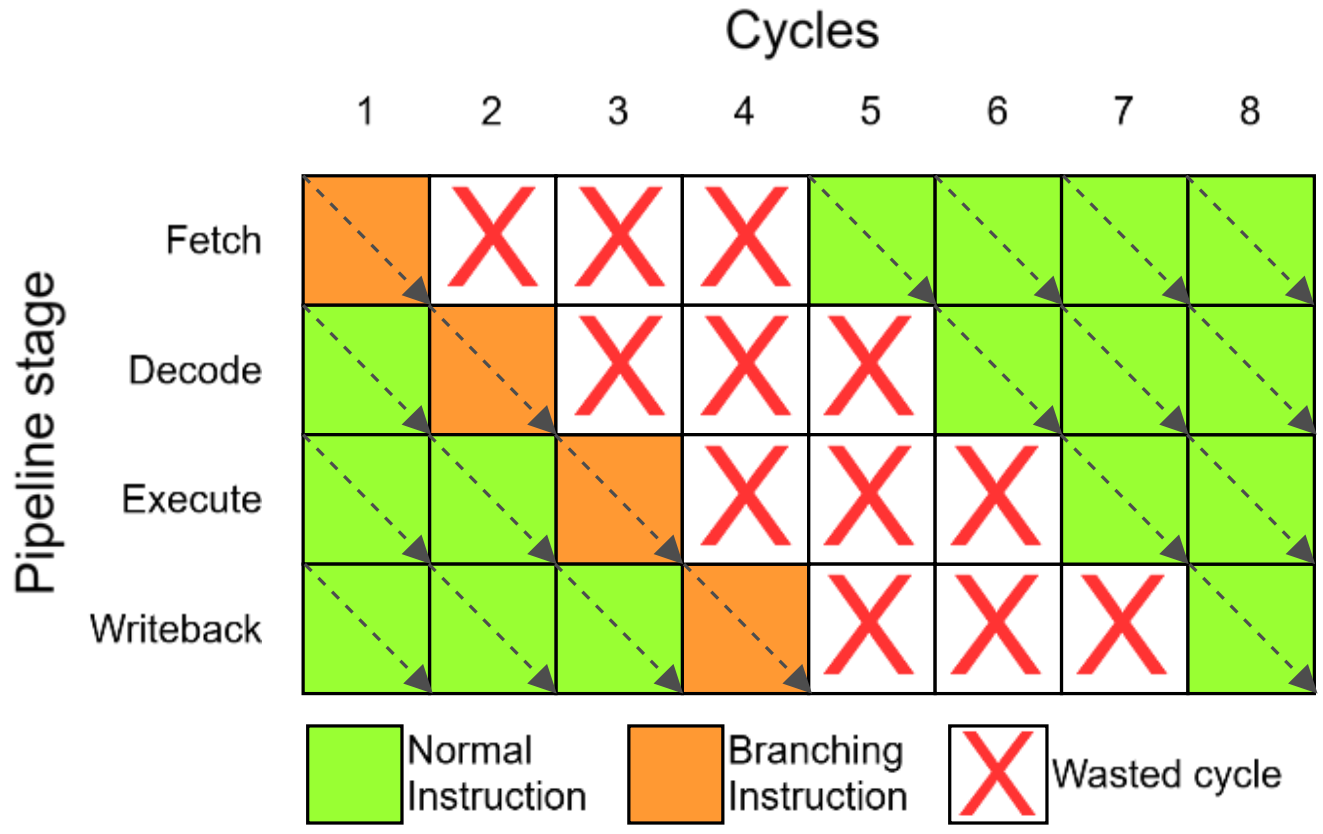

*Figure 7. diagram of example of pipeline control hazard arising from branching instruction.*

The pipeline bubbles resulting from control hazards can significantly degrade performance, especially in deep designs with many stage pipelines [2, 5, 28]. Branch prediction mitigates this by speculating on if the branch will be taken or not, allowing the instruction fetch logic to begin fetching the next instructions almost immediately. This is known as speculative execution [28] and can significantly increase software performance by way of higher execution resource utilisation by eliminating pipeline bubbles. However, if the branch is incorrectly predicted, known as a misprediction, the execution pipeline needs to be flushed and the results of the speculative execution discarded. This causes a significant penalty in wasted energy to execute the incorrect branch and the time required to walk back to the last valid execution pipeline state [2]. Hence, accurate BP is required both to mitigate control hazards arising from the pipelined implementation of CPU cores and to reduce the occurrence of mis-speculations [29].

Branch predictors are categorised into two broad categories of static and dynamic predictors [2]. Dynamic branch predictors, which are of interest to this project, all work on the principle of building a statistical model of branching behaviour [29]. Such as TAGE, considered state of the art in 2011, which uses a series of correlation tables increasing in length geometrically which are accessed by a hash of the instructions address [4]. Or neural methods such as perceptron, which work by encoding the branching patterns in the network's weights [5]. With the use of a Pattern History Table (PHT) also common to capture the history of recent branches (taken or not taken) alongside correlation tables [2, 5]. These large data structures result in physically large and complex implementations that while accurate, are slow and require significant chip area [2, 30]. With the increased latency of prediction often having a significant and adverse impact on software execution performance [30].

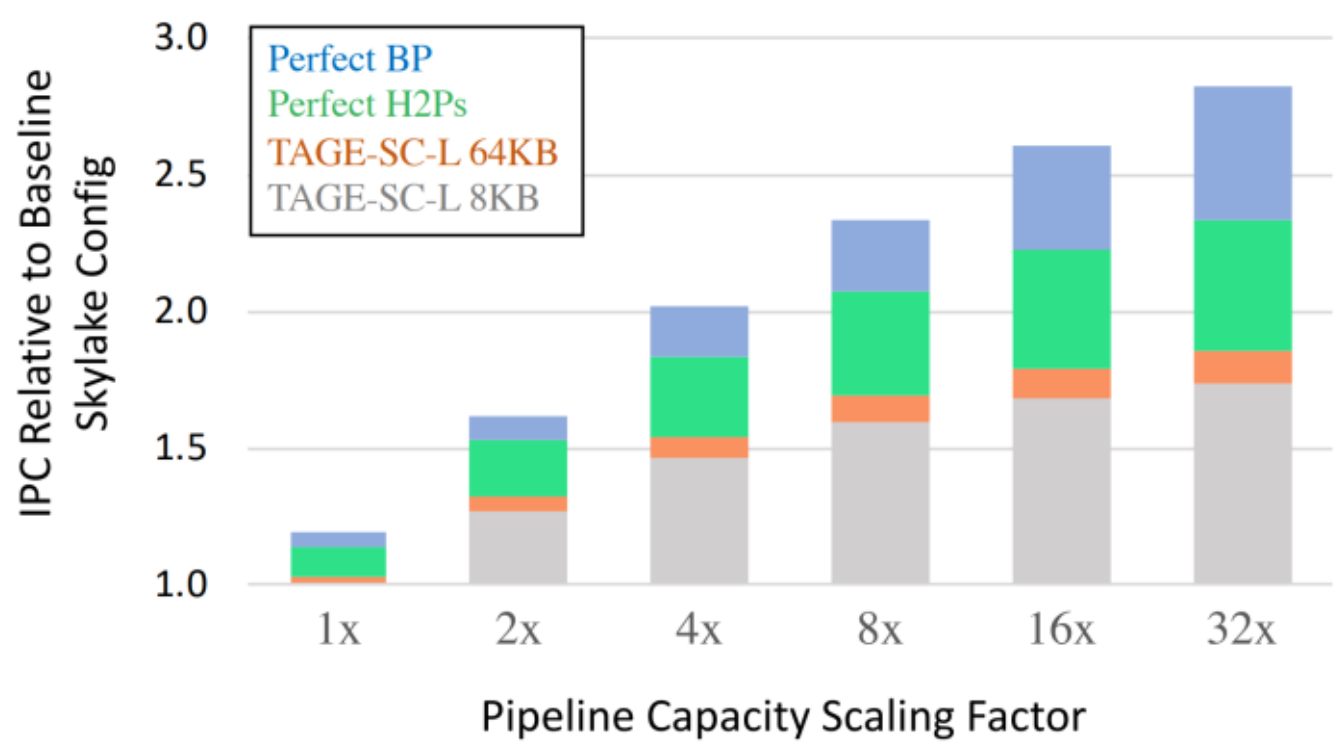

*Figure 8. relative IPC gain of different BP configurations with relative increase of width of CPU execution pipeline [29].*

Advanced methods like TAGE scale inefficiently with increased chip area usage. Resulting in mediocre software performance gains of 2.7% at the cost of increasing table capacity eight fold [29]. This lack of scaling has resulted in research turning to the application of more complex ML approaches to BP [29, 31]. With [29] investigating offline training methods to enhance real-time BP accuracy. Where it is demonstrated that the ability of advanced ML techniques to generalise from aggregated data results in enhanced prediction of hard to predict branches.

RC is hypothesized to be ideal for the workload of BP. Its ability to leverage the powerful compute provided by physical systems, such as nanoscale memristors, opens the door for area efficient and accurate BP designs. Additionally, the suitability of RC to temporal information processing makes it ideal for BP as a dynamic time series classification workload. RC is well positioned to bring the benefits of computationally sophisticated ML approaches to real-time BP implementation in the resource constrained environment of CPU core design.

# 3 Design and Implementation

This section presents the design and implementation of the RC BPU and its subcomponents, focusing on $\boldsymbol{D_1}$, $\boldsymbol{D_2}$, $\boldsymbol{D_3}$, and $\boldsymbol{D_4}$. First, a Verilog-AMS behavioural memristor model is developed and characterised to inform the proposed RC BPU design, $\boldsymbol{D_1}$. The initial proposed RC system is then outlined, followed by the development, implementation, and characterisation of a reservoir design framework, $\boldsymbol{D_2}$. Next, the trainable output layer, $\boldsymbol{D_3}$, is implemented in SystemVerilog. Finally, all subcomponents are integrated into a standardised RC BPU module, which is evaluated using a sequence detection workload to assess its ability to process temporal information through non-linear dynamics, $\boldsymbol{D_4}$.

## 3.1 Memristor

### 3.1.1 Background

A memristor is a 2-terminal device that exhibits a variable resistance, termed memristance (M), that changes over time as a response to stimuli. The ideal relationship proposed by Leon Chua [20] related flux linkage and the time integral of current:

$$M(q(t)) = \frac{\mathrm{d}\Phi/_{\mathrm{d}t}}{\mathrm{d}q/_{\mathrm{d}t}} \tag{2}$$

$$V(t) = M(q(t))\, I(t) \tag{3}$$

The term memristor has since been generalised to cover multiple categories of devices, such as SiOx ReRAM devices which are the focus of device modelling for this thesis. SiOx ReRAM devices are physically similar to metal-insulator-metal capacitors, Fig.9, and have two main switching methods; intrinsic and extrinsic [15]. Intrinsic switching occurs when oxygen vacancies coalesce under an applied electric field to form conductive filaments which charge carriers can flow through [15, 32]. Extrinsic switching occurs when an active electrode is used which contributes ion species to the insulating layer to form a metallic nanobridge through it [15, 11, 10]. The metallic nanobridge is formed via a reversible redox reaction and allows the flow of current [15, 11, 10], thus switching the devices resistance. Both switching mechanisms result in highly stable changes to the switching layer which can persist at a minimum for hours or days [9, 18, 33]. SiOx ReRAM devices have also proven to scale positively down to nanoscale device sizes. Exhibiting high ON/OFF ratios of up to $10^4$, endurance exceeding $10^7$ switching cycles, and operation at CMOS friendly voltages below 3.3V, even down to 0.5V [11, 10, 15]. With Wanjun Chen et al. able to demonstrate 15ns switching time for device set and reset using an Sn electrode material [18].

One of the main challenges for device modelling is switching noise as switching in both intrinsic and extrinsic devices is a stochastic process [34, 18]. Deriving an adequate representation of this will be challenging due to the lack of consensus on actual switching mechanism operation [15, 35]. However, [11] suggests this is addressable through device shrinkage as smaller devices exhibit increased uniformity and lower switching noise. This is attributed to the SiOx junction and filament area being comparable. [10] reinforces this by demonstrating significantly lower switching noise as the switching oxide layer thickness is shrunk from 6.5nm to 1nm. With the reduction in oxide thickness also delivering significantly lower voltage switching and hence higher speed switching at a given voltage [9, 11, 10]. This significant mitigation of switching noise provided by nanoscale devices makes switching noise modelling mostly optional while still fulfilling deliverable requirement $\boldsymbol{M_1}$.

Another challenge of device modelling is the varying conduction mechanisms. With extrinsic devices specifically exhibiting different conduction mechanisms in their high resistance state (HRS) and low resistance state (LRS) respectively. It has been observed that in the LRS conduction is predominantly ohmic, while in the HRS conduction is predominantly Schottky emission [15, 19]. However, this is addressed through $\boldsymbol{M_4}$, $\boldsymbol{M_5}$, and $\boldsymbol{M_6}$. Where the differing conduction mechanisms were taking into account as part of the work to explicitly model device geometries and temperature dependent behaviour.

Additionally, the challenge of modelling the different switching behaviours of intrinsic and extrinsic devices is addressed. As both intrinsic and extrinsic devices switch under an applied electric field, they can be modelled very similarly. Allowing the consideration of both device types for RC by memristor model tuning, fulfilling deliverable requirement $\boldsymbol{M_4}$.

### 3.1.2 Design

A modified version of the VTEAM [16] model has been developed to model SiOx devices. VTEAM is a voltage threshold version of the TEAM model developed by Kvatinsky et al [36]. VTEAM is chosen as it manages to accurately abstract complex device behaviours using mathematically simple equations which are fast to compute, $\boldsymbol{M_2}$. Providing a significant simulation speed advantage over more complex models [36] while achieving an MSE as low as 0.09% compared to more accurate models. Demonstrating its ability to accurately replicate device functionality, $\boldsymbol{M_1}$. Additionally, the threshold switching behaviour VTEAM provides is more consistent with the switching mechanisms observed in both intrinsic and extrinsic devices.

VTEAM uses the parameters $k_{off}, k_{on}, v_{off}, v_{on}, \alpha_{off}, \alpha_{on}, R_{LRS}$ $, R_{HRS}, m, and\ D_l$ and variables $w$ and $x$ to control its behaviour, providing a large degree of model configurability. To start with, VTEAM assumes the basic relationship given in (4) and (5).

$$\frac{dx}{dt} = f(x, v) \tag{4}$$

$$i(t) = G(x, v) \times v(t) \tag{5}$$

Where $v(t)$ is the voltage across the device, $i(t)$ is the current through the device, $G(x, v)$ is the device conductance, and $f(x, v)$ is a general function representing the derivative of the state variable $x$. The expanded modified form of state variable derivative is (6):

$$\frac{dx(t)}{dt} = \begin{cases} k_{off} \cdot \left(\frac{v(t)}{v_{off}} - 1\right)^{\alpha_{off}} \cdot W(i, x), & v < v_{off} < 0 \quad (6a) \\ 0, & v_{off} < v < v_{on} \quad (6b) \\ k_{on} \cdot \left(\frac{v(t)}{v_{on}} - 1\right)^{\alpha_{on}} \cdot W(i, x), & 0 < v_{on} < v \quad (6c) \end{cases}$$

Where $W(i, x)$ is a windowing function (8). $k_{off}$ & $k_{on}$ are tunable constants dictating the rate of change of $w$. $v_{off}$ & $v_{on}$ are the on and off voltage thresholds for device switching. $\alpha_{off}$ & $\alpha_{on}$ are tunable constants representing the nonlinearity of the devices switching behaviour. The variables $D_l$ and $w$ represent the thickness of the oxide layer and the device switching state expressed as thickness using (7).

$$w = x \cdot D_l \qquad \{x: \in \mathbb{R}, 0 < x < 1\} \tag{7}$$

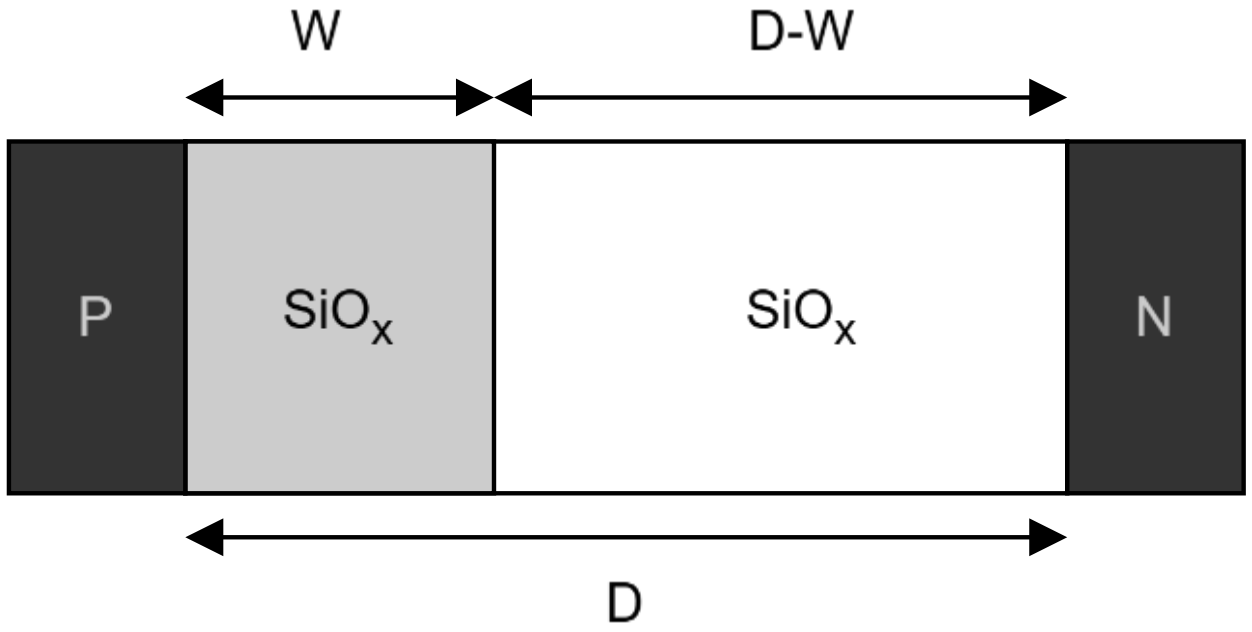

*Figure 9. simplified SiOx memristor structure.*

$W(i,x)$ is used to constrain the values that $x$ can take while also modelling the non-linear switching speed resulting from the Redox reaction that performs device switching [34, 35, 15]. The windowing function also prevents the state variable from becoming stuck at the boundaries by incorporating sign of $i$, giving the response shown in Fig. 10.

$$W(i,x) = 1 - \left(x - \text{sgn}(-i)\right)^{2m} \tag{8}$$
$$\{m: \in \mathbb{Z}, 1 < m < \infty\}$$

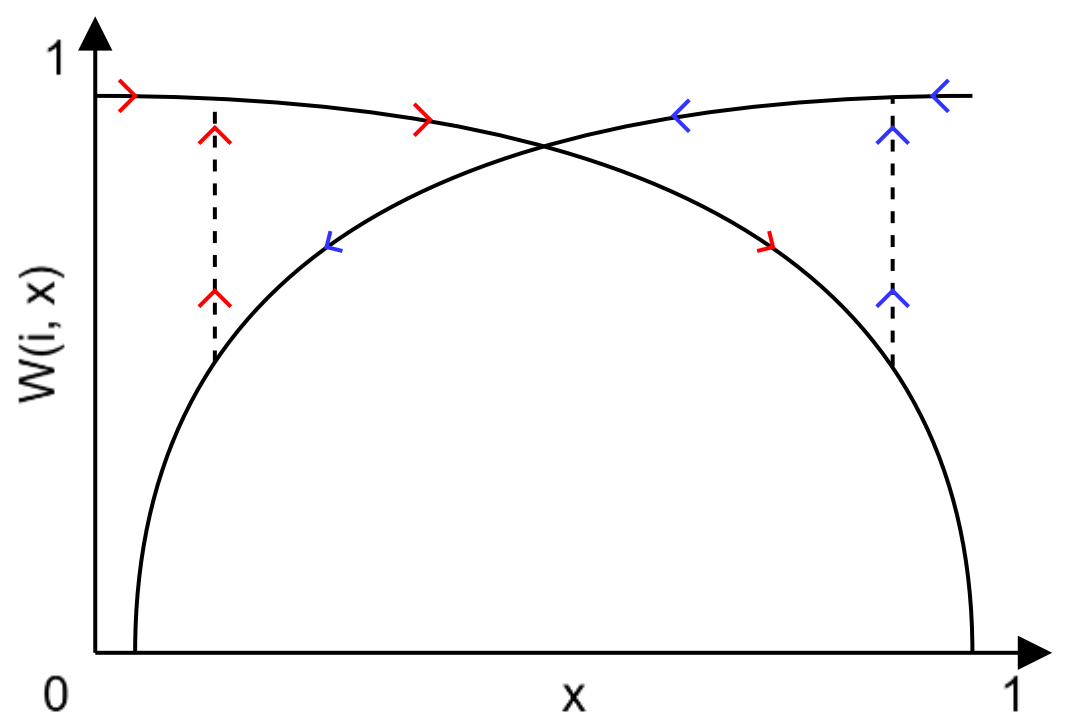

*Figure 10. window function W(i, x) sketch based on fig4 [36].*

The device I-V relationship remains unchanged from the original VTEAM paper and is defined as (9) with $\lambda$ representing a derived parameter from the HRS and LRS resistances [16].

$$i(t) = \frac{v(t)}{R(t)}, \quad R(t) = \left(\frac{e^{\lambda \cdot x}}{R_{LRS}}\right)^{-1}, \quad \lambda = \ln\left(\frac{R_{HRS}}{R_{LRS}}\right) \tag{9}$$

(9) creates an exponential relationship between the device state $x$ and the current $i(t)$. Explicitly modelling the non-linear I-V relationship with device state.

$$i(t)_{x\to 0} = \frac{1}{R_{LRS}} \cdot v(t) \;\rightarrow\; v(t) = i(t)R_{LRS} \tag{10}$$

$$i(t)_{x\to 1} = \frac{1}{R_{HRS}} \cdot v(t) \;\rightarrow\; v(t) = i(t)R_{HRS} \tag{11}$$

Using $e^{-\ln(x)} = \frac{1}{x}$ to solve (11) and $e^{-0} = 1$ for (10).

One of the required simplifications for modelling as a consequence of (9) is the assumption of ohmic conduction. As covered in Section 3.2.1, both extrinsic and intrinsic switching devices exhibit non ohmic conduction in their HRS [33, 15, 19]. This inaccuracy is acceptable due to the project requirements focussing on replication of switching behaviour being the priority, $\boldsymbol{M_1}$. The assumption of ohmic conduction in all states also contributes to the computability of the numerical model by keeping it simple, $\boldsymbol{M_2}$.

The next enhancement to the VTEAM model is the consideration of device noise sources as part of the deliverable requirement $\boldsymbol{M_1}$. The noise sources considered are shot noise for both device switching and conduction, and Johnson-Nyquist for conduction.

For the purpose of conduction noise modelling, the assumption of ohmic conduction in all device states is maintained. This allows the use of (12) to calculate the expected RMS noise voltage resulting from the thermal agitation of charge carriers, Johnson-Nyquist noise. Which results from the approximation of the noise model by band limiting to between 0Hz and 100GHz.

$$V_{RMS} = \sqrt{4 \cdot k_B \cdot T \cdot R(t) \cdot \Delta f} \tag{12}$$
$$k_B \approx 1.38 \cdot 10^{-23} \qquad \Delta f = 10GHz$$

$$I_{Thermal}(t) = \frac{V_{RMS}}{R(t)} \tag{13}$$

Where $k_B$ is the Boltzmann constant, T is the temperature in Kelvin, R(t) is the instantaneous resistance of the memristor device from (8), and $\Delta f$ is the frequency band limit. The instantaneous noise voltage can then be sampled from a normal distribution using $\sigma_x = V_{RMS}$ and $\mu_x = 0$ as it can be approximated to AWGN [37]. The Norton transform of the RMS voltage source is used to calculate the thermal noise as a current value (13).

Modelling of shot noise for conduction is done by sampling a Poisson distribution to realise the instantaneous current from the ideal device current (9).

The expected current value from (9) is converted to a flow rate of electrons per second by division by the elemental charge of the electron. After this, a Poisson is sampled and the electron flow rate is converted back into a current value in Amperes (14). The total non-ideal current flow is then given by (15) as the summation of the instantaneous current value, $I_{Shot}(t)$, and thermal noise current, $I_{Thermal}(t)$.

$$I_{shot}(t) \sim \frac{Poisson\left(\frac{v(t)}{R(t)} \times e^{-1}\right)}{e^{-1}}, \quad e \approx 1.602 \times 10^{-19} \tag{14}$$

$$I_{total}(t) = I_{Thermal}(t) + I_{Shot}(t) \tag{15}$$

Device switching noise is also modelled as part of deliverable requirement $\boldsymbol{M_1}$. The redox reaction that drives switching is modelled as a two step process with the expected rate given by (6) and its reciprocal interpreted as the expected interval. This allows the use of the Erlang distribution to realise the interval [38], which is the reciprocal of the realised switching rate. The Erlang distribution is sampled to give the instantaneous switching rate,

$$\frac{dw_{inst}(t)}{dt} \sim \left(Erlang\left(2, \left(\frac{dw(t)}{dt}\right)^{-1}\right)\right)^{-1} \tag{16}$$

The variation of HRS and LRS resistances with temperature is modelled as a modification to the VTEAM model as part of the deliverable stretch requirement $\boldsymbol{M_6}$. As the conduction mechanism in HRS and LRS may not be the identical and thus vary differently with temperature, each is assigned a temperature coefficient of resistance (TCR). This allows the linear approximation of resistance variation with temperature over a small range of temperatures. Such as the increasing current flow with increasing temperature of Schottky emission, or the decreasing current flow with increasing temperature of ohmic conduction. (17) is the general form of a temperature dependent resistance with a linear dependence, where $\alpha$ is the TCR of resistance $R_0$ measured at a temperature $T_0$ and $T$ is the current temperature in kelvin. Experimental evidence from [15] indicates that although Schottky emission has dependence on the square of the temperature, it can be approximated linearly over the range of interest (0C – 115C). This allows (18) to be used to model the temperature variations in HRS and LRS reference resistances. Where $R_{0_LRS}$ and $R_{0_HRS}$ are the resistances of the LRS and HRS state respectively at 20C. And $TCR_{LRS}$ and $TCR_{HRS}$ are temperature coefficients of resistance for the LRS and HRS state respectively, measured at 20C.

$$R = R_0\left(1 + TCR(T - T_0)\right), \qquad \alpha\Delta T \ll 1 \tag{17}$$

$$R_{LRS} = R_{0_LRS} \times \left(1 + TCR_{LRS} \times (T - 293.15)\right) \tag{18a}$$
$$R_{HRS} = R_{0_HRS} \times \left(1 + TCR_{HRS} \times (T - 293.15)\right) \tag{18b}$$

A further addition to the VTEAM model to meet the deliverable stretch requirement $\boldsymbol{M_5}$ is the modelling of varying device geometry. This allows the model to account for varying oxide thickness[10] and cross sectional area using the parameters $D_l$, $D_a$, $\rho_{on}$, $\rho_{off}$. Changes to the bulk resistance of the device in the HRS and LRS states based on varying geometry are calculated by (19). Using the device cross sectional area, $D_a$, and oxide thickness, $D_l$, in addition to the bulk resistivity of in each resistance state, $\rho_{LRS}$ and $\rho_{HRS}$.

$$R_{0_LRS} = \rho_{LRS} \times \left(\frac{D_l}{D_a}\right) \quad (19a)$$

$$R_{0_HRS} = \rho_{HRS} \times \left(\frac{D_l}{D_a}\right) \quad (19b)$$

Additional to this, (20) is used to set the switching threshold voltages $v_{on}$ and $v_{off}$ based on the threshold electric field required to initiate switching, $\varepsilon_{on}$, $\varepsilon_{off}$. This directly models the electric field driven redox switching mechanism that drives both extrinsic and intrinsic devices [9].

$$V_{on} = \varepsilon_{on} D_l \quad (20a)$$

$$V_{off} = \varepsilon_{off} D_l \quad (20b)$$

### 3.1.3 Implementation

The memristor is implemented in Verilog-AMS as a 2-port module, expressed as a set of differential equations using equations from Section 3.2.2. The module has the fitting parameters given in Table 1 which allow it to model many different device physical parameter variations. The module also has the ability, through the setting of Verilog-AMS *localparams*, to modify HRS and LRS conduction mechanisms in addition to the toggling of noise source modelling.

Verilog-AMS is used as it allows human readable description of continuous valued analogue systems and discrete event driven systems. Verilog-AMS use of System Verilog style syntax and support for parameterised modules enables deep model configurability through module parameterisation. Contributing to deliverable requirements $\boldsymbol{M_4}$, $\boldsymbol{M_5}$, and $\boldsymbol{M_6}$. Verilog-AMS built in system functions for time integrals, derivatives, and statistical distribution sampling enable optimised numerical simulation. With the use of these functions enabling accurate behavioural modelling with variable time step simulation to decrease simulation runtimes. Contributing to deliverable requirements $\boldsymbol{M_1}$, $\boldsymbol{M_2}$ and $\boldsymbol{M_3}$.

The model is defined using a single analogue block which computes the device behaviour each simulation time step in two stages. The first stage is to update the device internal state variable, after which the model defines it's I-V relationship.

The code first calculates the idealised state variable derivative using (6), (7), and (8). This is done by first sampling the windowing function using the current device state (8). Then the state variable derivative is computed based on the voltage across the device (6). After this, the state variable derivative is realised using the *$rdist_erlang()* function to sample the Erlang distribution (16). Next, the sampled state variable derivative is used to update the state variable by integration using the *$idt()* system function. *$idt()* is a simulator native time series integral method which allows time series integrals in simulation even with variable length time steps.

Once the state variable is updated, the I-V relationship is defined. First, (9) is used to compute the updated device resistance and then ideal device current using the simulator provided voltage value. Next, the Johnson-Nyquist noise is sampled using the *$rdist_normal()* and *$sqrt()* system functions to compute the instantaneous RMS noise voltage from (12). The resulting noise current is subsequently calculated by division by the device's resistance. Next, shot noise is calculated. To do this, the current is converted into a flow rate of electrons per second using the *$abs()* system function and division by the elemental charge value. The absolute flow rate of electrons per second is then used to sample the shot noise from a normal distribution using *$rdist_normal()*. This is done due to a bug in Siemens Symphony AMS simulator. The Poisson process can be approximated to a normal distribution for current flows > 1fA, which is sufficient for device modelling. Finally, the total device current is computed by summing the shot noise and Johnson-Nyquist noise currents (15).

Equations (18), (19), and (20) are precomputed at module instantiation out of the simulation "hot loop" as to have no impact on simulation performance. This also applies to the calculation of fitting parameter $\lambda$ from (9) and $4 \cdot k_B \cdot T \cdot \Delta f$ from (12), which are computed only once at module instantiation.

*Table 1. memristor Verilog-AMS behavioural model parameters.*

| Parameter | Meaning | Units |
|---|---|---|
| $D_l$ | Device length | m |
| $D_A$ | Device Area | $m^2$ |
| $\rho_{LRS}$ | On state bulk resistivity | Ωm |
| $\rho_{HRS}$ | Off state bulk resistivity | Ωm |
| $\varepsilon_{on}$ | Switching threshold field | V/m |
| $\varepsilon_{off}$ | Switching threshold field | V/m |
| $K_{on}$ | Switching speed scalar | X |
| $K_{off}$ | Switching speed scalar | X |
| $a_{on}$ | Switching speed exponent | X |
| $a_{off}$ | Switching speed exponent | X |
| T | Temperature | K |
| $TCR_{LRS}$ | Temperature Coefficient of Resistance | 1/K |
| $TCR_{HRS}$ | Temperature Coefficient of Resistance | 1/K |
| $W_{start}$ | Starting state | m |

### 3.1.4 Testing

To better understand memristor device behaviours and verify the behavioural model functionality, the developed Verilog-AMS model is subjected to a variety of tests. Testing seeks to develop an intuitive understanding of device switching behaviour and device to device variation modelling. In addition to verifying the device model against physical device behaviours from literature. This is important as replicating the key characteristic device behaviours is foundational to the aim of assessing memristor based RC implementation.

The initial model configuration, Table 2, is chosen to represent a theoretical high switching speed copper electrode extrinsic device. Providing a suitable baseline model configuration with which to verify against experimental observations from literature. A positive $TCR_{LRS}$ is used to model the metallic nanobridges temperature dependent resistance, and a negative $TCR_{HRS}$ to model the SiOx dielectrics temperature dependent resistance. With $TCR_{HRS}$ calculated based off results from [15]; however, it is typically highly variable and based on specific manufacturing techniques, lattice structure, and temperature range [15].

*Table 2. Memristor model configuration parameters for I-V sweep.*

| $D_l$ | $D_a$ | $\rho_{LRS}$ | $\rho_{HRS}$ | $\varepsilon_{on}$ | $\varepsilon_{off}$ | $K_{on}$ | $K_{off}$ | $a_{on}$ | $a_{off}$ | T | $W_{start}$ | $TCR_{LRS}$ | $TCR_{HRS}$ |
|---|---|---|---|---|---|---|---|---|---|---|---|---|---|
| 5, 7.5, 10 | 1E-14 | 1E-2 | 1E2 | 1E7 | -1E7 | 5E-5 | -5E-5 | 7 | 7 | 293.15 | $0.5D_l$ | 3.8E-3 | -4.4E-3 |

#### 3.1.4a Device Switching Behaviour

The first test is to perform an I-V sweep of the memristor device. This is done by applying a triangular voltage waveform across the device which sweeps from +3.3V to -3.3V with a period of 1 ns, Fig.11a. The resulting current waveform can be seen in Fig.11b where the device parameters from Table 2 are used with $D_l$ of 10nm with noise sources disabled. Device switching behaviour results in transitions from its HRS to its LRS. The device can be seen to switch into the LRS at 0.3ns, evidanced by the sharp increase in current to -0.33uA at 1ns. Changing the device resistance to 10kΩ. The device can then be seen to reset and switch back into its HRS when a negative voltage is applied across the device. This transition is apparent by the sudden reduction in current at 0.8ns where the memristor transitions to a resistance of 10MΩ. This demonstrates the correct switching behaviour of SiOx devices, which the switching speed is exponential with increase in applied voltage [9].

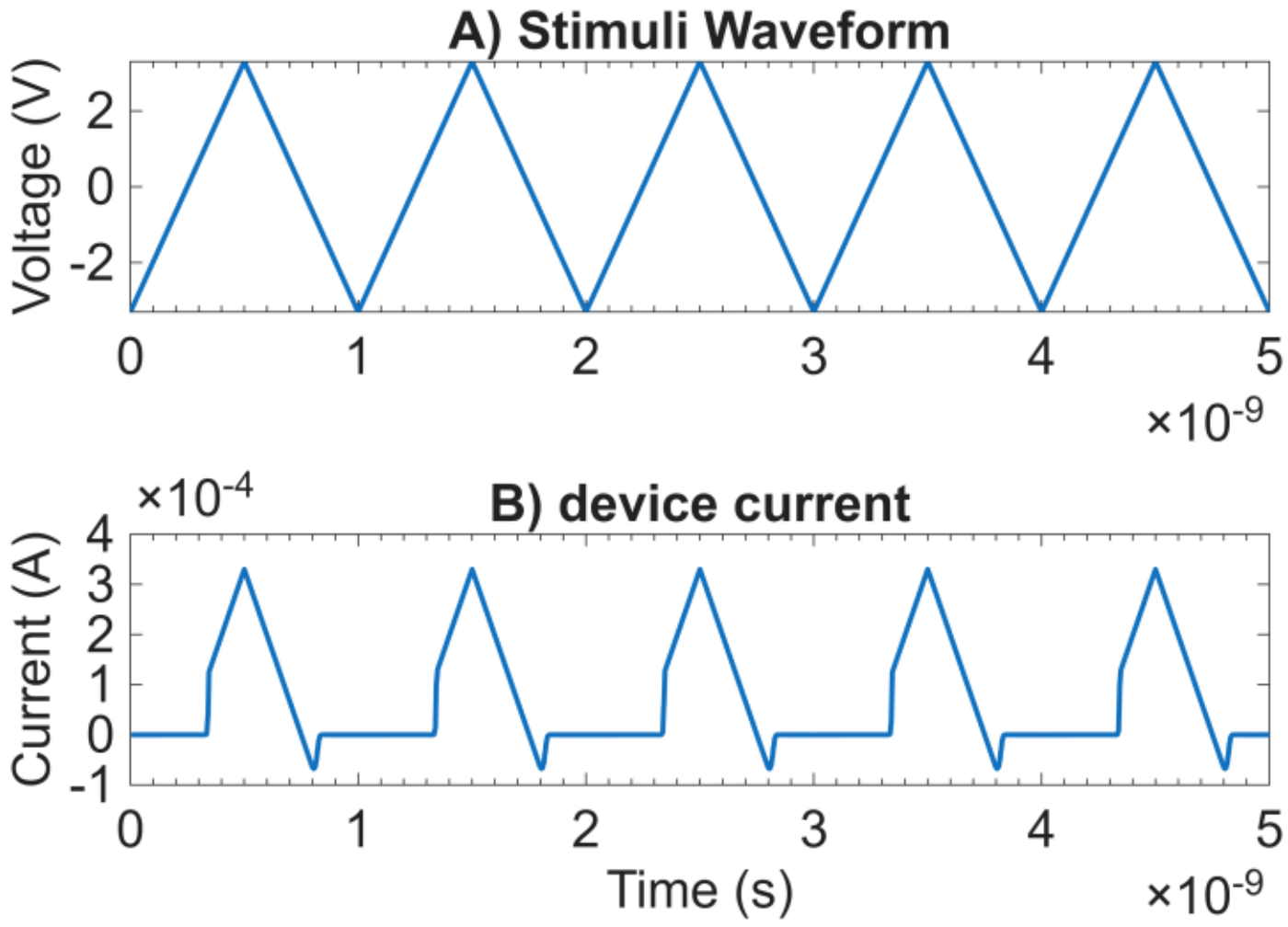

*Figure 11. voltage waveform (a), and device current (b).*

The I-V sweep results from Fig.11b are plotted as an I-V plot in Fig.12, showing the I-V relationship for 1 million device switching cycles. Here it can be seen that the device switches symmetrically with $V_{set} = 1.26V$ and $V_{reset} = -1.26V$, consistent with the models symmetrical switching parameters, Table 2. Fig.12 also shows the sudden switching which exponentially increases with voltage in line with physical device switching shown in Fig.14b [11, 18, 10, 34]. The main difference between the developed model and physical device I-V shapes can be attributed to the model's assumption of ohmic conduction. Otherwise, the developed model has demonstrated the successful modelling of the non-linear switching behaviour of SiOx ReRAM devices.

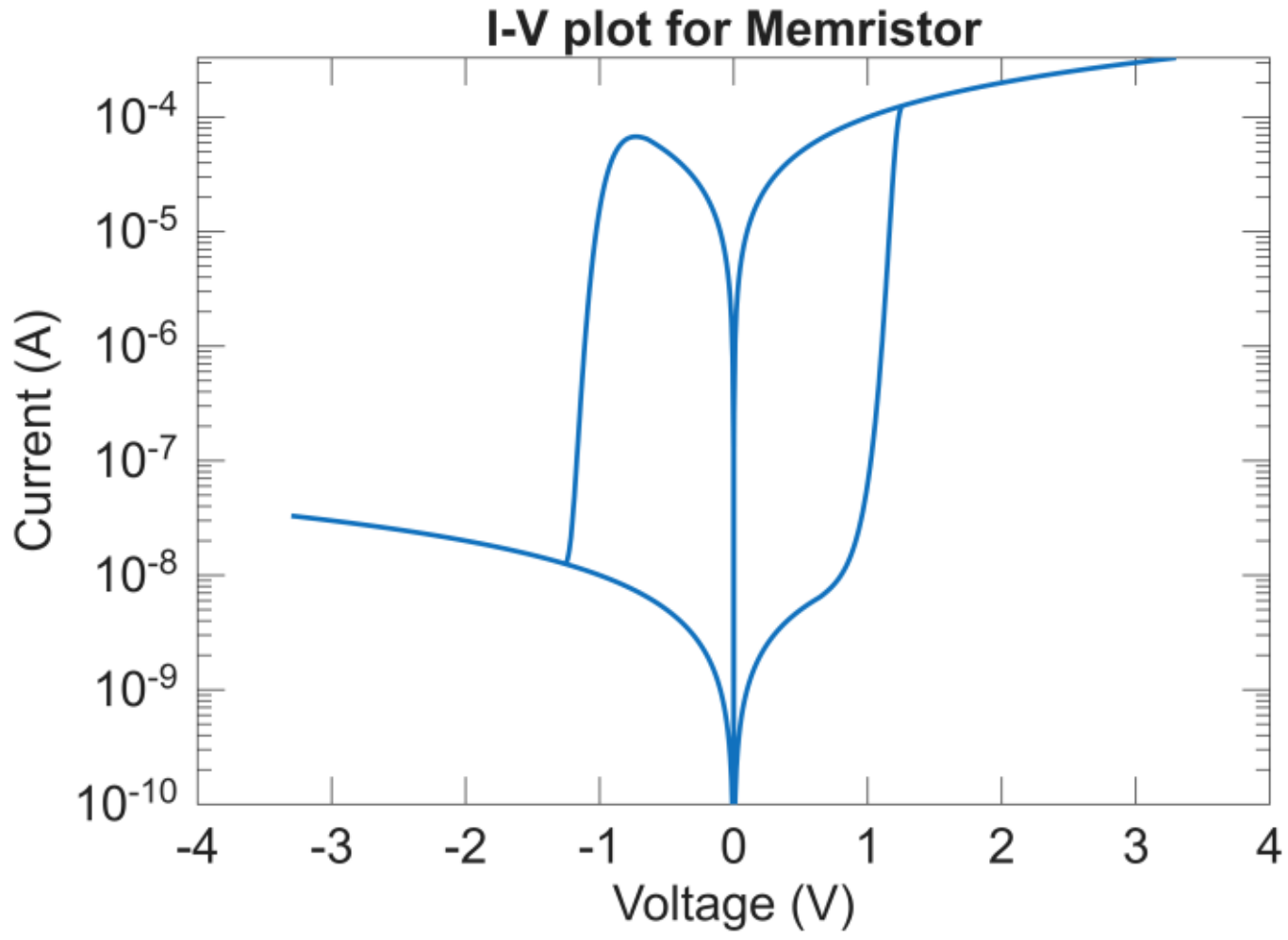

*Figure 12. I-V plot of Fig.11b of 1 million device switching cycles.*

#### 3.1.4b Device Oxide Thickness Variations

The model's ability to replicate oxide thickness variations is tested next, with this being an important functionality for modelling device to device variance for reservoir implementation. This is measured by using the device configurations from Table 2 and performing the same I-V sweep as described in Section 3.1.4a. The results of this are presented in Fig.13, which shows a clear variation in both LRS and HRS resistances and switching behaviour.

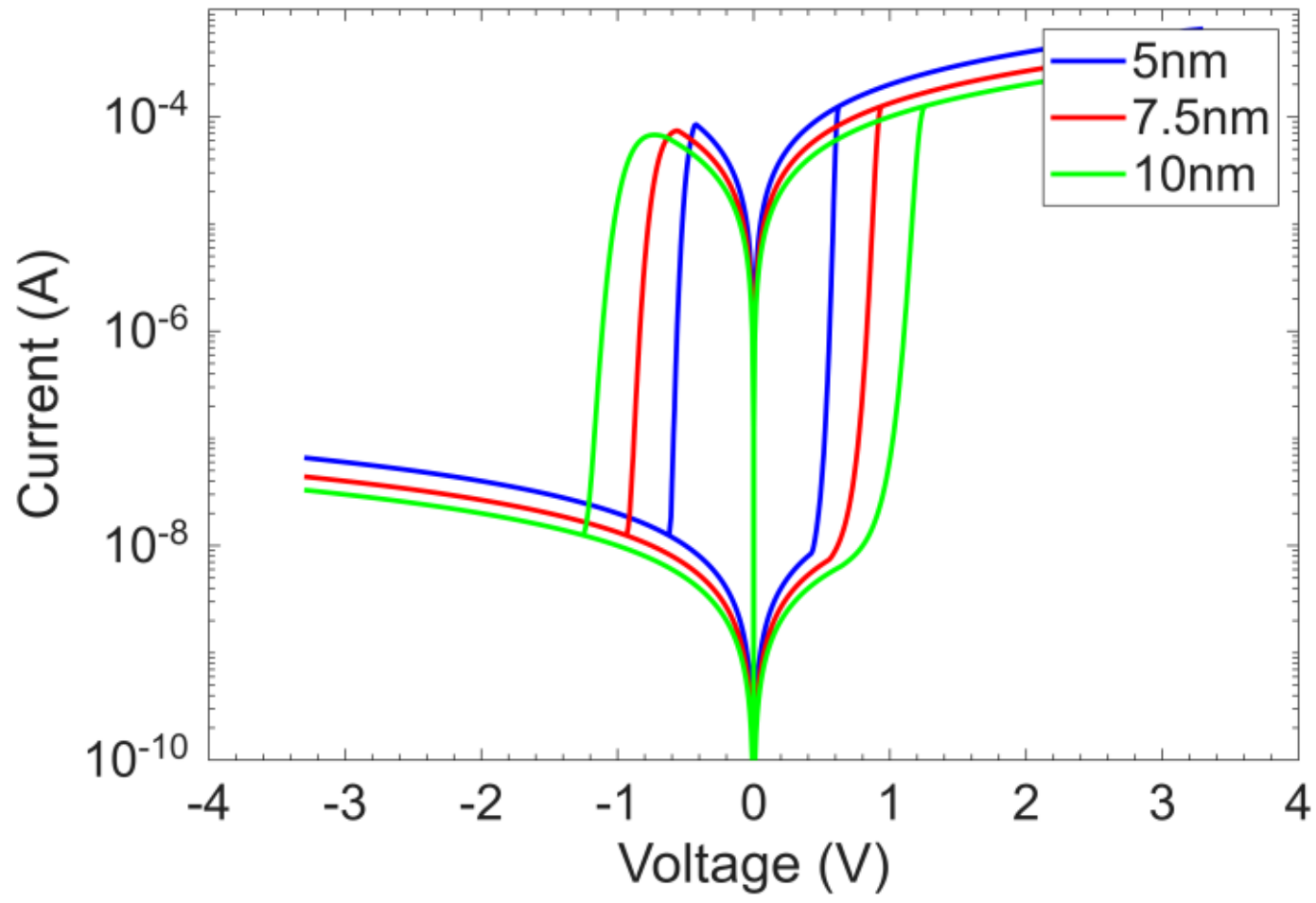

*Figure 13. I-V plot for device configurations from table 2 with oxide thicknesses of 5nm, 7.5nm, and 10nm.*

The decrease in switching voltage requirement with oxide thickness can be attributed to the E field driven switching [10, 9]. The thinner oxide results in a stronger applied E field for a given voltage, and thus the redox reaction that drives switching is faster. This can be seen in Fig.14 where a reduction in oxide thickness results in a reduction of required switching voltage. The developed model manages to replicate the general scaling, with a reduction of $V_{set}$ by 50% for an equivalent reduction in oxide thickness. However, Fig.14 indicates physical devices show non-linear scaling with reduction in oxide thickness, consistent with E field driven switching. Despite this, the models thickness variation effects remain accurate over small ranges of $\pm10\%$ device thickness. Which is adequate for device to device variation modelling for reservoir design, fulfilling the deliverable requirement $\boldsymbol{M_5}$.

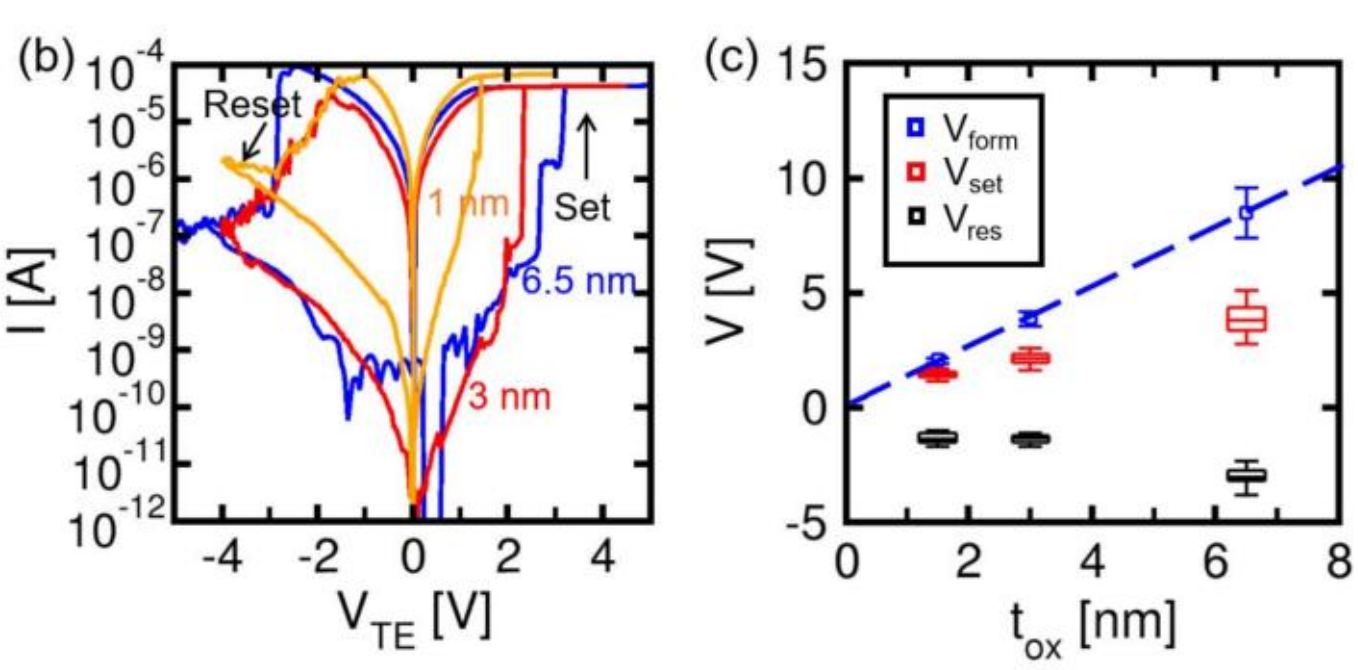

*Figure 14. figures 2b and 2c showing I-V sweep with varying oxide thickness (b) and plot of oxide thickness vs Vset and Vreset (c) for manufactured SiOx devices [10].*

#### 3.1.4c Temperature Variation Modelling

Modelling the temperature effects on device operation is part of the deliverable stretch requirement $\boldsymbol{M_6}$. This is because it is important to properly investigate memristor operation across the range of expected operating temperatures for modern CPUs to comprehensively assess RC BPU applicability.

Temperature related device behaviour variation is measured using the same I-V sweep from Section 3.1.4a from +3.3V to -3.3V

with a 1ns period. The device configuration from Table 2 is used with $D_l = 10\ nm$ and all noise sources disabled. The model is tested at the temperatures of 0C, 75C, and 115C; which are representative of the expected range of operating temperatures in commercial CPU designs.

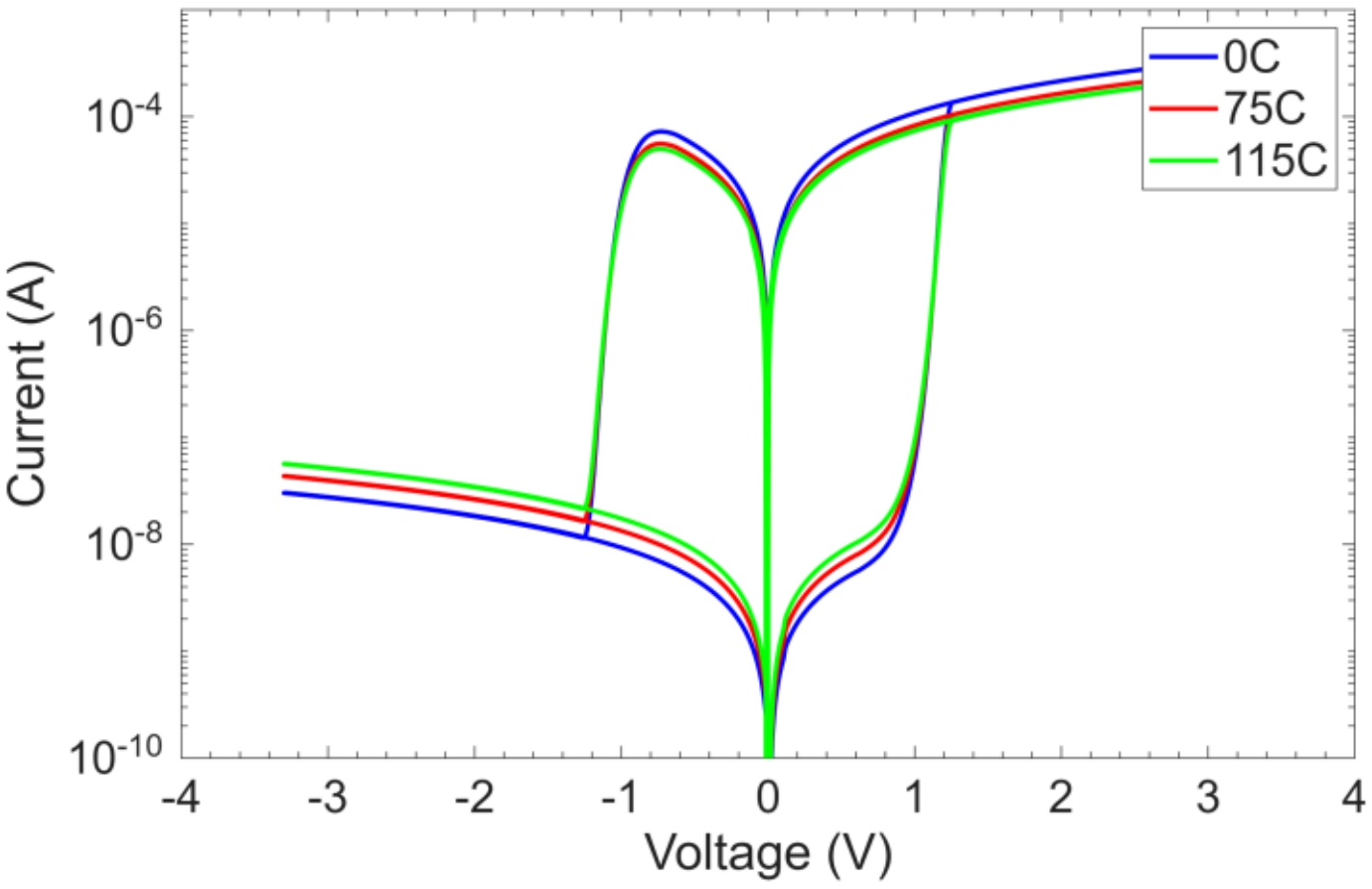

*Figure 15. I-V plot for device configuration from Table 2 with operating temperatures of 0C, 75C, and 115C.*

The behavioural model demonstrates the intended temperature dependence on HRS and LRS resistances, achieving deliverable requirement $\boldsymbol{M_6}$. The HRS resistance reduces as temperature increases due to conduction being predominantly through defects in the SiOx structure (Schottky emission). While the LRS resistance increases with temperature as conduction is predominantly through the metallic nanobridges formed during device set operation (ohmic). This is consistent with experimental evidence from [15, 10], which shows an increase in LRS resistance due to ohmic conduction and a decrease in HRS resistance due to Schottky emission and trap assisted tunnelling. The variation in LRS and HRS resistances with temperature is significant and likely to be a significant barrier for memristor RC implementation in actual devices due to its variability.

#### *3.1.4d Noise Source Modelling*

The model's ability to replicate noise sources is important to fulfil deliverable requirement $\boldsymbol{M_1}$ and to build an intuitive understanding of the requirements for a low noise memristor device. Too much noise from either device switching or conduction would necessitate operation of devices parallel, resulting in increased area usage and power consumption. Which would make RC less favourable for implementation.

Switching and conduction noise is investigated by performing an I-V sweep as described in Section 3.1.4a between +3.3V and -3.3V with a period of 1ns. The model config from Table 2 is initially tested with $D_l = 10\ nm$, all noise sources enabled, and a frequency bandwidth of 10GHz for thermal noise.

The results, Fig.16, show that switching noise is noticeable but has a minimal impact on set and reset voltages as $V_{set}$ and $V_{reset}$ are consistent to within $\pm 6\%$ of 1.26V for $V_{set}$ and -1.26V for $V_{reset}$. Conversely, conduction noise is substantial; especially in the HRS state where thermal noise results in significant device current fluctuations despite being band limited to 10GHz. This could have a large impact on reservoir performance, necessitating the use of ReRAM devices with a significantly lower HRS resistance.

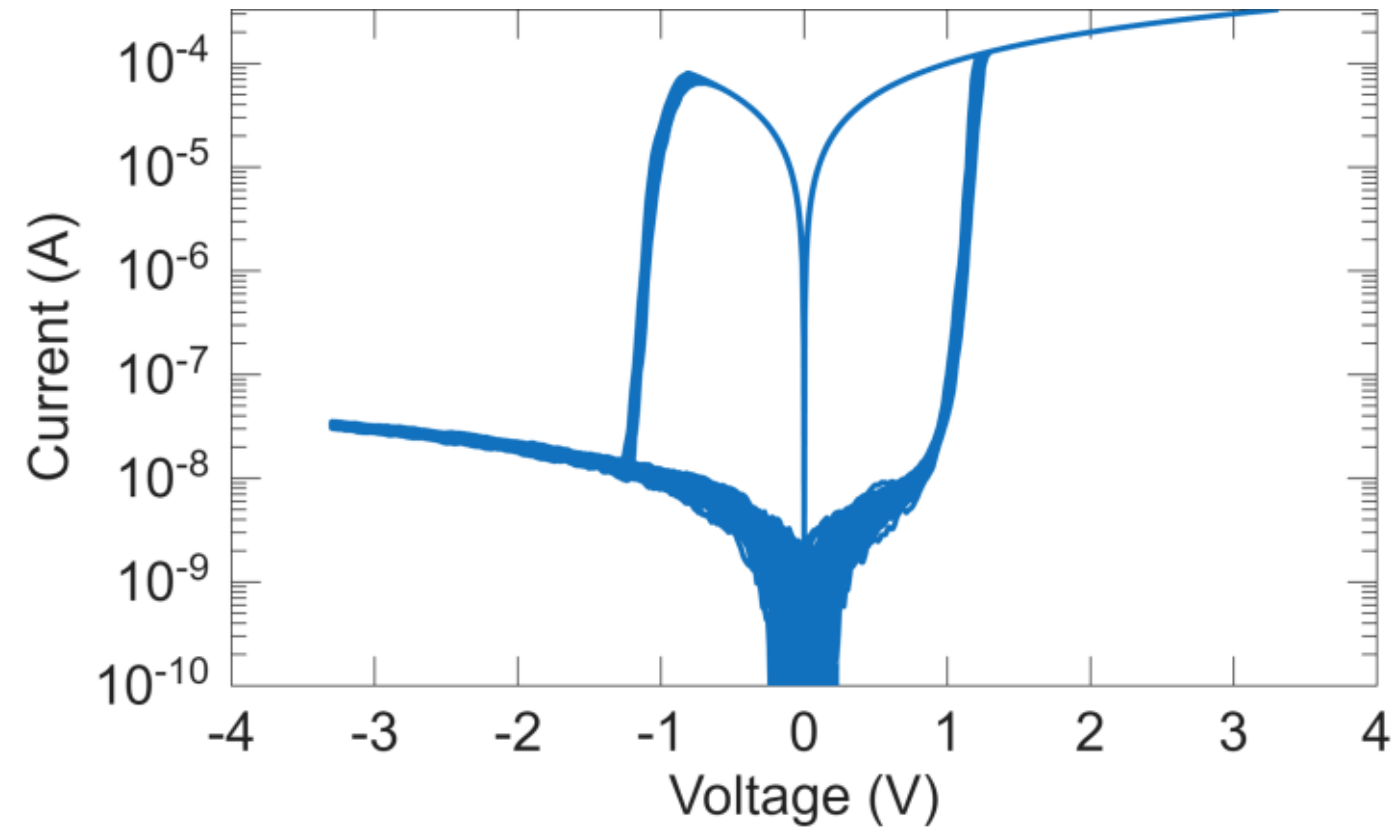

*Figure 16. I-V plot for the model configuration from Table 2 over 1000 switching cycles.*

Noise both in device switching and conduction can be reduced by reducing oxide thickness [11]. Additionally, the resistance of the HRS can be reduced to mitigate the impact of thermal noise at the cost of reduced ON/OFF ratio. An alternative model configuration is given in Table 3, which is designed to reduce both the switching noise and conduction noise. Switching noise is reduced by decreasing the thickness of the oxide layer, and thermal conduction noise is reduced by decreasing the resistance of the HRS.

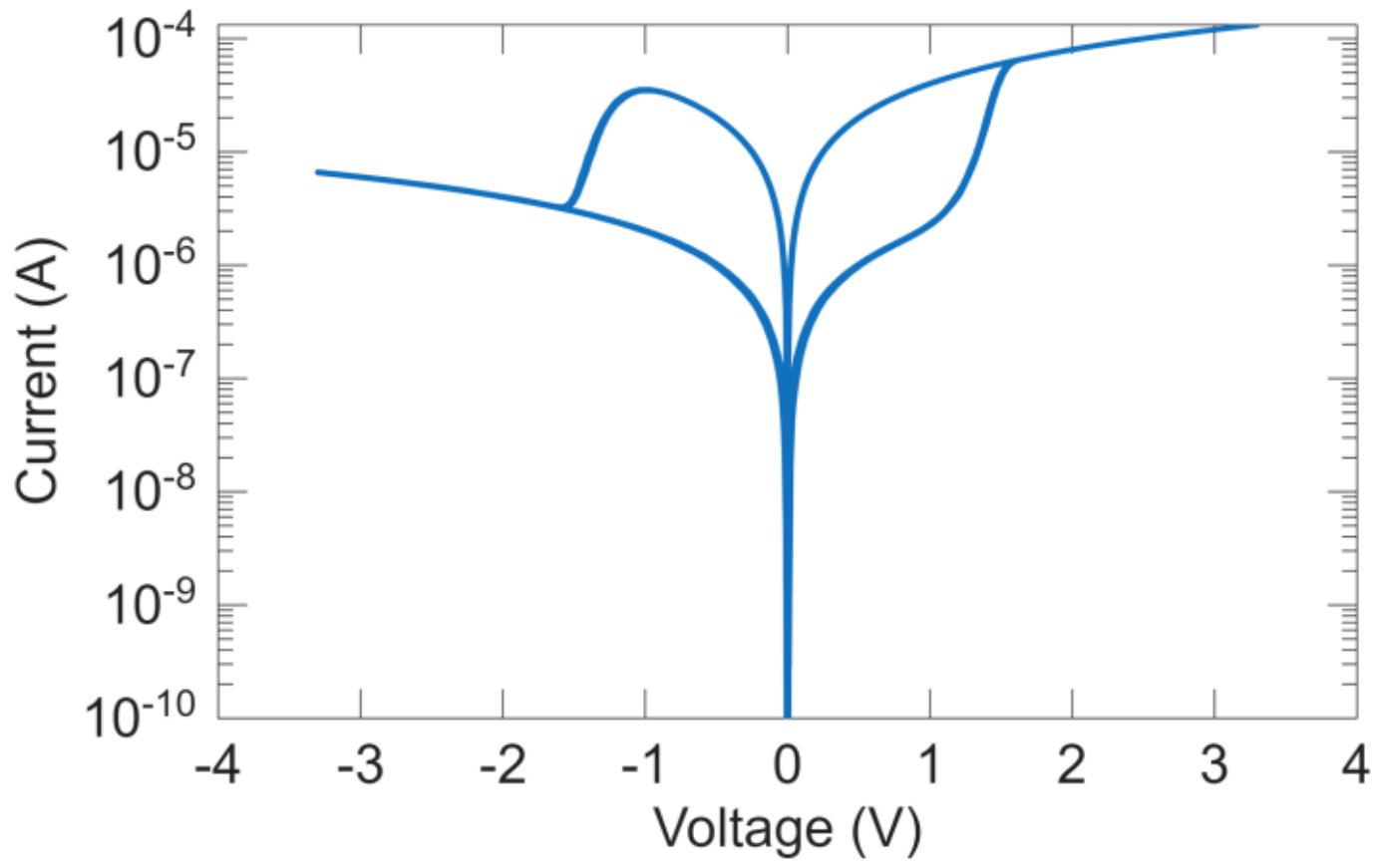

*Figure 17. I-V plot for the model configuration from Table 3 over 1000 switching cycles.*

The modified memristor model configurations shows a significant reduction in both conduction and switching noise, Fig.17. $V_{set}$ and $V_{reset}$ are highly consistent to within $\pm 2.7\%$ of 1.59V for $V_{set}$ and $\pm 2.7\%$ of -1.59V for $V_{reset}$. Representing an over 2x increase in consistency of device switching compared to the initial noise model, highlighting the importance of device engineering to achieve low noise. The conduction noise is also significantly reduced as a result of the lower HRS resistance. Providing highly regular and consistent current flows at equivalent voltages in comparison to the original noisy model.

These results indicate that it is more than possible for device engineering to achieve highly consistent switching behaviour as indicated in [11]. Paving the way for implementation of RC using memristor devices as the building block for reservoir implementation.

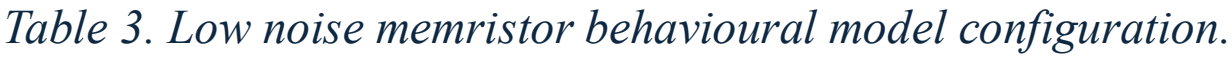
*Table 3. Low noise memristor behavioural model configuration.*

| $D_l$ | $D_a$ | $\rho_{LRS}$ | $\rho_{HRS}$ | $\varepsilon_{on}$ | $\varepsilon_{off}$ | $K_{on}$ | $K_{off}$ | $a_{on}$ | $a_{off}$ | T | $W_{start}$ | $TCR_{LRS}$ | $TCR_{HRS}$ |
|---|---|---|---|---|---|---|---|---|---|---|---|---|---|
| 1 | 1E-14 | 2.5E-1 | 5 | 1E7 | -1E7 | 5E-14 | -5E-14 | 7 | 7 | 293.15 | $0.5D_l$ | 3.8E-3 | -4.4E-3 |

#### *3.1.4e Operating Point Sweep*

The next step is to understand the devices switching behaviour across a range of operating frequencies and voltages. This gives an understanding of the voltage amplitude and hold time required to fully switch the device state (full switching). And the range of voltage amplitudes and hold times that result in the device only partially switching (partial switching). Understanding the partial and full switching regions of operation is necessary to inform reservoir design. With the eventual reservoir design necessitating device operation in the partial switching region to minimise instability.

Two different device configurations are proposed, Table 4, based on prior testing. The first configuration emphasises full switching and the second emphasises partial switching. Both are optimised for noise immunity with low ON/OFF ratios of 20, HRS resistance of 550kΩ, and oxide thickness of 1nm. The low oxide thickness is intended to improve cycle to cycle variations in switching as evidenced by [10, 11] while the low resistivity of HRS and LRS is intended to reduce the impact of thermal noise.

*Table 4. memristor Verilog-AMS behavioural model parameters.*

| Parameter | Digital Switching | Analogue Switching |
|---|---|---|
| $D_l$ (nm) | 1 | 1 |
| $D_A$ ($m^2$) | $1 \times 10^{-14}$ | $1 \times 10^{-14}$ |
| $\rho_{LRS}$ | $2.5 \times 10^{-1}$ | $2.5 \times 10^{-1}$ |
| $\rho_{HRS}$ | 5 | 5 |
| $\varepsilon_{on}$ | $1 \times 10^{7}$ | $1 \times 10^{7}$ |
| $\varepsilon_{off}$ | $-1 \times 10^{7}$ | $-1 \times 10^{7}$ |
| $K_{on}$ | $5 \times 10^{-14}$ | $2 \times 10^{-5}$ |
| $K_{off}$ | $-5 \times 10^{-14}$ | $-2 \times 10^{-5}$ |
| $a_{on}$ | 7 | 3 |
| $a_{off}$ | 7 | 3 |
| T | 293.15 | 293.15 |
| $TCR_{LRS}$ | 0.0038 | 0.0038 |
| $TCR_{HRS}$ | -0.0044 | -0.0044 |
| $W_{start}$ | 0.5 | 0.5 |

Operating point sweep testing is performed by measuring the ON/OFF ratio achieved at different combinations of operating voltage amplitude and frequency. A smoothed square wave (21) is used to switch the memristor device, which is chosen as it minimises discontinuities for improved simulation performance. It also matches more closely with real world high frequency square waves which are often non-ideal due to rise and fall time.

$$V(t) = \tanh(\sin(t) * 5) \quad (21)$$

The operating point sweep is performed by a python script based verification flow as part of the project's methodology. For each voltage amplitude and pulse duration combination, a transient simulation is ran using Siemens Symphony AMS simulator. The devices are then driven by the smoothed square wave while the maximum and minimum observed resistance is recorded. The python script based verification flow then aggregates this information from each simulation run into a series of .csv files. The simulation results can then be processed using MATLAB to generate a matrix of observed ON/OFF ratios across a range of voltage amplitude and pulse durations.

ON/OFF ratio testing is performed at 330 different voltages from ±0.01V to ±3.30V. With 100 different pulse widths tested, ranging from 1ns to 100ns at each different voltage. The result is plottable as a surface mesh, Fig.19, to visualise the devices switching across a range of operating voltages and frequencies to provide insight into partial and full switching regions.

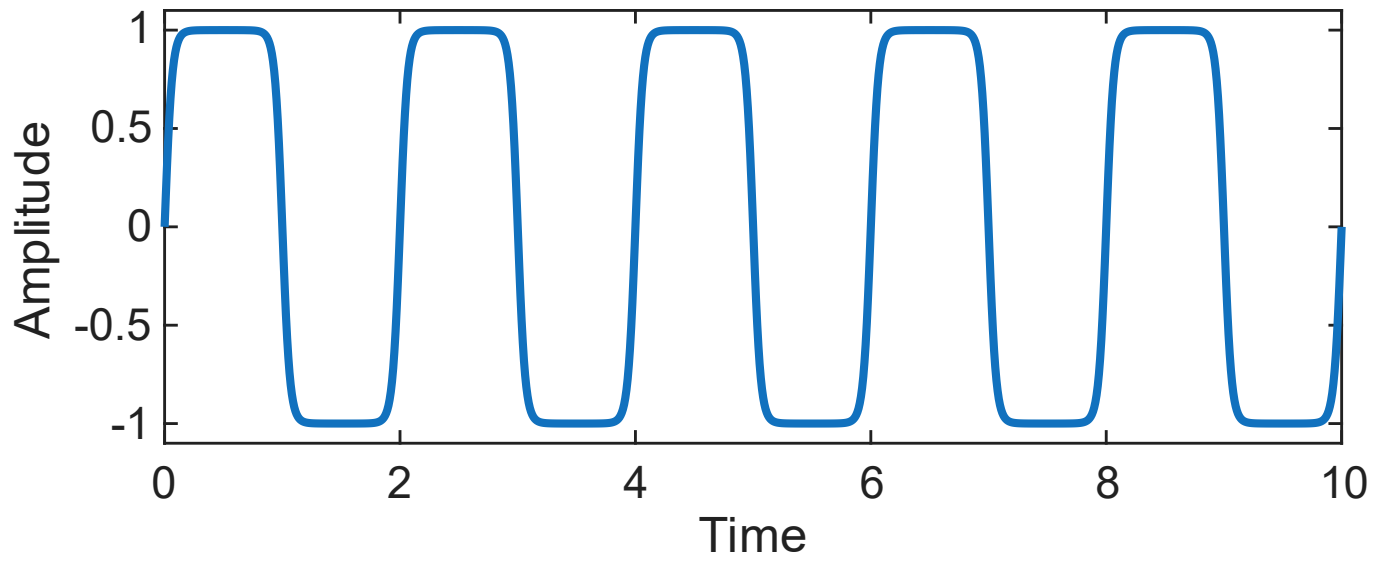

*Figure 18. Smoothed square wave from (21).*

The threshold voltages required for complete device set and reset exhibit an inverse relationship with the square-wave pulse width. In the digital switching configuration, this relationship is approximately proportional to the reciprocal of the pulse width (Fig. 19c). This behaviour is expected because switching in this regime is primarily voltage-dependent rather than pulse-width-dependent, as indicated by the high $a_{\text{on}}$and $a_{\text{off}}$ values. In contrast, the analogue switching configuration demonstrates a stronger dependence on pulse width, with the threshold voltage following a relationship closer to an inverse-square or inverse-cube trend (Fig. 19d). This behaviour arises from the comparatively lower $a_{\text{on}}$and $a_{\text{off}}$ values, together with the higher $k_{\text{on}}$and $k_{\text{off}}$ values. Which increase the influence of pulse width on state switching.

Both model configurations exhibit distinct operating regions corresponding to partial and full switching (Figs. 19c and 19d). The digital switching configuration demonstrates lower full-switching voltages across all pulse widths, while also exhibiting higher partial-switching threshold voltages compared to the analogue switching configuration. For both configurations, the partial-switching threshold voltage follows the same general dependence on pulse width as the full-switching voltage.

The primary difference between the two configurations lies in the voltage range over which partial switching occurs. In the digital switching configuration, the partial-switching region is relatively narrow, maintaining an approximately constant width of 0.25 V for pulse widths between 100 ns and 10 ns. At shorter pulse widths, the width of the partial-switching region increases, reaching approximately 0.45 V at a pulse width of 1 ns. In contrast, the analogue switching configuration exhibits a substantially wider voltage range for partial switching (Fig. 19d). For pulse widths between 100 ns and 20 ns, the partial-switching region remains highly consistent, extending from 0.06 V to 0.76 V, corresponding to a width of 0.7 V. Below 15 ns, the width of the partial-switching region increases rapidly with decreasing pulse width, expanding from 0.7 V to approximately 3 V at a pulse width of 1 ns. This provides a significantly broader operating voltage range over which switching remains partial when compared with the digital switching configuration.

The analogue switching configuration is preferable for use as part of the reservoir design. This is due to its significantly wider range of partial switching voltages, in addition to its over 4x lower partial switching threshold voltage. Allowing it to operate at lower voltages and higher speeds in comparison to the digital switching configuration.

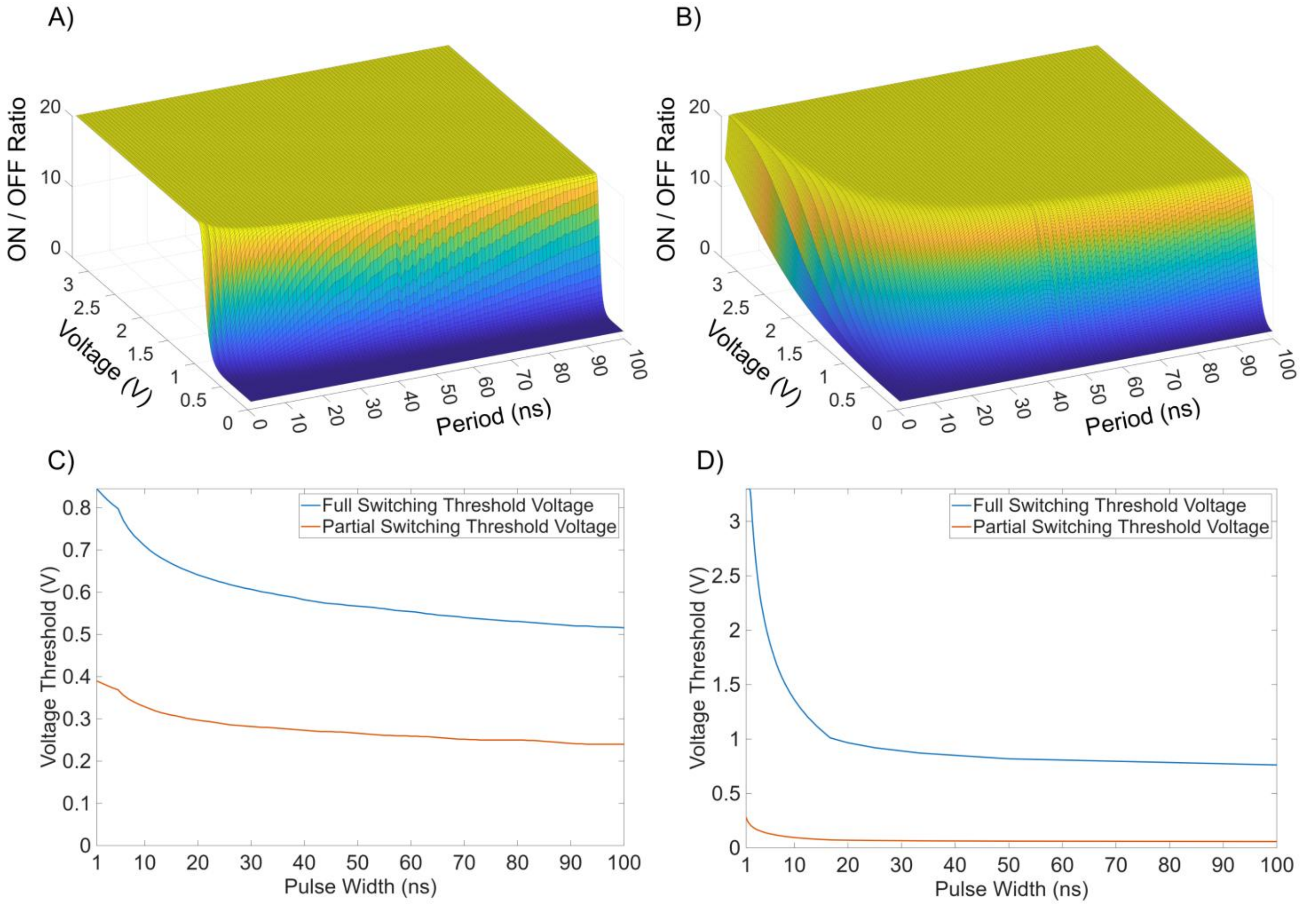

*Figure 19. ON/OFF ratio measured by voltage frequency sweep for digital switching device configuration (a) and analogue switching device configuration (b). And the partial and full switching threshold voltages of the digital switching device configuration (c) and the analogue switching device configuration (d).*

## 3.2 RC BPU Development

A novel RC architecture, Fig.22, comprised of four main subcomponents has been developed for the task of BP. The design can be grouped into two distinct sections. The analogue logic comprising the reservoir and its associated interfaces, and the digital logic comprising the perceptron readout layer and BPU control logic.

The developed RC BPU has a two stage prediction pipeline with the reservoir as the first stage and the perceptron readout layer as the second stage. During the first stage, the binary encoded input to the RC BPU is converted to analogue voltages by the reservoir input layer and is input into the reservoir. During the second stage, the reservoir output voltages are read by the reservoir output layer ADCs and the readout layer perceptron processes this to form a prediction. The timing diagram, Fig.20, shows how the RC BPU produces a prediction from input data.

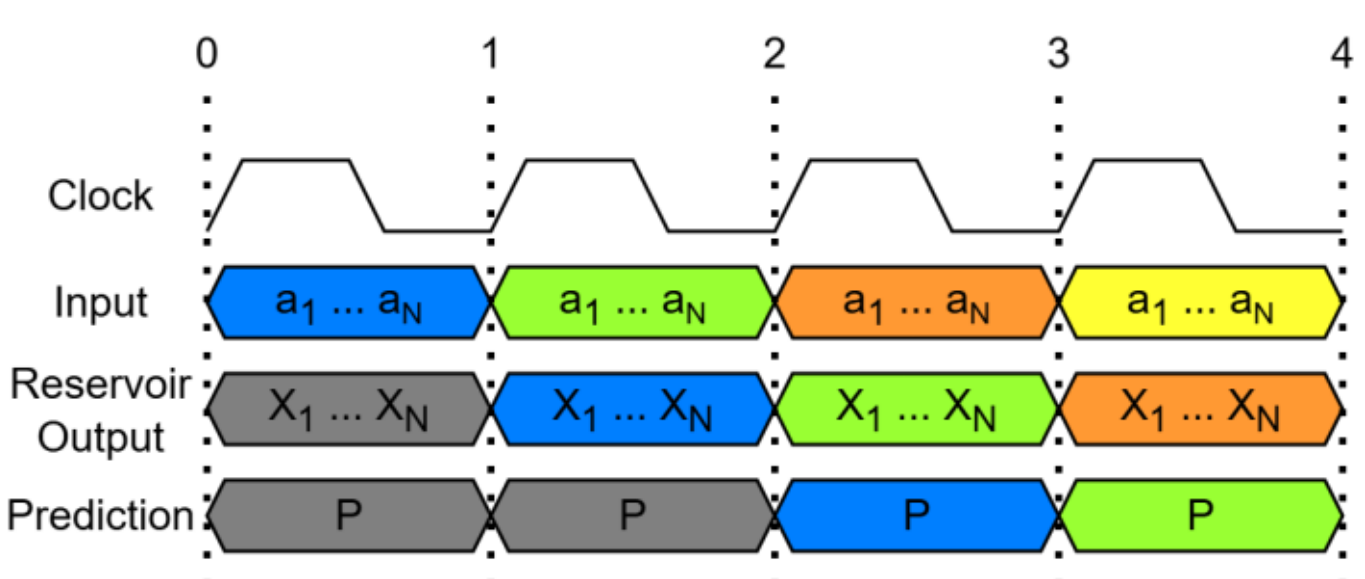

*Figure 20. Timing diagram for RC BPU prediction pipeline.*

The developed RC BPU design also has an in-situ training capability to correct the perceptron's weights in real time in the case of a misprediction. The RC BPU uses a feedback signal to prompt training of the perceptron's weights on signal assertion. When asserted, the perceptron's weights are updated using the previous cycles prediction inputs. This is described in detail in Section 3.4.

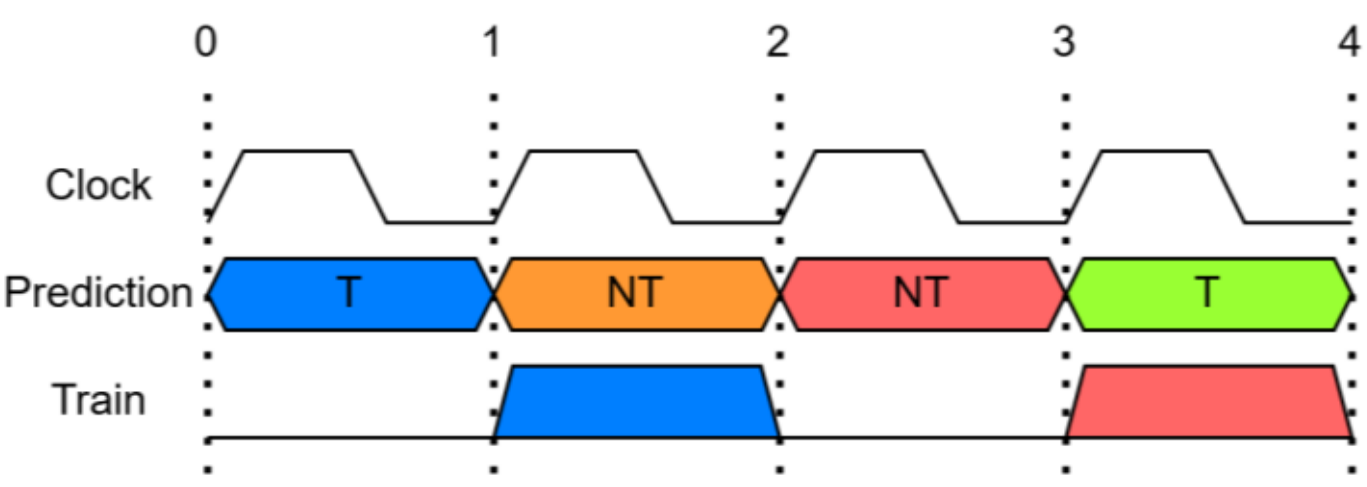

*Figure 21. Timing diagram for RC BPU training stage.*

RC BPU training can be briefly understood by Fig. 21 where both the blue and red predictions are incorrect and result in the assertion of the RC BPU train enable signal. The signal is asserted on the positive clock edge after prediction output from RC BPU, with training complete on the subsequent clock edge. Training thus acts as an optional extra stage in the prediction pipeline which happens after prediction is complete.

The developed design represents a compromise between the original project requirements of 1ns prediction latency and the feasibility of future design realisation. This resulted in the adjustment of project aims, forming $\boldsymbol{P_1}$ and $\boldsymbol{P_2}$ in its place. Both of which are achieved by the pipelining of prediction in the developed design.

### 3.2.1 Reservoir Input

Data input to the RC BPU prediction pipeline takes the form of binary encoded integers, which can be thought of as bit vectors. The reservoir input layer is responsible for taking the bit vector inputs and converting them into an appropriate series of input voltages at each input node of the reservoir.

To simplify the design and better position it for high speed operation, a direct mapping of the input bit vector to a reservoir input is chosen. Each binary value in the input bit vector is directly converted to a voltage at a corresponding reservoir input node using 1-bit DACs. Binary values are converted from logic values of 0 or 1 to a non-return to zero (NRZ) encoded analogue signal of either 0V or $V_{DD}$. The bit width of the RC BPU prediction pipeline input is thus determined by the number of reservoir input nodes.

The simple direct mapping of digital inputs to NRZ analogue signals is a key design decision to enable high frequency operation, driven by the project aims $\boldsymbol{P_1}$ and $\boldsymbol{P_2}$.

### 3.2.2 Reservoir Design

The reservoir is the dynamical system responsible for the bulk of information processing in the first stage of the prediction pipeline. Its non-linear dynamics are used to map the prediction pipelines input data into a higher dimensional space to enhance linear separability.

The proposed reservoir architecture takes the form of a potential divider network of individual memristor devices, Fig.22. The reservoir is driven by the reservoir input layer, Section 3.2.1, and its output read by the reservoir output layer, Section 3.2.3. The reservoir operates through the collective nonlinear behaviour of interconnected voltage driven memristor devices to process inputs as described in Section 2.1.

The simplicity of the reservoir design as a potential divider network aims to simplify physical implementation by enabling the use of dense crossbar arrays. Contributing to the project aim $\boldsymbol{P_4}$. The full reservoir design is covered in Section 3.3 alongside design decision justifications with respect to the projects overall aims.

### 3.2.3 Reservoir Output

The reservoir output layer constitutes the first component of the second stage of the prediction pipeline. It samples and digitises the voltages at each reservoir output node at every positive clock edge using 4-bit flash ADCs.

In general, the number of reservoir outputs is expected to be significantly greater than the number of inputs. For instance, an 8-input reservoir containing 100 memristors in a low-density configuration could provide up to 40 output nodes. This introduces a limitation as the large number of outputs required to capture a sufficiently high-dimensional representation of the reservoir state would result in unacceptable chip area usage.

To address these constraints, flash ADCs are proposed for future implementation. They provide an effective compromise between the conversion speed required for 1 GHz operation and the cumulative chip area overhead associated with integrating tens of ADCs.

Compared with alternative ADC architectures, such as successive approximation register (SAR), sigma-delta, and dual-slope converters, flash ADCs enable substantially higher operating speeds. This performance advantage arises from their use of a resistor ladder and parallel comparators, which allow near-instantaneous conversion. The primary disadvantages of flash ADCs are their relatively high power consumption and large chip area requirements. In this work, these limitations are mitigated by constraining the reservoir output voltage sampling to 4-bit quantisation, reflecting the low-resolution operation required for high-speed area efficient silicon implementation. Fulfilling the projects aim to assess the physical implementation and application as accurately as possible in simulation. In addition to providing a practical compromise to reach the project aims $\boldsymbol{P_1}$ and $\boldsymbol{P_2}$.

### 3.2.4 Trainable Perceptron

The trainable perceptron forms the final stage of the prediction pipeline. Its primary function is to perform binary classification by computing the weighted sum of the digitised reservoir output voltages and determining the output prediction from the sign of the resulting value. In addition to classification, the perceptron provides the in-situ training capability of the RC BPU, enabling the prediction model to adapt dynamically during operation.

The perceptron operates on the quantised outputs generated by the reservoir output layer. Each reservoir node contributes to the final prediction according to an associated trainable weight. To support online learning, the perceptron incorporates input buffering from the previous clock cycle. This allows it to retain the required reservoir state information from the previous prediction for weight updates in the event of an incorrect prediction, further described in Section 3.4.

A perceptron-based classifier was selected due to its comparatively low hardware complexity and suitability for high-speed implementation. The weighted-sum operation maps efficiently onto digital hardware and introduces minimal latency, making it well suited to integration within the timing constraints of a branch prediction pipeline, $\boldsymbol{P_1}$. Furthermore, the linear nature of the perceptron enables straightforward and resource-efficient online training.

The inclusion of in-situ training capability is essential for adapting to variations in program execution behaviour over time. By updating the perceptron weights directly in hardware following incorrect predictions, the RC BPU can continuously refine its predictive accuracy without requiring offline retraining or external processing resources. This approach enables adaptive learning while maintaining a compact and computationally efficient architecture. Contributing to the project aims $\boldsymbol{P_3}$ and $\boldsymbol{P_4}$.

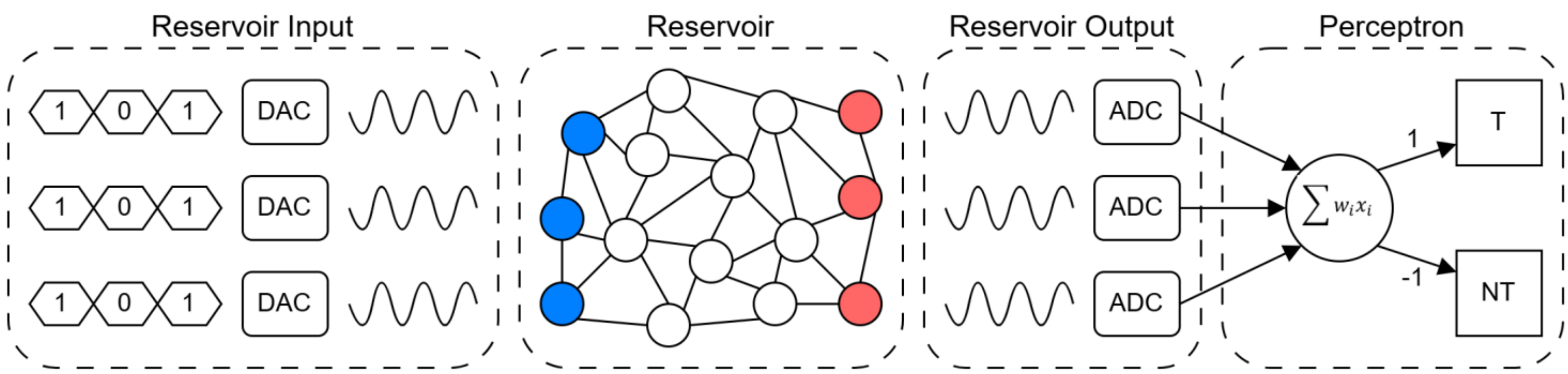

*Figure 22. proposed RC BPU architecture.*

## 3.3 Reservoir

### 3.3.1 Background

The reservoir is the key component of reservoir computing and aptly its namesake. Providing the mechanics and responses that both encodes information into a higher state and "remembers" past inputs. A reservoir can be any system capable of mapping an input signal to a higher dimensional space via fixed non-linear dynamics [12]. This means that for a given starting state and series of inputs, the response of the reservoir should be consistent and identical. Almost any sufficiently complex natural system is capable as a reservoir, with even a bucket of water being usable. Demonstrated by Chrisantha Fernando et al [25], a bucket of water is sufficient as a reservoir, where they applied RC to solve an XOR problem with a single perceptron neuron.

The reservoir can be thought of as a state machine with a state $\vec{X}_k$ at time point k. Its next state is determined by (22), where $\vec{u}_k$ is the input to the reservoir (excitation / stimuli) at time point k. (22) shows that the state of the reservoir at k+1 is determined by its state at k and the input to it at k, explicitly defining the memory of the reservoir.

$$\vec{X}_{k+1} = \vec{f}(\vec{X}_k, \vec{u}_k) \tag{22}$$

The key properties of a good reservoir are possession of many readable states, reproducible and consistent fixed dynamics, and fading memory [12]. The requirement of a large number of readable states is derived from the intended function of the reservoir to project a series of inputs into a higher dimension. More quantifiable states that a reservoir can possess leads to higher the effective computational capacity [12] and is a key to performance [8]. Reproducibility and consistency are also necessary, stemming from the requirement for fixed dynamics [12, 24]. This ensures a "smooth" projection of inputs $\vec{U}_N$ into $\vec{X}$, where changes in $\vec{X}$ are directly proportional and consistent with changes in $\vec{U}_N$. This ensures that the output layer is able to generalise properly as a result of the consistency of inputs projections into the reservoir state [12]. The final property of a good reservoir is fading memory. This is expressed as the requirement that two identical reservoirs with differing states $\vec{X}_k^1$ and $\vec{X}_k^2$ which are given the same input stimuli should converge over time. Such that after n inputs $\vec{X}_{k+n}^1 = \vec{X}_{k+n}^2$, where n is the memory capacity of the reservoir. This means that the influence of the initial reservoir conditions fades over time [12]. This is required to allow generalisation, with the ideal memory capacity being comparable to the workload sequence length for best results [12].

Memristor reservoirs have already been explored in literature, such as Chao Du et al. [26] who utilised volatile memristor devices to perform digit recognition. Achieving 92.1% accuracy of digit classification on MNIST digit dataset using a memristor crossbar reservoir design with 88 devices which are multiplexed. Demonstrating that memristor devices are experimentally promising for reservoir implementation.

Wenxuan Sun et al. [7] demonstrated 90% accuracy for gesture recognition utilising 3D memristor RC, where each memristor serves as a virtual reservoir. Also demonstrating ultra compact 3D vertical memristor devices could perform 21 million predictions per second.

Effectiveness of memristor RC compute at significantly lower speeds than required for this project has been demonstrated to be both feasible and accurate. Ultra high-speed operation is yet to be explored by literature. Reservoir implementation as a memristor network is also yet to be properly explored, with many existing approaches partitioning data. Such as [26] where multiple memristors are use but multiplexed and read chunks of image data instead of the entire image.

### 3.3.2 Design

The reservoir is implemented as the novel configuration of a resistor network of memristors. With nets assigned as input, output, and intermediate. An example design is shown in Fig.23, with 2 input nodes and 6 output nodes. The potential divider network architecture is chosen in part due to the advantages it provides for future reservoir design realisation. It is physically realisable as a memristor crossbar array, which is a highly optimised and dense physical structure that provides high area efficiency and device consistency. Contributing to the project aim $\boldsymbol{P_4}$ and the deliverable aim $\boldsymbol{R_3}$.

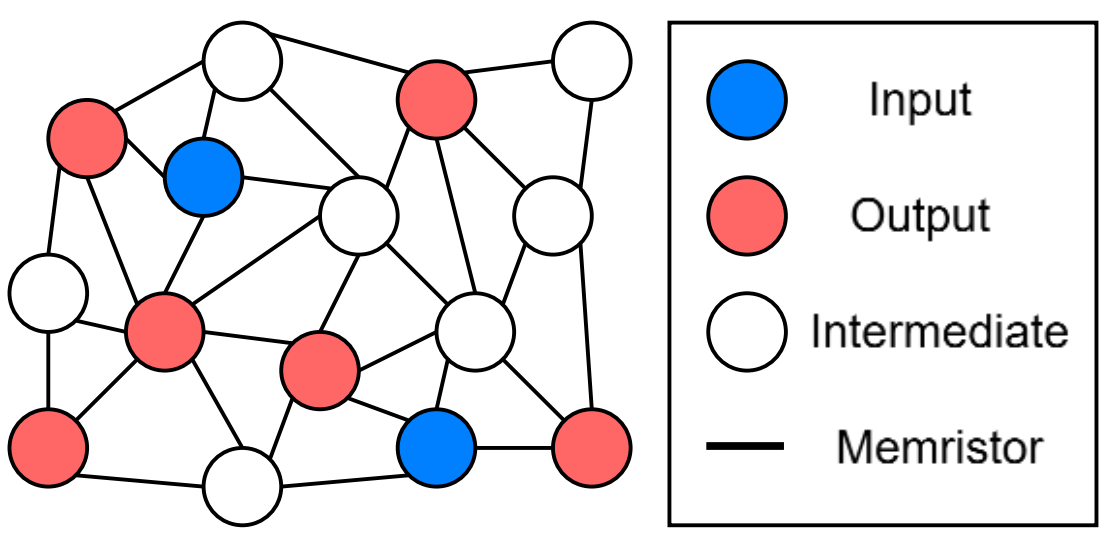

*Figure 23. Reservoir design graph view.*

Most conventional implementations of RC in literature focus on data partitioning; driving many individual memristors in isolation [26, 7, 8]. Where complex data encoding, such as [7] which uses virtual nodes, is used to drive the individual memristor devices. It is hypothesised that a potential divider network architecture will prove to be an enhancement over individual discrete memristor reservoirs. The interconnected system of memristors is hypothesised to increase the number of distinguishable states through the interactions of constituent devices. Resulting in more complex reservoir level non-linear behaviour and additional distinguishable states.

The reservoirs design as a potential divider network results in its switching being driven by the voltage gradient formed across it by the input nodes. This necessitates a minimum of 2 input nodes required to form the voltage gradient. The design targets the use of 1-bit DACs, utilising non return to zero (NRZ) encoding for data where a value of 0 = 0V and a value of 1 = VDD. This significantly simplifies future design realisation, where each input node can be driven by a single transistor. Contributing the project aim $\boldsymbol{P_4}$ and the deliverable requirement for high speed operation, $\boldsymbol{R_1}$.

The reservoirs state is represented by the voltage of each of the output nodes, which is a combination of the constituent memristors resistances and the current input stimulus. As such, the reservoir relaxes to a quiescent state if there is no present input stimuli, a hallmark of a good reservoir per [12]. State readout is performed by measuring the voltage of each output node while under stimulus. This is performed in parallel with reservoir input stimuli, eliminating the need for readout voltage pulse and decreasing the effective latency of the reservoir stage, $\boldsymbol{R_1}$. This is also hypothesised to benefit the richness of reservoir states as it puts a large emphasis of the reservoir state on the current input [12]. The reservoir design is to target state readout using 4-bit flash ADCs. Providing a balance between the requirement for high speed operation for $\boldsymbol{P_1}$ and the ability to practically distinguish a large number of reservoir states.

The reservoir design is also targeting an operating voltage of 3.3V at 1GHz, $\boldsymbol{R_2}$ and $\boldsymbol{R_1}$. 3.3V is chosen as it allows the use of existing on chip power planes that are used for powering external input/output pins, such as for PCIe. This reduces the need for additional on chip infrastructure for future physical design integration per the projects aim $\boldsymbol{P_4}$.

### 3.3.3 Implementation

The reservoir is implemented in VAMS as a behavioural model which is generated by the *resgen* tool, which is described further down, in accordance with configuration parameters. The reservoir module is comprised of three main sections. The first is the net definitions and module port connections. The second is the instantiation of memristor models. And the third is the instantiation of capacitors between each net and ground.

Each node from the graph representation of the reservoir, Fig. 23, is represented by a corresponding *electrical* type net in VAMS. The *electrical* datatype is the simulators native representation of paths of conduction used to simulate analogue electrical behaviours. The edges from the graph representation of the reservoir, the memristor devices, are then instantiated and connected to the *electrical* nets. Each memristor is fully configured with instantiation parameters based on reservoir configuration provided to the *resgen* tool. The final component of the reservoir behavioural model is the capacitors added between each net and ground. These are included as they serve as a mitigation against thermal noise by band limiting the reservoir. As such, their impact on its operation is important for modelling to adequately represent a possible implementation.

A software tool written in C++, called *resgen*, has been developed to enable seeded constraint-based generation of reservoirs. This allows many different reservoir configurations to be easily generated; varying size, sparsity, input node selection, and output node selection. *resgen* also supports per device randomised memristor configuration using constraints of average and standard deviation to generate device parameters to replicate manufacturing variance. Random number generation (RNG) seeds can also be specified for reproducibility of exact configurations. The structure generation configuration parameters of *resgen* are outlined in Table 5 and memristor device instantiation parameter of *resgen* are outlined in Table 6.

*Table 5. configuration variables for reservoir structure generation using resgen tool.*

| Parameter | Purpose |
|---|---|
| N | Number of Memristors |
| $n_I$ | Number of Input nodes |
| $n_O$ | Number of Output nodes |
| $\rho_{min}$ | Minimum connections per node |
| $\rho_{max}$ | Maximum connections per node |
| $S_{struct}$ | RNG seed for structure generation |
| $S_{init}$ | RNG seed for initial reservoir state |
| $D_I$ | Input node allocation algorithm |
| $D_O$ | Output node allocation algorithm |
| $a_I$ | Anchor node for Input |
| $a_O$ | Anchor node for Output |

*Resgen* produces the reservoir code in five stages. The first stage is the base structure generation, shown in Pseudocode 1, where *resgen* creates graph nodes which it randomly connects edges to. The input and output nodes are assigned after base structure generation, shown in Pseudocode 2, based on the configuration provided to *resgen*. Then the graph edges are populated with model configuration parameters for the memristor behavioural model. Memristor behavioural model parameters are set by sampling parameters from normal distributions described by the configuration variables provided to *resgen*. The model parameters $a_{on}$, $a_{off}$, $TCR_{on}$, and $TCR_{off}$ are constant across all devices so are statically assigned. Resgen then outputs reservoir module code as *<config.design_name>.vams* and the structure generation result as *<config.design_name>nodes.csv*.

*Pseudocode 1. resgen base structure generation algorithm.*

```
procedure generate_base_structure(N, ρ_min, ρ_max, seed)
  ports ← {0 … 2N}
  for each port ∈ ports do
    port ← −1
  end for
  M ← 0
  while -1 ∈ ports do
    C = random(ρ_min, ρ_max, seed)
    while C ≠ 0 and −1 ∈ ports do
      i ← RANDOM (0, |ports|, seed)
      if ports[i] = −1 and ports[i %2] ≠ M then
        ports[i] ← M
        C ← C − 1
      end if
    end while
    M ← M + 1
  end while
  return ports, M
end procedure
```

*Pseudocode 2. resgen node assignment function.*

```
procedure assign _nodes(M, N_i, N_o, a_i, a_o, strategy, seed)
  nodes ← {0 … M} ∈ {input, output, intermediate}
  for each node ∈ nodes do
    node ← intermediate
  end for
  if a_i ≠ −1 then
    nodes[a_i] ← input
  end if
  if a_o ≠ −1 then
    nodes[a_o] ← output
  end if
  nodes ← ASSIGN_INPUT_NODES(nodes, N_i, strategy, seed)
  nodes ← ASSIGN_OUTPUT_NODES(nodes, N_o, strategy, seed)
  return nodes
end procedure
```

There are two different algorithms for both input and output node assignment. The input node assignment algorithms, Pseudocode 3, consist of evenly distributing the inputs across the reservoir structure and clustered placement about an anchor node. The output node assignment algorithms, Pseudocode 4, consist of even distribution across the reservoir structure and maximum distance from input nodes.

*Pseudocode 3. Input node assignment algorithm.*

```
procedure assign_input_nodes(nodes, N, strategy)
  if strategy = EVEN do
    for 0 ≤ i < N, i ∈ ℤ do
      inputs ←{x ∈ nodes ∩ type(x) = input}
      C ← MAX(DISTANCE(nodes, inputs))
      type(C) ←input
    end for
  end if
  if strategy = CLUSTERED do
    for 0 ≤ i < N, i ∈ ℤ do
      inputs ←{x ∈ nodes ∩ type(x) = input}
      C ← MIN(DISTANCE(nodes, inputs))
      type(C) ←input
    end for
  end if
  return nodes
end procedure
```

*Pseudocode 4. Output node assignment algorithm.*

```
procedure assign_output_nodes(nodes, N, strategy)
  if strategy = EVEN do
    for {i : x ∈ ℤ, 0 ≤ x < N} do
      outputs ← {x ∈ nodes | type(x) = output}
      C ← MAX(DISTANCE(nodes, outputs))
      type(C) ← output
    end for
  end if
  if strategy = MAX_DIST do
    for {i : x ∈ ℤ, 0 ≤ x < N} do
      inputs ← {x ∈ nodes | type(x) = input}
      C ← MAX(DISTANCE(nodes, inputs))
      type(C) ← output
    end for
  end if
  return nodes
end procedure
```

Constituent memristors are individually parameterised by *resgen* using the configuration variables from table 6. Memristor model parameters $D_l$, $D_A$, $\rho_{on}$, $\rho_{off}$, $E_{on}$, $E_{off}$, $K_{on}$, and $K_{off}$ are set on a per device basis and sampled from normal distributions. Memristor model parameters $a_{on}$, $a_{off}$, $TCR_{on}$, and $TCR_{off}$ are explicitly set for all devices. The initial memristor state, $W_{start}$, for each device is sampled from a uniform distribution to allow the variation of the reservoirs initial state set by the $S_{init}$ seed. This allows variation of $\rho_{off}$ and $\rho_{on}$ to represent device to device variations in amorphous SiOx switching layer resistivity. In addition to modelling physical geometry variations by varying $D_l$, $D_A$ and variations in ion mobility which would manifest as variation in switching speed by varying $K_{on}$ and $K_{off}$.

*Table 6. configuration variables for memristor instantiation using resgen tool.*

| Parameter | Purpose | Units |
|---|---|---|
| $\mu_{D_l}$ | Device Length Average | m |
| $\sigma_{D_l}$ | Device Length S.D | m |
| $\mu_{D_A}$ | Device Area Average | $m^2$ |
| $\sigma_{D_A}$ | Device Area S.D | $m^2$ |
| $\mu_{\rho_{LRS}}$ | LRS Resistivity Average | Ωm |
| $\sigma_{\rho_{LRS}}$ | LRS Resistivity S.D | Ωm |
| $\mu_{\rho_{HRS}}$ | HRS Resistivity Average | Ωm |
| $\sigma_{\rho_{HRS}}$ | HRS Resistivity S.D | Ωm |
| $\mu_{E_{on}}$ | On Field Threshold Average | V/m |
| $\sigma_{E_{on}}$ | On Field Threshold S.D | V/m |
| $\mu_{E_{off}}$ | Off Field Threshold Average | V/m |
| $\sigma_{E_{off}}$ | Off Field Threshold S.D | V/m |
| $\mu_{K_{on}}$ | K On Average | X |
| $\sigma_{K_{on}}$ | K Off S.D | X |
| $\mu_{K_{off}}$ | K Off Average | X |
| $\sigma_{K_{off}}$ | K off S.D | X |
| $a_{on}$ | A on | X |
| $a_{off}$ | A off | X |
| $TCR_{LRS}$ | TCR of LRS | 1/K |
| $TCR_{HRS}$ | TCR of HRS | 1/K |

### 3.3.4 Testing

To understand the reservoirs dynamics and operation, testing is performed. Testing seeks to verify the presence of the key properties of non-linear dynamics, fading memory, and the richness and quantity of detectable states. In addition to verifying the reservoirs' ability to operate at the required frequencies and voltages set forth in the deliverable's specification requirements.

*Table 7. resgen memristor instantiation parameters used for all testing.*

| Parameter | Value | Units |
|---|---|---|
| $\mu_{D_l}$ | $1\times10^{-9}$ | m |
| $\sigma_{D_l}$ | $5\times10^{-11}$ | m |
| $\mu_{D_A}$ | $1\times10^{-14}$ | $m^2$ |
| $\sigma_{D_A}$ | $1\times10^{-16}$ | $m^2$ |
| $\mu_{\rho_{LRS}}$ | $2.5\times10^{-1}$ | Ωm |
| $\sigma_{\rho_{LRS}}$ | $1\times10^{-2}$ | Ωm |
| $\mu_{\rho_{HRS}}$ | 5 | Ωm |
| $\sigma_{\rho_{HRS}}$ | $1\times10^{-1}$ | Ωm |
| $\mu_{E_{on}}$ | $1\times10^{7}$ | V/m |
| $\sigma_{E_{on}}$ | $1\times10^{4}$ | V/m |
| $\mu_{E_{off}}$ | $-1\times10^{7}$ | V/m |
| $\sigma_{E_{off}}$ | $-1\times10^{4}$ | V/m |
| $\mu_{K_{on}}$ | $2\times10^{-5}$ | X |
| $\sigma_{K_{on}}$ | $5\times10^{-7}$ | X |
| $\mu_{K_{off}}$ | $-2\times10^{-5}$ | X |
| $\sigma_{K_{off}}$ | $5\times10^{-7}$ | X |
| $a_{on}$ | 3 | X |
| $a_{off}$ | 3 | X |
| $TCR_{LRS}$ | $3.8\times10^{-3}$ | 1/K |
| $TCR_{HRS}$ | $-4.4\times10^{-3}$ | 1/K |

#### 3.3.4a Fading Memory Characterisation

Fading memory is an important property of an effective reservoir as discussed in Section 3.3.1 as it directly determines the temporal visibility of the reservoir structure. As such, it is important to understand and characterise it.

Fading memory characterisation is performed by a long time transient simulation. Two identical reservoirs with different initial states are fed the same input sequence and their output voltages recorded for each time step of the simulation. After the simulation, the data is processed by MATLAB where the difference between the two reservoirs output voltages is calculated by subtraction. The difference between the two reservoirs states at each time step is then calculated by the norm of the difference between their output voltages. This gives a measure of distance between the two reservoirs states for each time step of the simulation.

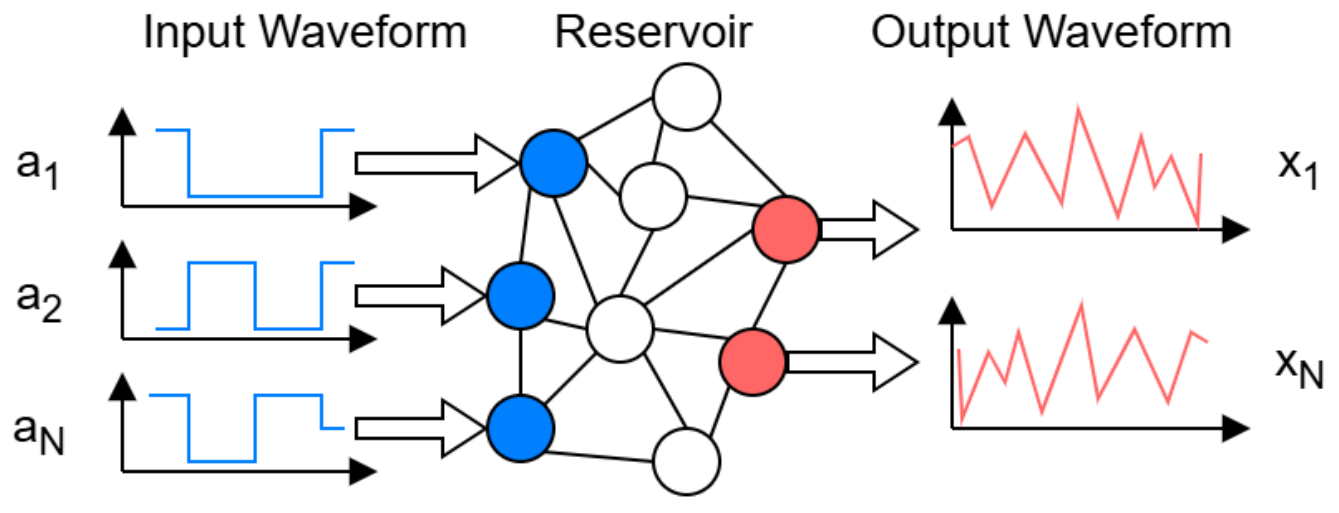

*Figure 24. Fading memory characterisation testing apparatus visualisation.*

The characterisation is implemented using a python script based autonomous flow. The *resgen* tool is first used to generate two identical reservoirs with different initialisation states. The flow then configures the simulation environment, specifying the time period for input hold time, operating voltage, and input sequence length. After which it runs the simulation and aggregates the output data for visualisation in MATLAB.

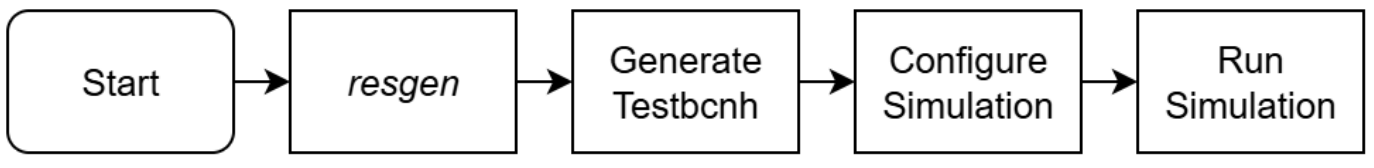

*Figure 25. Python automated testing flow.*

The fading memory of the reservoir architecture is tested for various different reservoir sizes and densities using a sequence of 500 generated inputs with an input hold time of 1ns.

The results, Fig.26, demonstrate that the reservoir possesses the required fading memory as all configurations display a general reduction of the difference between outputs over time. Following a second order exponential decay function in the form,

$$\|V(t)\| = ae^{-bt} + ce^{-dt} \quad (23)$$

Where $\|V(t)\|$ is the norm of the difference in output voltages between the two reservoirs at time t. The results of fitting the data to (23) are shown in Fig.27 and Table 8.

*Table 8. Second order exponential decay function fitting parameters.*

| VDD | a | b | c | d |
|---|---|---|---|---|
| 1.8V | 0.6226 | 0.07866 | 0.1825 | 0.004756 |
| 3.3V | 1.47 | 0.2344 | 0.2831 | 0.02251 |

The fit second order exponential decay functions show an approximate 4x reduction in fading memory half-life with an increase in voltage from 1.8V to 3.3V. This is indicative of the non-linear voltage driven switching of the memristor devices and is a desired expression of non-linearity indicative of an optimal reservoir design. The 4x reduction in the fading memory half-life can be equivalently understood as a 4x increase in the effective memory window of the reservoir.

The expected effective increase in memory window from changing operating voltage from 3.3V to 1.8V is 8x. Derived from a combination of 4x higher half-life and 2x higher ADC voltage resolution. The actual effective memory window of 20ns for 3.3V and 152ns for 1.8V is obtained from Fig.24 and gives an increase of 7.6x. Approximately in line with the 8x increase implied by the increase in half-life and increase in ADC resolution. The deviation from expected increase results from the noise inherent to the memristors switching and conduction.

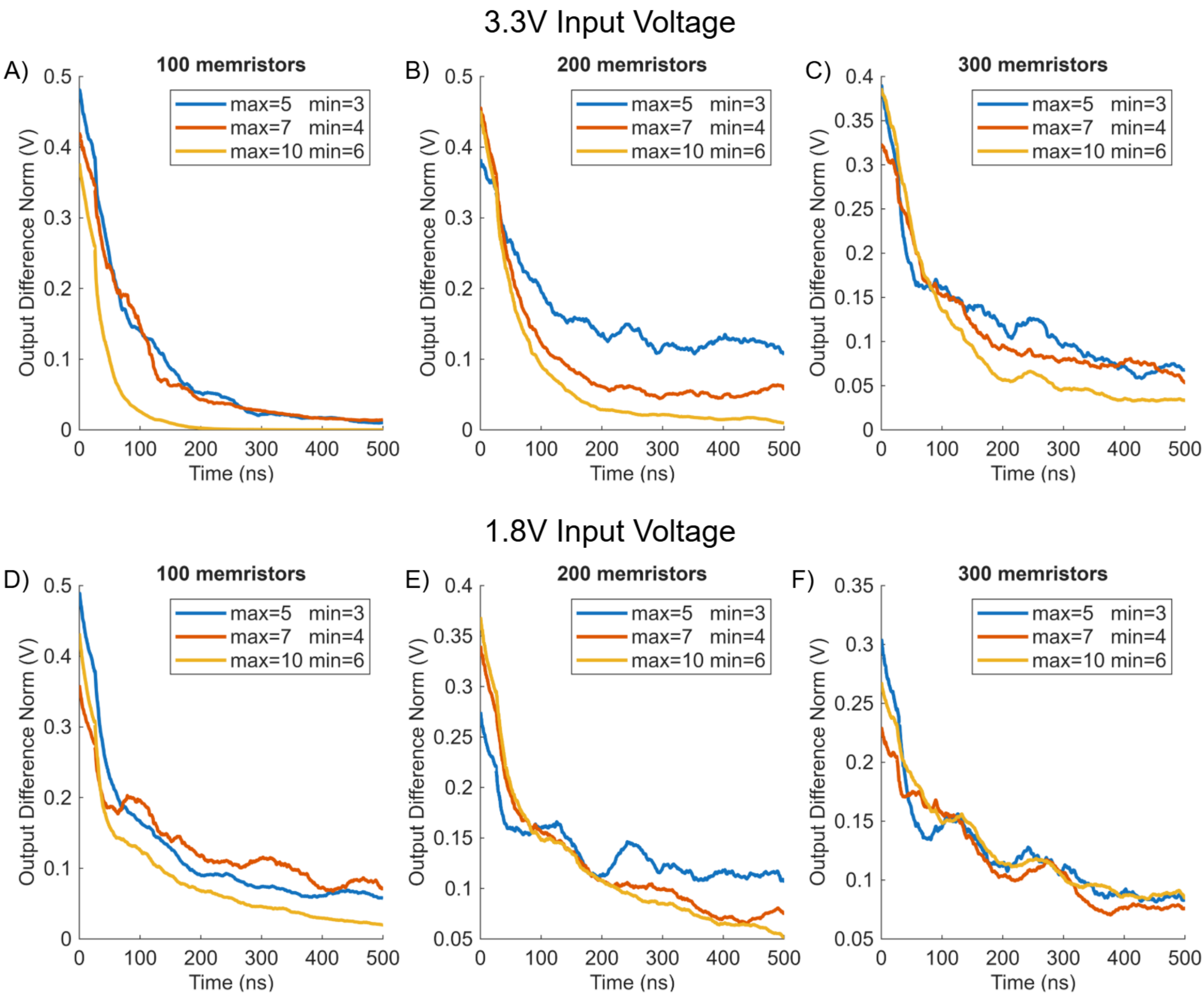

*Figure 26. Plots of vector norm of the differences in reservoir outputs over time with stimuli frequency of 1GHz and varying reservoir density. With 3.3V 100 memristors (a), 3.3V 200 memristors (b), 3.3V 300 memristors (c), 1.8V 100 memristors (d), 1.8V 200 memristors (e), and 1.8V 300 memristors(f).*

The consistency of the fading memory can also be seen to be heavily influenced by reservoir density and size, Fig.26. Reservoirs with more voltage nodes, either through sparsity or a larger number of memristors, present less consistent fading memory. Fig.26b shows this where the less dense configuration (the blue line) exhibits far more noise than the denser configuration (the yellow line). This is attributed to both the increased behaviour complexity of less dense systems and the impact of noise on measured voltages. The inconsistency in fading memory exhibited in Fig.26 is likely to be a significant constraint on reservoir performance. As it will affect the ability of the resultant RC system to properly generalise due to unstable behaviour. As such, it is important to address this for full system integration.

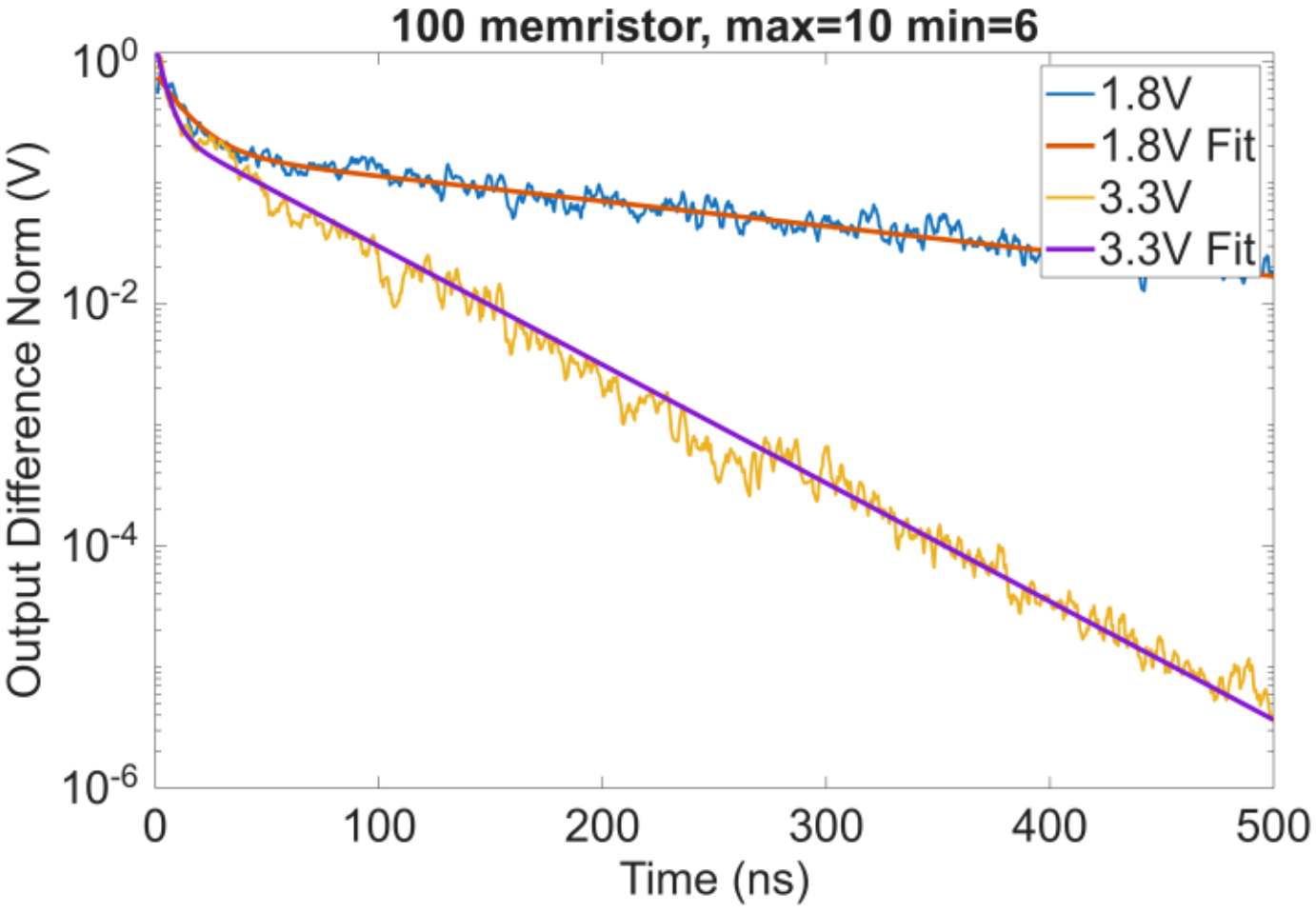

*Figure 27. Plot of the norm of the difference in output voltages for 100 memristor reservoir configuration with maximum of 10 connections per node and minimum of 6 connections per node at operating voltages of 1.8V and 3.3V.*

### 3.3.4b Quantizable State Testing

Reservoir readout is performed with low resolution ADCs, placing a potentially significant limitation on the readable states. As such, the reservoir needs to be optimised to make full use of the readout ADCs dynamic range to maximise the number of distinguishable states. The theoretical number of unique distinguishable states is given by,

$$S_N = \max\{S^N \in \mathbb{B}^{Q_i \times N}\} \quad (24)$$

Where $S_N$ is the maximum number of theoretically detectable states and $S^N$ is the set of detectable states with N outputs with $Q_i$ bits per output. For 32 outputs with 4-bit quantisation this gives $\approx 3.40 \times 10^{38}$ detectable states. However, this is for an ideal case assuming the reservoirs output voltages are uniformly distributed between $V_{DD}$ and 0V. In reality, it's more likely that the reservoir outputs will not be uniformly distributed and hence the actual number of unique distinguishable states will be far less.

*Table 9. Reservoir structure configuration for quantizable states testing.*

| Parameter | Value |
|---|---|
| Memristor Count | 200 |
| Maximum Node Connections | 10 |
| Minimum Node Connections | 6 |
| Number of Inputs | 8 |
| Number of Outputs | 32 |

Exhaustively testing the number of distinguishable states in simulation would require in excess of $S_N$ timesteps and is infeasible in any rational amount of time. Instead, testing focuses on assessing the distribution of reservoir output node voltages over a finite time simulation. A transient simulation is performed during which the reservoir is fed 10 million randomly generated bit vectors and its output node voltages are recorded. Each input is a randomly generated bit vector in equal length to the number of input nodes on the reservoir and is held at the reservoirs input nodes for 1ns before the next input is presented. The resulting reservoir output node voltage waveforms are processed using MATLAB to visualise the cumulative probability distribution (CDF) of output node voltages, Fig.30.

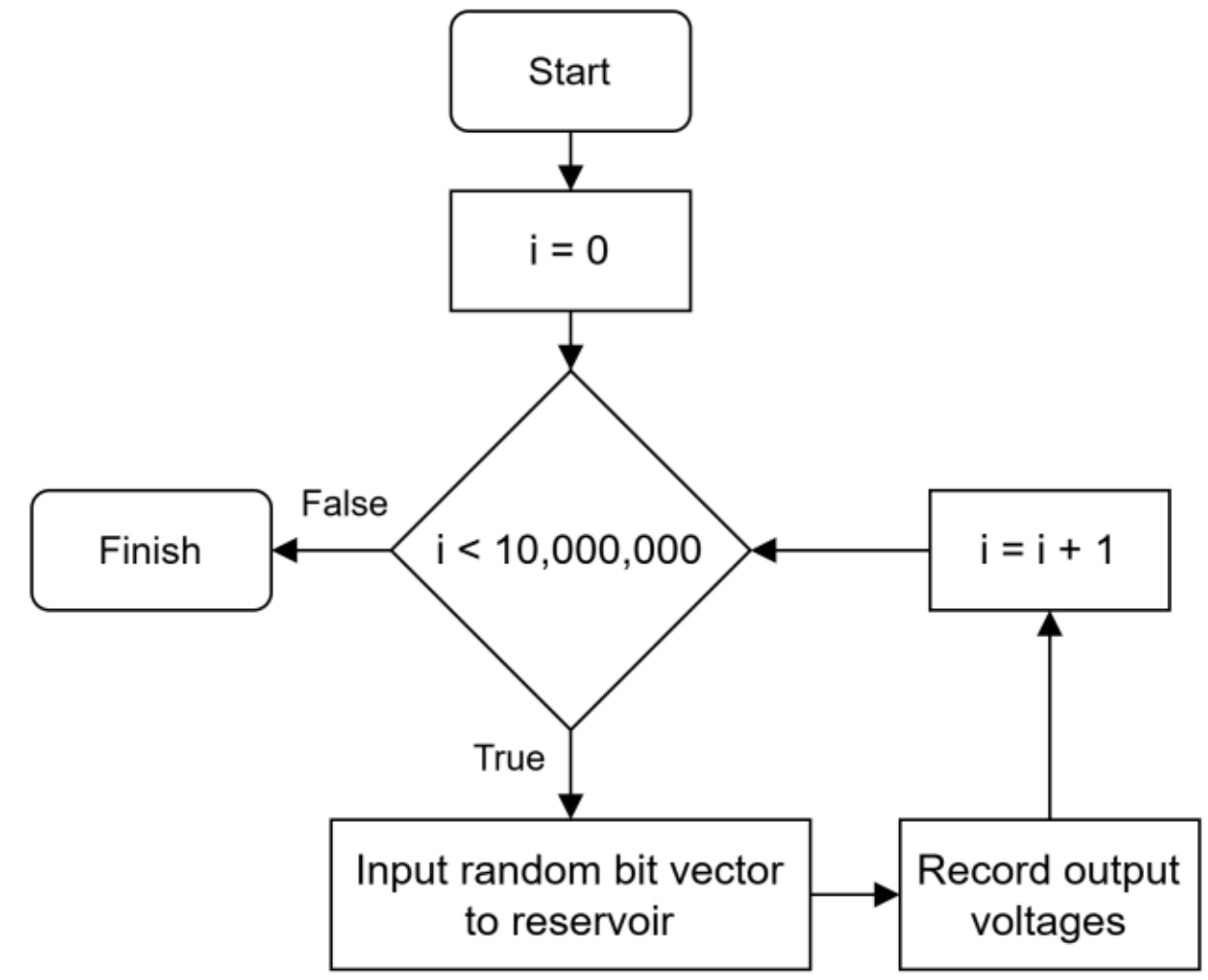

*Figure 28. Flow chart of the transient simulation procedure to measure reservoir ADC dynamic range usage.*

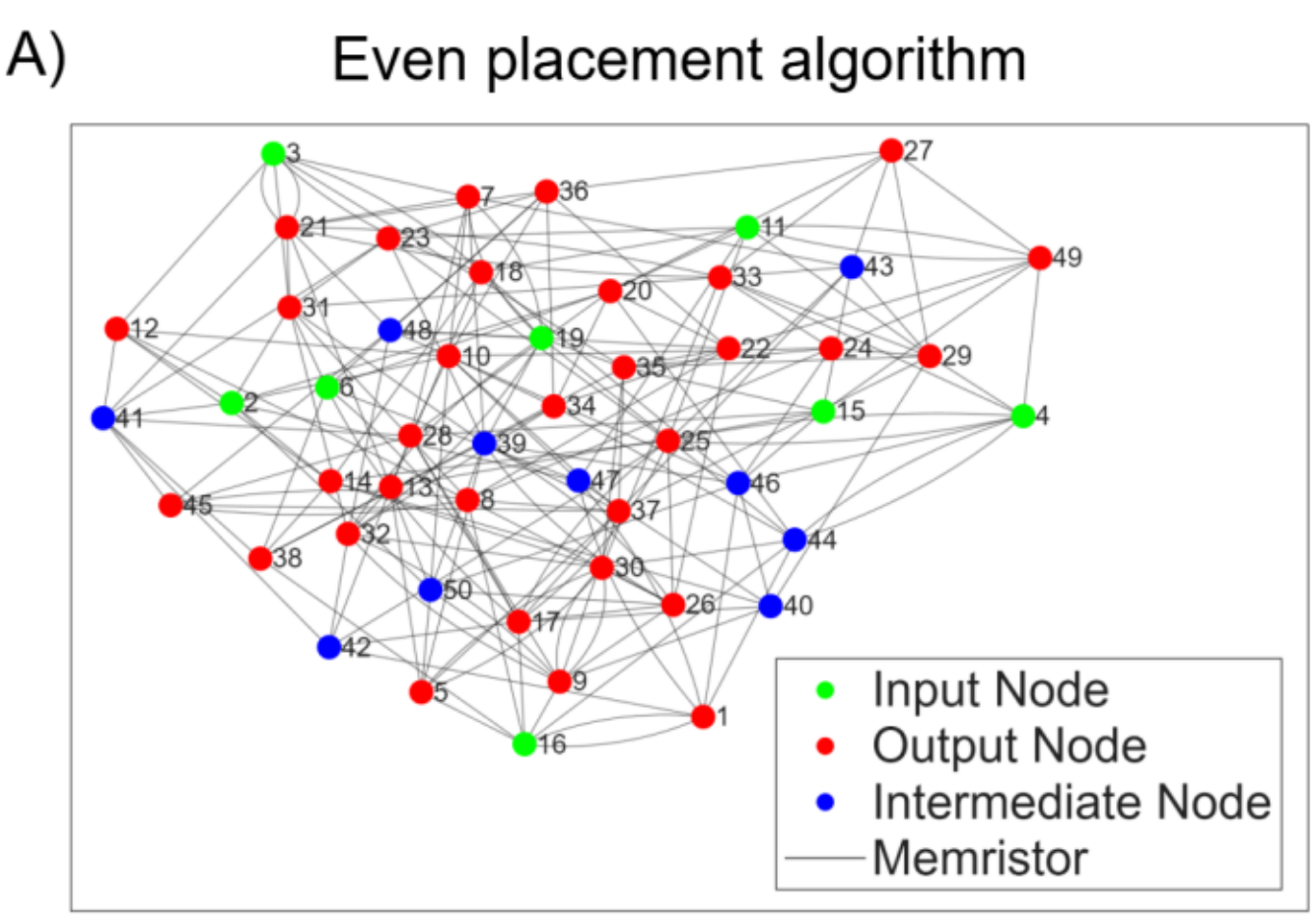

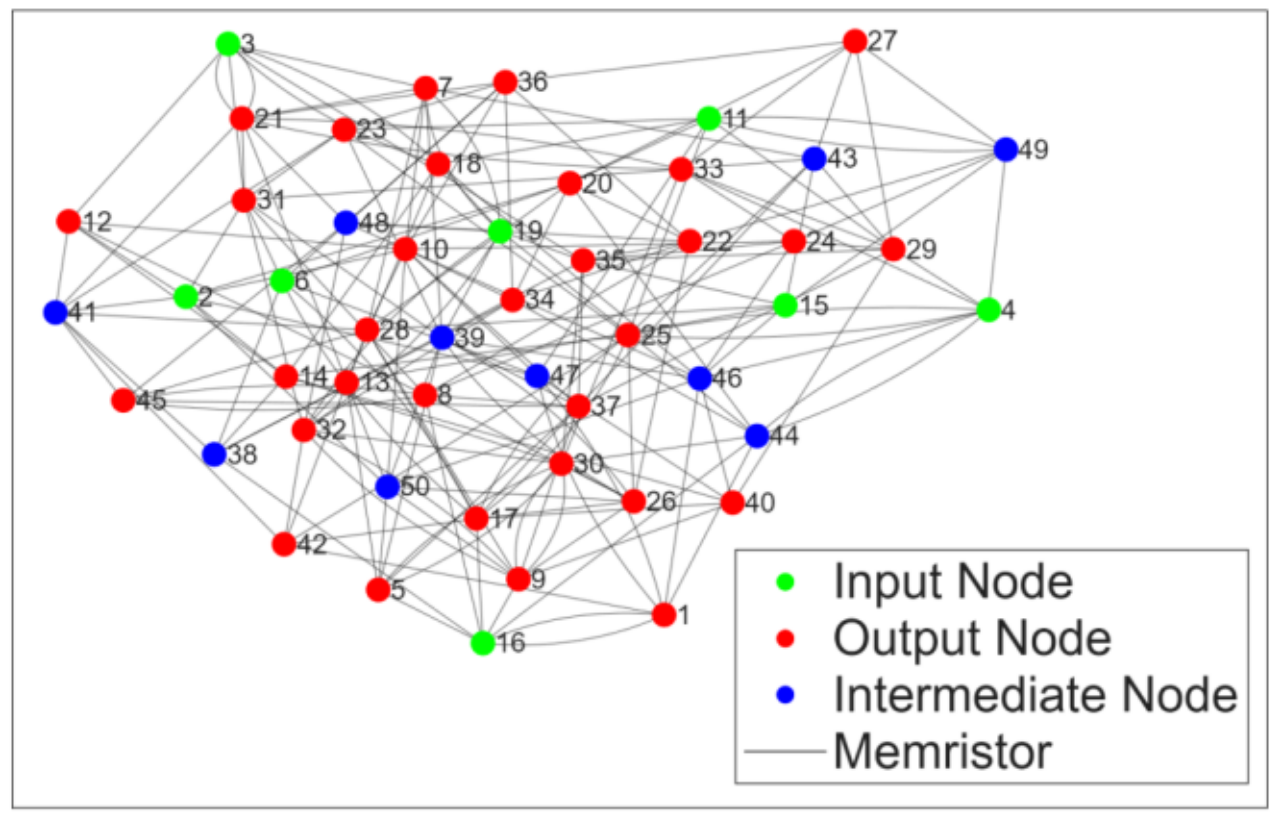

*Figure 29. Reservoir structure and node assignment for even output node assignment (a) and maximum distance from input node assignment (b).*

Testing is implemented by using python scripting to generate the reservoir VAMS code in addition to the test bench that is used to provide input to the reservoir and record its output voltages. The python script based flow follows the same process as for fading memory testing, Fig.25. First it performs code generation, then the simulation environment is configured and finally the simulation is run until 10 million input samples have been processed.

The results, Fig.30, show that the output voltages of the reservoir are normally distributed about a mean of ½ $V_{DD}$ for all operating voltages and output node configurations. There is only a marginal difference in output voltage distribution between the two output node assignment configurations. With even distribution of output nodes yielding a slight decrease in the average output voltage and a slight increase in standard deviation, Table 10. However, this is marginal and has negligible impact on the effective use of ADCs dynamic range. This is attributable to the similarity of the output node placement between the two algorithms, Fig.29, due to the comparable number of available nodes limiting variety in placement.

*Table 10. Reservoir output voltage mean and standard deviation at different operating voltages and output node assignments.*

| $V_{DD}$ | Output Assignment | $\mu$ | $\sigma$ |
|---|---|---|---|
| 3.3V | Max Distance | 1.69 | 0.6027 |
| 3.3V | Even | 1.67 | 0.6172 |
| 1.8V | Max Distance | 0.92 | 0.3236 |
| 1.8V | Even | 0.91 | 0.3321 |

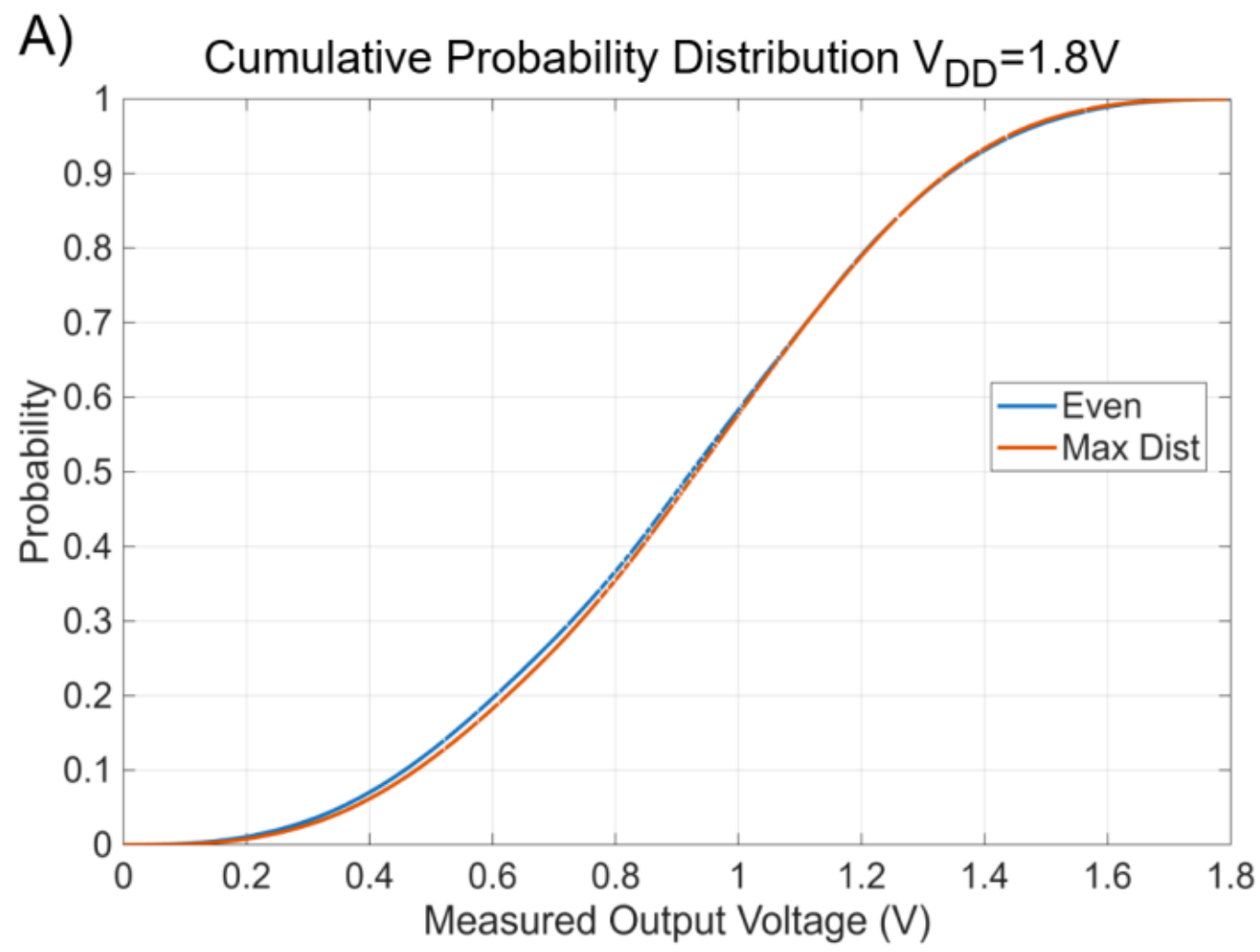

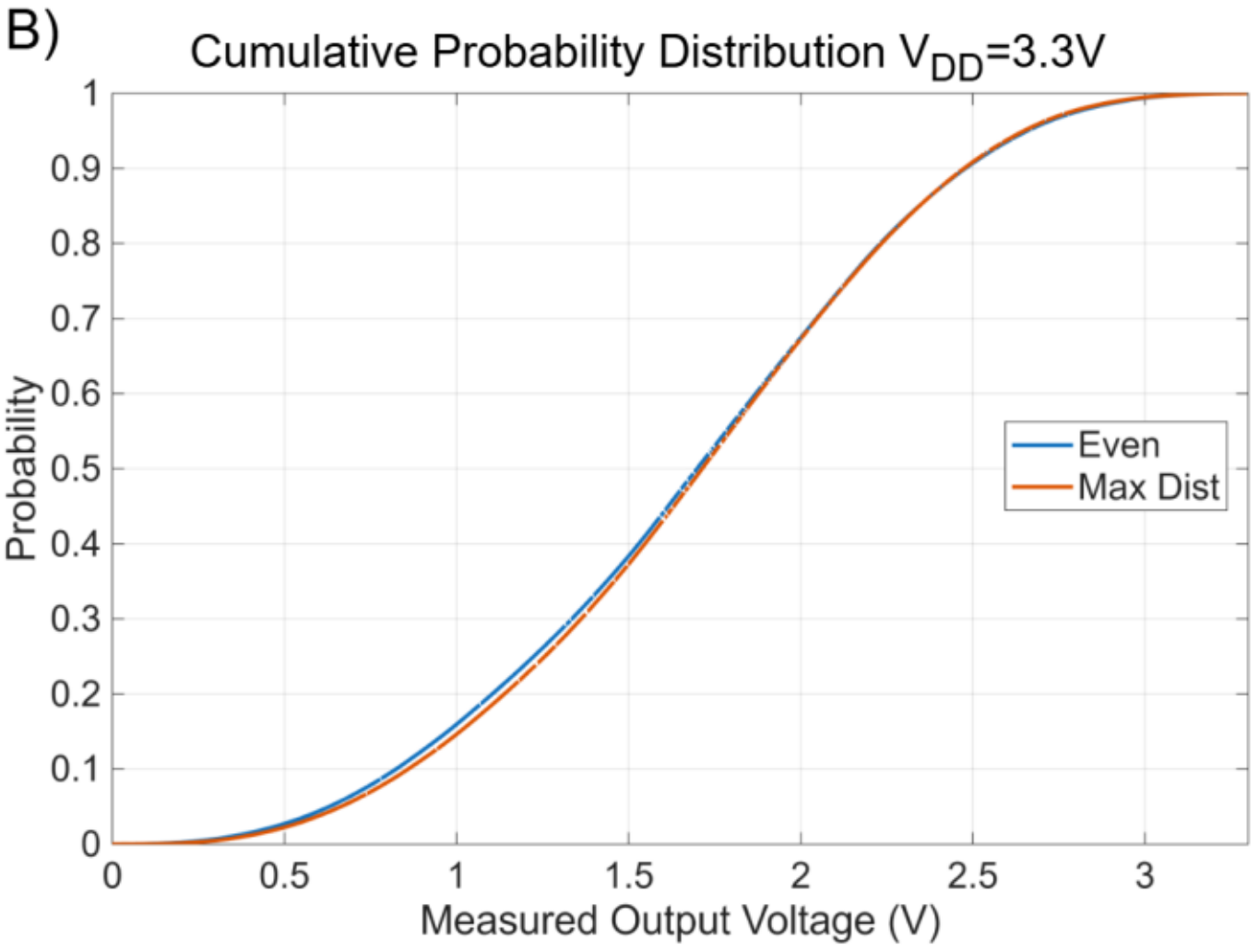

*Figure 30. Cumulative Probability Distribution (CDF) plot for reservoir output voltages at 1.8V operation (a) and 3.3V operation (b).*

The high standard deviation and minimal skew of the output voltage distribution means that the ADCs dynamic range is used sufficiently to provide a large number of distinguishable states. Disregarding states outside of $2\sigma$ of the mean gives a usable dynamic range of 0.44V to 2.90V for 3.3V operation with even distribution output node assignment. The histogram plot, Fig.31, shows that 95% of the measured voltages fall within the centre 12 bins.

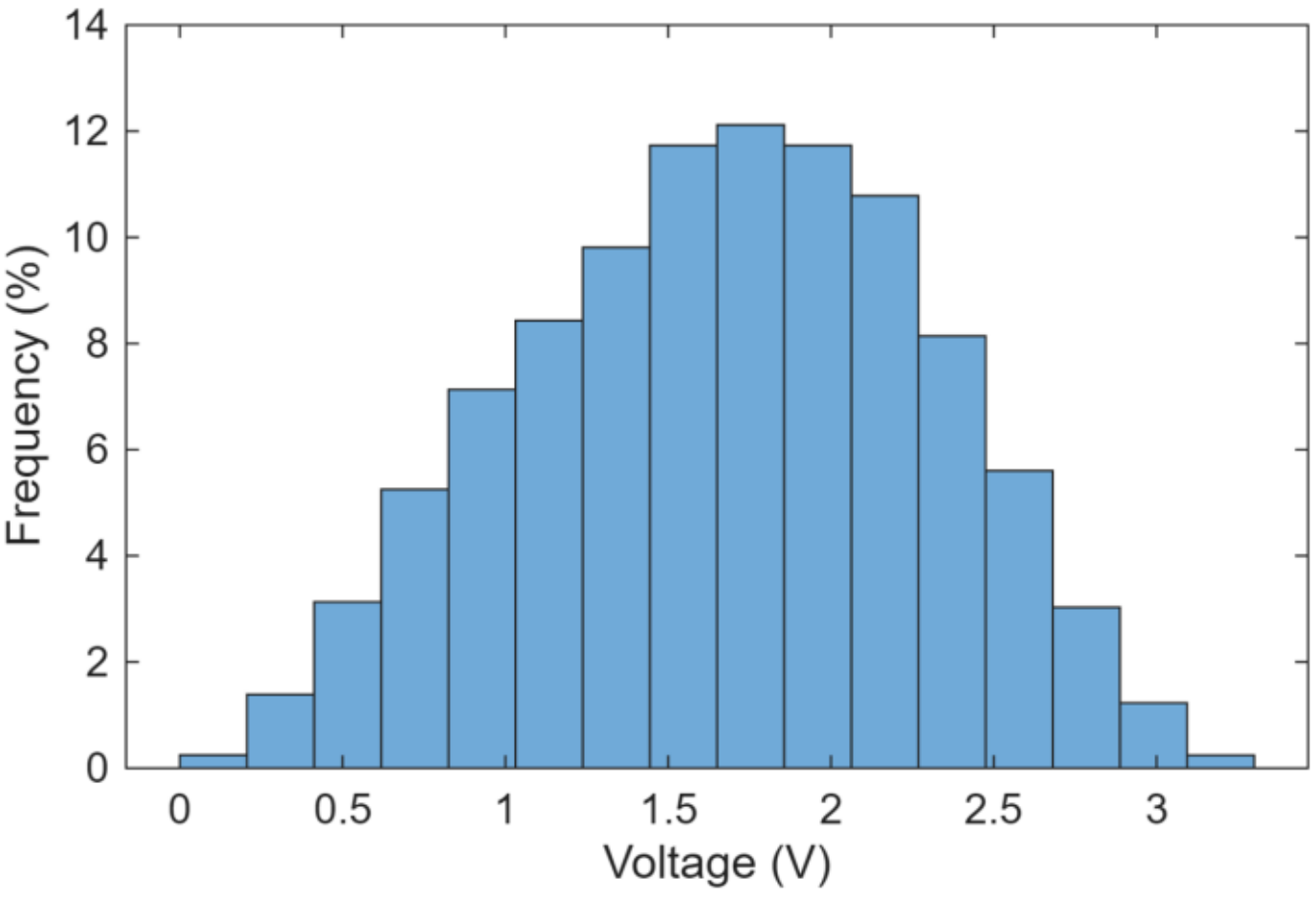

*Figure 31. Histogram plot for 3.3V $V_{DD}$ even output node assignment reservoir output voltages.*

Considering only the voltage range where the gradient of the CDF is approximately linear, meaning that it is approximately uniformly distributed, gives a range of 1.2V to 2.2V at 3.3V operation. This gives a minimum of $1.845 \times 10^{19}$ distinguishable states, which is likely sufficient.

### *3.3.4c Nonlinearity Test*

Non-linearity is a vital requirement for reservoir dynamics as discussed in Sections 3.3.1 and 2.1. Non-linearity is the key property that allows the reservoir to usefully project its inputs to a higher dimension. It is vital for problem simplification to increase the likelihood of linear separability at the output layer. As such, it's important to verify the reservoirs dynamics are non-linear for optimal reservoir performance.

The test for non-linearity is performed by driving the reservoir with sinusoidal voltage waveform and recording the reservoir output voltages at each time step. This follows the testing setup shown in Fig.32 where only a single input node is driven by the sinusoidal voltage and all other input nodes are tied to ground. At the same time, all the output voltages are recorded for each time step alongside the sinusoidal input voltage.

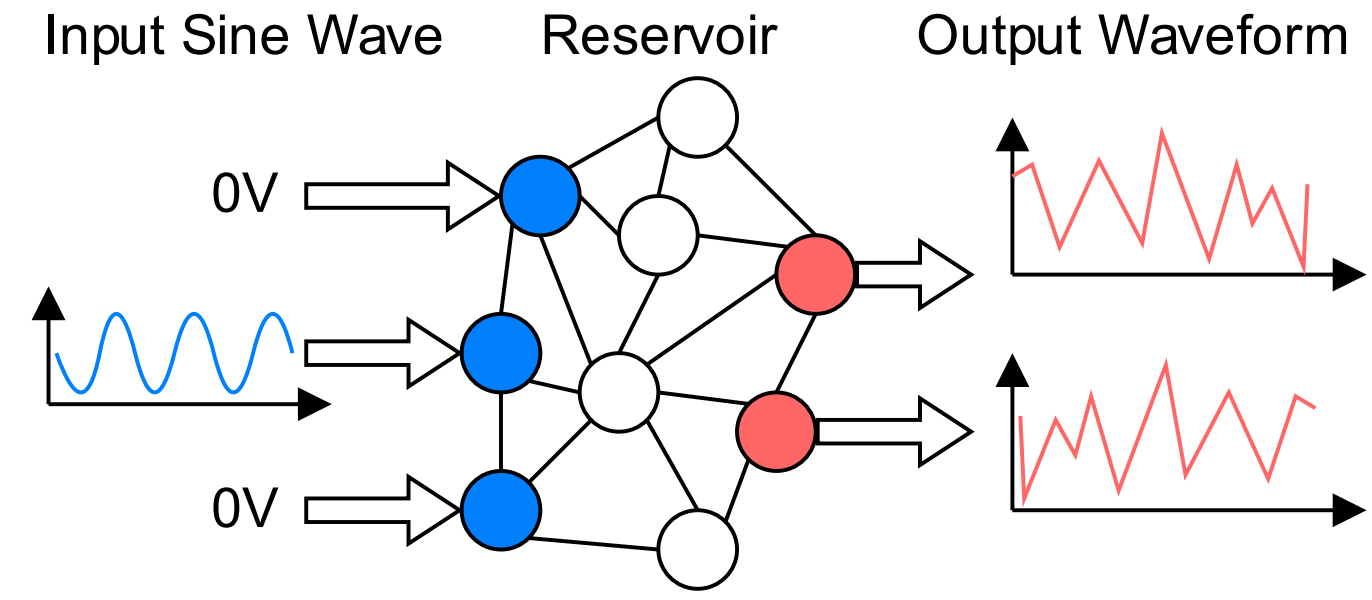

*Figure 32. Reservoir non-linearity test apparatus visualisation.*

Non-linearity of reservoir dynamics is tested using an automated python script driven testing flow which runs a long time transient simulation. First, the reservoir code for simulation is generated using the *resgen* tool alongside the configuration of a template test bench for reservoir stimulus and data capture. After this, the python script sets up the transient simulation and runs it until completion. After which the output results are visualised in MATLAB.

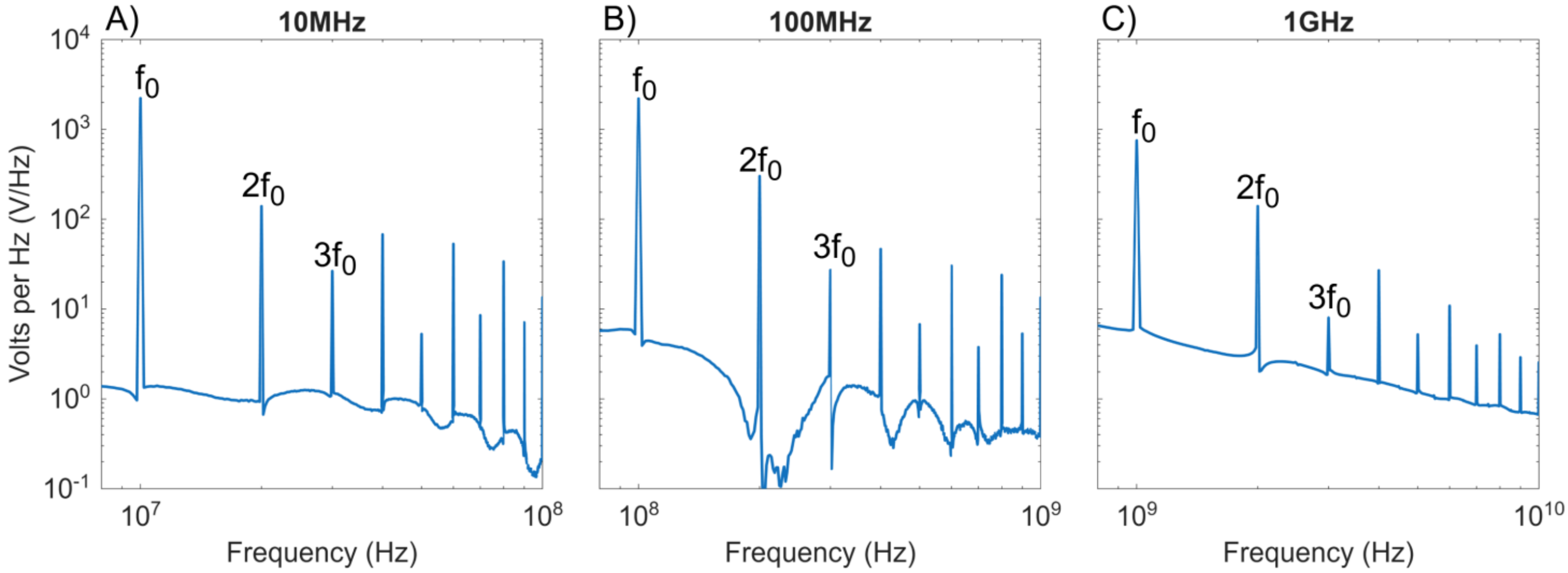

*Figure 33. Fourier transform of reservoir output waveform for 3.3V 10MHz input sine wave (a), 100MHz input sine wave (b), and 1GHz input sine wave (c).*

The reservoir was tested with 3.3V amplitude sine waves of 10MHz, 100MHz, and 1GHz for 50 wave cycles each. The output waveform data is processed using MATLAB to calculate the Fourier transform, Fig. 33, of the linear superposition of all the reservoir output nodes. The results show that additional harmonics are introduced, labelled $2f_0$ and $3f_0$, for all input sine wave frequencies. The additional harmonics amplitudes decrease with frequency as expected, with harmonics of 1GHz attenuating significantly due to low pass filtering effect of the reservoir capacitors.

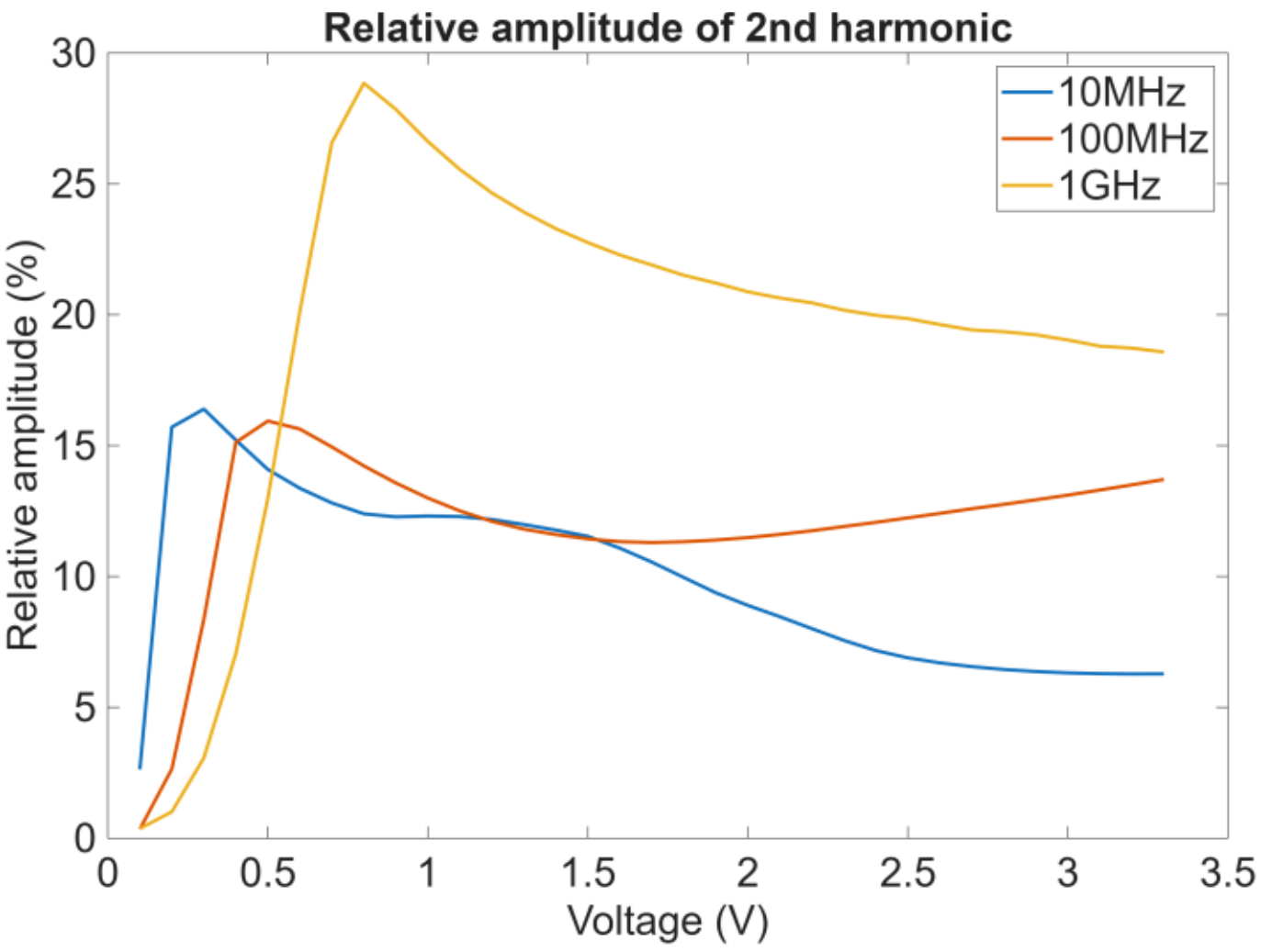

*Figure 34. Plot of the relative amplitude of 2nd harmonic to 1st harmonic across voltage sweep from 0.1V to 3.3V at frequencies of 10MHz, 100MHz, and 1GHz.*

Each sine wave input frequency was tested for non-linearity across a range of input voltages from 0V to 3.3V. The results, Fig.34, shows the relative amplitude of the 2$^{nd}$ harmonic to the first harmonic at the reservoir output nodes across the range of input signal amplitudes.

The relative amplitude of the 2$^{nd}$ harmonic initially increases with voltage. This is caused by the exponentially faster switching of the memristor devices as voltage increases. The relative amplitude then begins to decrease after a certain voltage. This is attributed to the threshold nature of memristor switching as the higher input voltage results in the activation of state change switching in more devices deeper in the reservoir. This results in additional harmonic generation and mixing which spreads energy out across the frequency spectrum and into higher frequencies. This can be seen in Fig. 35 where the spectral power concentrated outside the first harmonic increases with input sine wave amplitude.

While the relative amplitude of the 2$^{nd}$ harmonic peaks at 0.8V, Fig.34, the relative spectral power outside the 1$^{st}$ harmonic increases up to 2V, Fig.35, for 1GHz input frequency. This verifies that further mixing and harmonic generation is taking place inside the reservoir as the higher voltage activates more memristors. After 2V the relative spectral power outside the 1$^{st}$ harmonic begins to decrease, this is a direct result of the limitation in sampling frequency of the simulation. The increased voltage results in spectral power beginning to be concentrated at frequencies higher than what are measurable during data logging. This is because the reservoir output voltages are logged only 1000 times per input wave cycle for performance reasons.

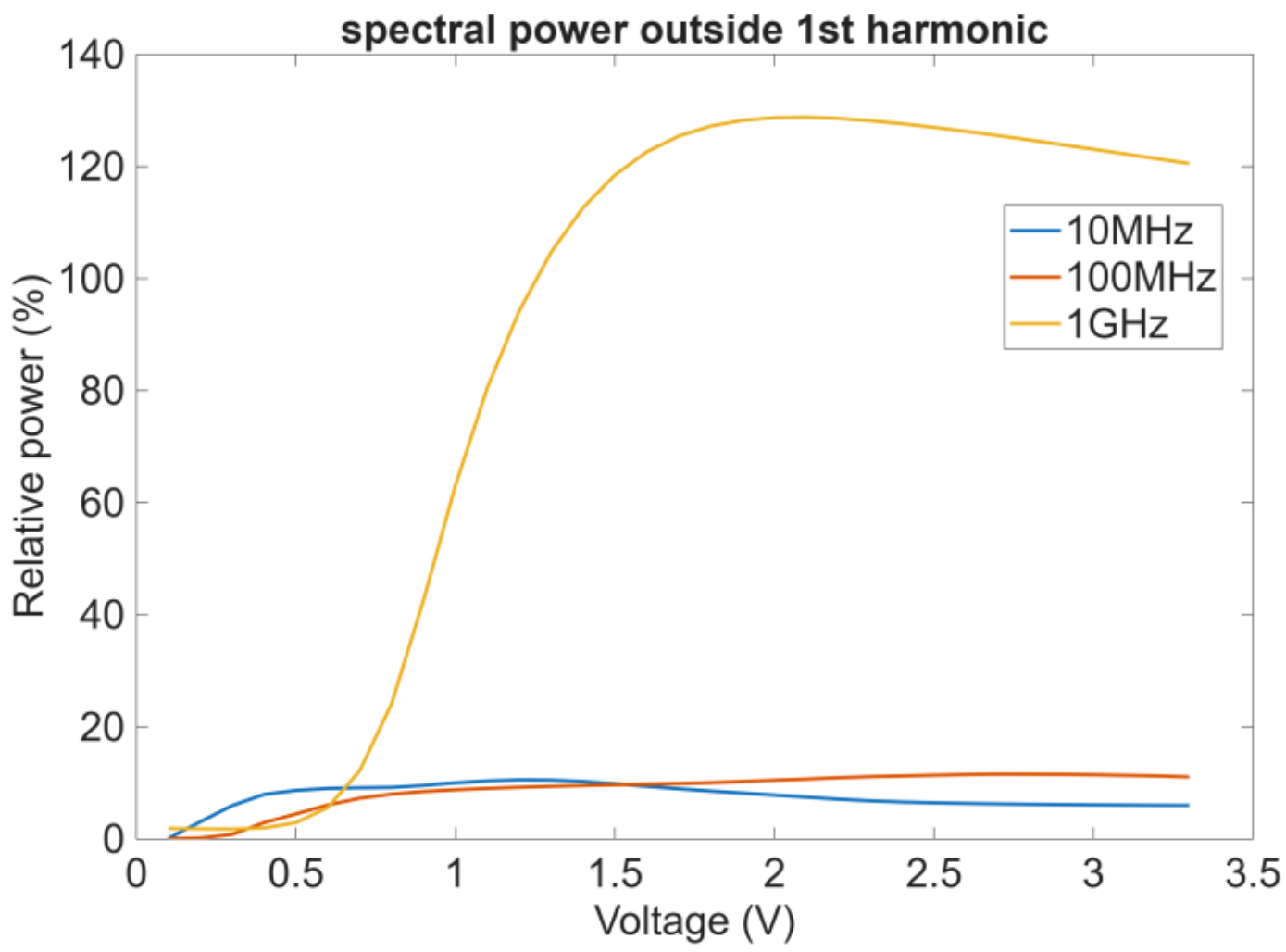

*Figure 35. Plot of PSD of entire frequency spectrum relative to PSD of first harmonic.*

For an input frequency of 100MHz, the reservoir demonstrates peak relative 2$^{nd}$ harmonic amplitude, Fig. 34, at a lower voltage compared to 1GHz input frequency. This is a result of the increased time period from 1ns to 10ns causing a larger change in resistance of each memristor due to prolonged switching. Resulting in the 2$^{nd}$ harmonics relative amplitude peaking at a lower voltage compared to 1GHz input sine wave. The 10MHz input follows this trend, with 2$^{nd}$ harmonic relative amplitude peaking at 0.3V compared to 0.5V for 100MHz and 0.8V for 1GHz.

However, the 100MHz input sine wave shows a partial recovery in the relative amplitude of 2nd harmonic, Fig. 34, as voltage approaches 3.3V.

### *3.3.4d Kernel Rank Testing*

Kernel Rank (KR) is a measure of the number of truly independent dimensions in the reservoir output and provides a measure of the reservoir dynamics ability to non-linearly transform its inputs into a higher dimensional space [39]. Poor performing reservoirs with N dimensional output will have less than N truly independent outputs, manifesting as a rank deficiency in KR testing. As such, higher KR means more truly independent dimensions exist in the reservoir output.

KR testing is performed by presenting the reservoir with time series input sequences and recording the output voltages at the end of each sequence. An $m$ x $n$ matrix can then be constructed, where columns are the output voltages at the end of each sequence. Singular value decomposition (SVD) is then used to perform an eigen decomposition of the matrix into a set of orthogonal basis vectors and singular values. The number of non-zero singular values thus represents the number of independent dimensions. The number of non-insignificant singular values represents the degree of independence of each dimension, with more non-insignificant values indicating richer high dimensional reservoir dynamics.

The testing setup from Section 3.3.4a is re used for KR testing. A continuous sequence of random inputs is presented to the reservoir at a specified frequency determining each inputs hold time. Instead of recording the output voltages for each input, the output voltages are recorded for every N inputs using a counter which triggers the sampling of reservoir output voltages on reset. This allows the continuous input stream to be sampled periodically, with each period being a distinct set of random inputs. Testing was performed on the reservoir structure configuration given in table 11. Testing was conducted with input voltages of 0.9V, 1.8V and 3.3V at 1GHz (1ns input hold time) for sequences of varying length from 5 to 100 samples. Each voltage and sequence length combination was tested for 500 unique input sequences.

The results of the singular value decomposition (SVD), shown in Fig. 36, indicate that sequence length has a negligible effect on the effective dimensionality of the reservoir. All tested sequence lengths exhibit consistent variance distributions across the SVD basis vectors, within the bounds of run-to-run variation. Furthermore, input voltage has minimal influence on the variance distribution among the SVD basis vectors, as demonstrated by the similarity in the standard deviation profiles shown in Fig. 36.

*Table 11. Reservoir structure configuration for KR testing.*

| Parameter | Value |
|---|---|
| Memristor Count | 200 |
| Maximum Node Connections | 10 |
| Minimum Node Connections | 6 |
| Number of Inputs | 8 |
| Number of Outputs | 32 |

The SVD decomposition of the KR testing result matrix further demonstrates that a single basis vector captures the majority of the output signal information. This behaviour arises from the resistor-divider network architecture of the reservoir, which produces approximately normally distributed output voltages, as discussed in Section 3.3.4b. Despite this dominant basis vector, the reservoir outputs are not rank deficient, as none of the singular values are equal to zero. This indicates that the true dimensionality of the reservoir output remains equal to the number of output nodes.

Although the SVD analysis indicates that the reservoir possesses an effective output dimensionality of 32, as the output space is not rank deficient, the number of practically useful dimensions is likely to be lower than the total number of output dimensions. This is demonstrated by the nearly two orders of magnitude difference in standard deviation between the second and thirty-second basis vectors. Consequently, while the reservoir output spans 32 dimensions, only a subset of these dimensions contributes significant information content, with the number of useful dimensions estimated to lie between 10 and 12 depending on the operating voltage.

The limited contribution of outputs beyond the estimated 10–12 meaningful dimensions may present a significant limitation to the effectiveness of the developed reservoir design. Consequently, further evaluation during full system integration and verification, described in Section 3.5, is required to determine the true impact of output dimensionality on overall system performance.

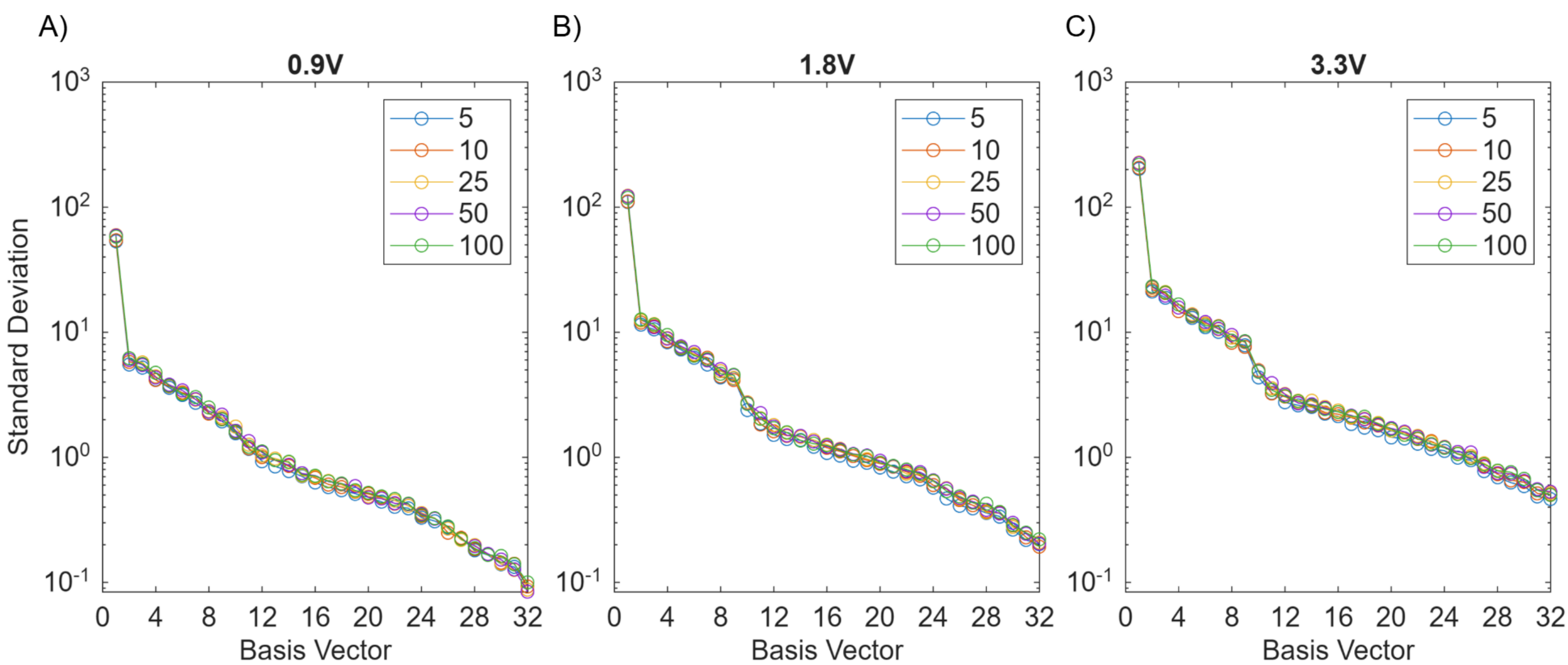

*Figure 36. SVD basis vector standard deviations for sequence lengths of 5, 10, 25, 50, and 100 samples at voltages of 0.9V (a), 1.8V (b), and 3.3V (c).*

## 3.4 Perceptron

This section covers the design and development of the perceptron neuron output layer for the RC BPU design. First, background on perceptron is explored, detailing its capabilities and function. Then, a chosen perceptron configuration is detailed to meet the deliverable and project requirements. Finally, the design is implemented and tested.

### 3.4.1 Background

Once the reservoir state is read, it needs to be classified into an output prediction of taken (T) or not taken (NT). To achieve this, a perceptron with a threshold activation function is used.

The perceptron is the simplest form of neuron and a type of supervised learning binary classifier. It works by multiplying a series of input values, $\underline{A}^N$, and bias with a weights vector, $\underline{W}^{N+1}$, which results in an activation state $X$. This activation state $X$ is transformed by a nonlinear activation function such as binary output, ReLU, sigmoid, etc.

$$X = [1, \underline{A}^T] \cdot \underline{W} \tag{25}$$

Here, a single layer perceptron is trained using the perceptron learning rule [40], a supervised training method which requires both the input and classification labels of the desired output.

A single layer perceptron can fit to any problem that is linearly separable. This requires the predictors (the inputs to the perceptron) belonging to each class to be disjoint, allowing a single hyperplane to separate them. Such as in Fig.37 where the red and blue class predictors are separable (dichotomised) by the single line $\underline{W}^T\underline{A} = 0$.

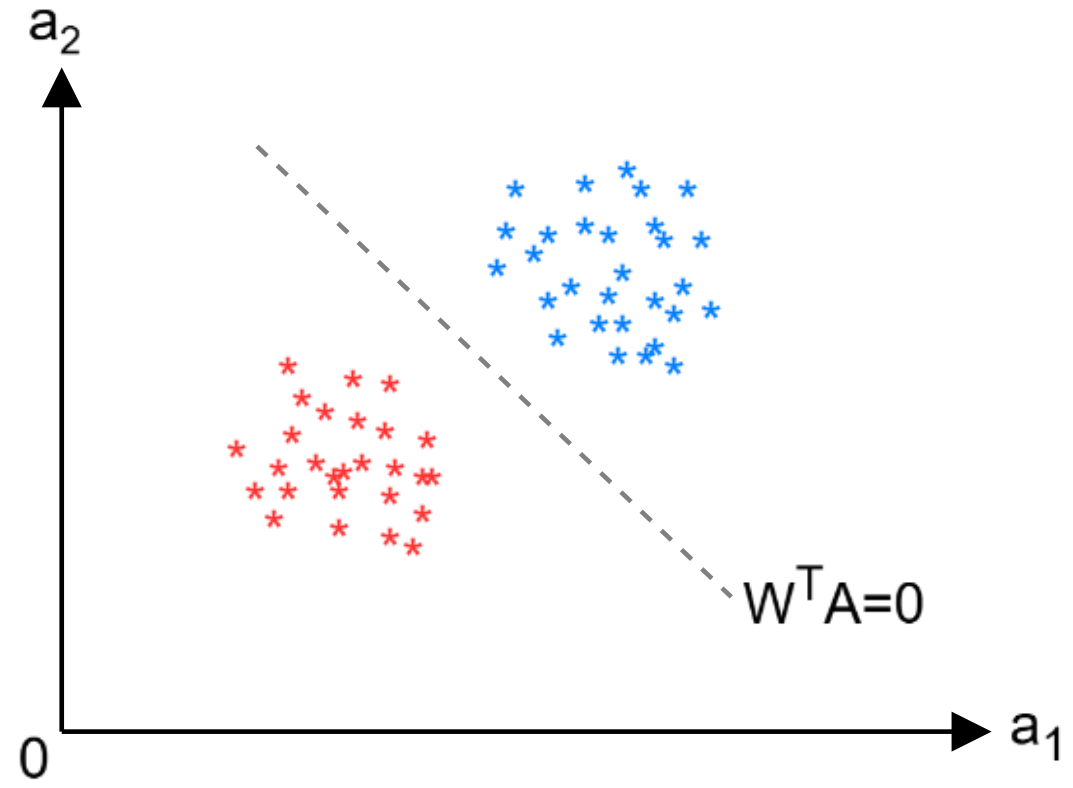

*Figure 37. example plot of disjoint classification regions, where red class predictors do not overlap with blue class predictors.*

The ability of the perceptron to fit to a problem and fully dichotomise is given by its capacity. The perceptron's capacity, given by Cover's theorem (26), gives a measure of the neurons capability to store information. Cover's theorem gives the count of possible linearly separable sets comprised of $N$ points in $d$ dimensional space. With a higher capacity indicating the neuron capability fit to more complex problems.

$$C(N,d) = 2\sum_{k=0}^{d-1}\binom{N-1}{k} \tag{26}$$

When a problem is too complex, the perceptron provides an optimal but not error free dichotomisation of the input space. Such as when class predictors are not linearly separable. Thus, the reservoirs non-linear projection of inputs to a higher dimensional space allows the application of a single perceptron neuron to the complex BP problem. This requires that a perceptron is developed that is capable of implementation with many inputs, Cover's theorem, and in-situ training capability with fast critical path.

### 3.4.2 Design

The perceptron is implemented following (25) to calculate activation state and (27) to calculate neuron output using threshold function. The neuron has $N$ inputs, $\underline{A}^N$, and N+1 weights, $\underline{W}^{N+1}$, and the neuron has a single output which is 1 or 0, $T$ or $NT$.

$$\varphi(X) = \begin{cases} 0, & X \le 0 \\ 1, & X > 0 \end{cases} \tag{27}$$

The use of a simple threshold activation function is to meet requirement $\boldsymbol{O_2}$ as the neuron output is set only by the sign of the activation state. This is simple compared to ReLU and sigmoid activation functions, which require multiplication and addition operations. Making them less ideal compared to a simple sign check when targeting fast critical path for high-speed operation.

The perceptron is trained by the perceptron learning rule,

$$\underline{W}_{K+1}^{N+1} = \begin{cases} \underline{W}_K^{N+1} + \eta \cdot e_k \cdot \underline{A}_K & , e_k < 0 \\ \underline{W}_K^{N+1} & , e_k = 0 \\ \underline{W}_K^{N+1} + \eta \cdot e_k \cdot \underline{A}_K & , e_k > 0 \end{cases} \tag{28}$$

For incorrect classification, the weights vector is stepped in a direction determined by the error, $e_k$, and magnitude of the previous input vector, $\underline{A}_K$, proportional to the learning rate, $\eta$. With the error computed using,

$$e_k = d_k - \varphi(X_k) \tag{29}$$

The perceptron learning rule is a simple learning algorithm, as it steps the weights by a fixed proportion of each input determined by learning rate, $\eta$. Its simplicity makes it ideal as it only utilises addition and multiplication operations. Which for implementation can use fast integer adders, and for multiplication either fixed coefficient multipliers or bit shift if $\eta$ a reciprocal of a power of 2. Fulfilling requirements $\boldsymbol{O_1}$, $\boldsymbol{O_2}$, and $\boldsymbol{O_3}$.

### 3.4.3 Implementation

The perceptron neuron is implemented in System Verilog unlike most other components of the RC BPU. System Verilog provides versatile syntax for implementation of digital systems; which is used to develop the perceptron as a parameterizable module. Meeting deliverable requirement $\boldsymbol{O_3}$.

The perceptron implementation is comprised of two parts, the prediction logic and the training logic, shown in Fig. 39. The prediction logic is combinatorial single cycle logic, using an array of multipliers and an adder tree to compute the activation state $X$. From which the neurons output is set as the inverse of the twos compliment sign bit of the activation state $X$.

The perceptron's training logic consists of a two stage pipeline, Fig. 39. The first stage computes the weights update value with both positive and negative sign. The second stage then selects the correctly signed weights update values based on $e_k$ and applies them to the weights vector using integer adders. This follows the timing diagram, Fig. 38. For an input presented at cycle $k$, the prediction is ready at cycle $k + 1$, and the weights are trained at cycle $k + 2$ using the neuron inputs from cycle $k$.

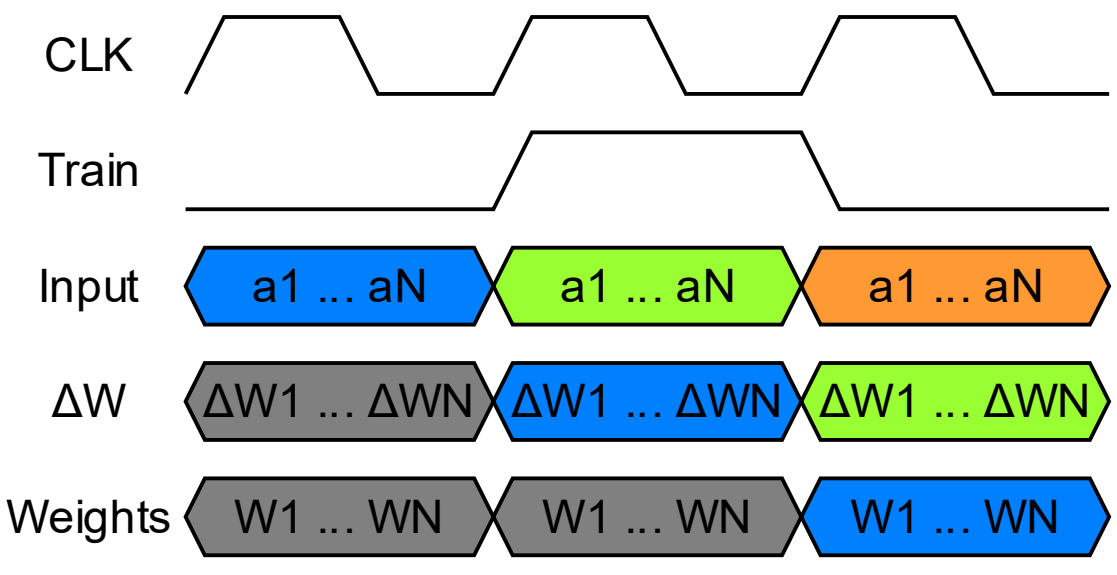

*Figure 38. Perceptron training logic timing diagram.*

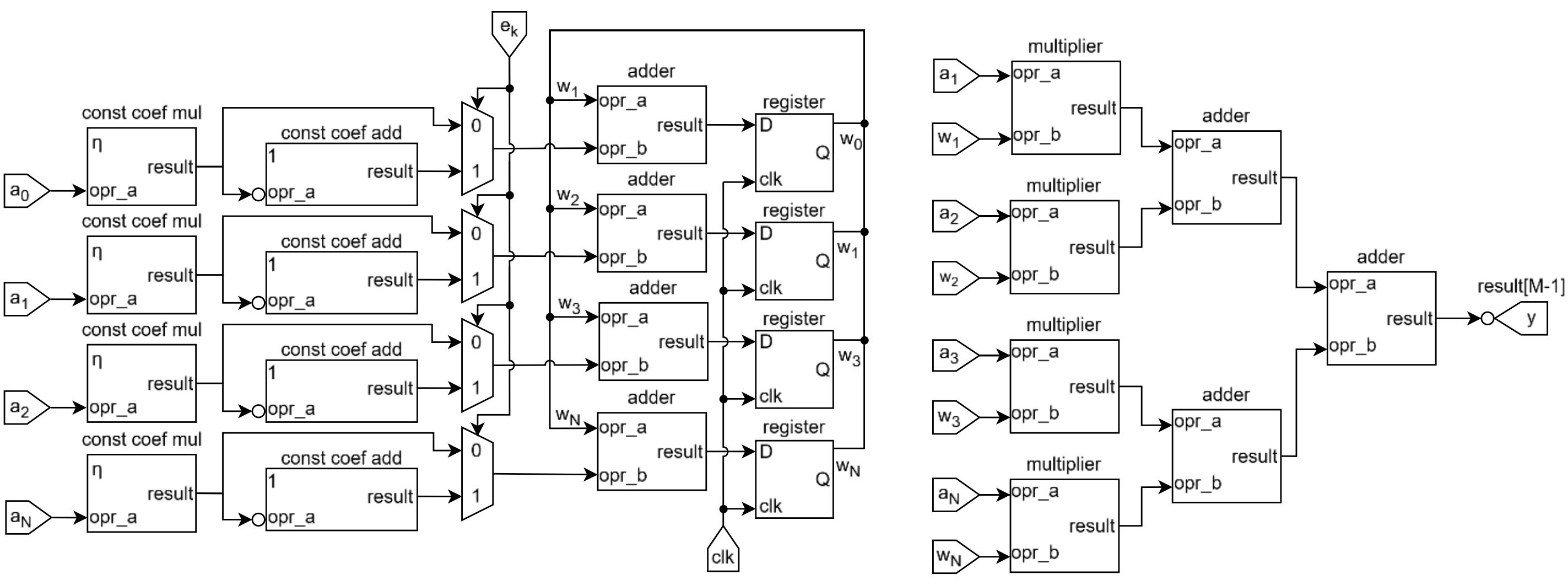

*Figure 39. perceptron prediction and training logic.*

The perceptron has two modes of operation, the first being real number modelling with FP64 values and the second being integer arithmetic. The FP64 operating mode acts as a control and is used to demonstrate basic functionality of the perceptron as part of the projects iterative design methodology. The integer arithmetic mode acts as the final realisation of the perceptron design. The operating mode is selected by module instantiation parameters, given in table 12. When configured to use integer arithmetic; parameters $Q_i$, $Q_w$, and $Q_a$ determine the bit depth of the inputs, weights, and accumulator respectively and are defined in (30).

$$\underline{W}^{N+1} \in \mathbb{B}^{Q_w \times (N+1)} \tag{30a}$$

$$\underline{A}^{N} \in \mathbb{B}^{Q_i \times N} \tag{30b}$$

$$X \in \mathbb{B}^{Q_a} \tag{30c}$$

*Table 12. module configuration parameters for perceptron neuron.*

| Parameter | Description |
|---|---|
| M | Operating Mode |
| N | Number of Inputs |
| $Q_i$ | Input quantisation (bits) |
| $Q_w$ | Weights quantisation (bits) |
| $R_w$ | Weights range |
| $Q_a$ | Accumulator quantisation |
| $\eta$ | Learning rate. |

### 3.4.4 Testing

Testing initially demonstrates the perceptron is capable of in-situ real time training at both FP64 and low bit depth quantisation of its weights, fulfilling deliverable requirement $\boldsymbol{O_1}$. Additionally, testing shows the importance of learning parameter tuning for fast adaptability and rapid convergence, fulfilling deliverable requirement $\boldsymbol{O_2}$. The achievable prediction latency is then assessed targeting implementation on the ASAP7 predictive 7nm PDK using OpenROAD EDA tools. The timing requirement of 1ns prediction latency is easily attainable for many configurations by targeting the super low voltage threshold libraries of ASAP7, fulfilling requirement $\boldsymbol{O_2}$. However, the stretch requirement of 0.5ns critical path, $\boldsymbol{O_4}$, is not achieved due to the cascaded adder tree presenting a significant constraint on achieving lower critical path timing.

#### *3.4.4a In-Situ Learning Testing*

The perceptron's ability to fit to an arbitrary function in operation has a direct impact on the accuracy of the resulting RC BPU design. Faster training directly results in increased adaptability of the perceptron to underlying branching pattern changes, which translates into increased prediction accuracy. As such, the perceptron's ability to perform binary classification on an arbitrary function is to be tested. To do this, the perceptron is trained to fit to (31) where inputs that sum to 20 or more are labelled as 1 and inputs that sum to less than 20 labelled as 0.

$$d = \begin{cases} 0, & a < 20 \\ 1, & a \geq 20 \end{cases} \qquad a = \sum_{i=1}^{4} x_i \tag{31}$$

The testing flow, Fig.40, first instantiates the perceptron with 4 4-bit inputs and random initial weights. Then successive input combinations are generated and fed to the perceptron. The resulting prediction is used to generate an error which the perceptron uses to update its weights. This test runs for 100,000 iterations; recording the perceptron's prediction as true positive (TP), false positive (FP), true negative (TN), and false negative (FN) at each step. From which the accuracy over time can be plotted, Fig.41. Perceptron configurations with FP64, INT10, and INT8 weights quantisation are tested with various learning rates to investigate their performance and ability to converge to solution weights.

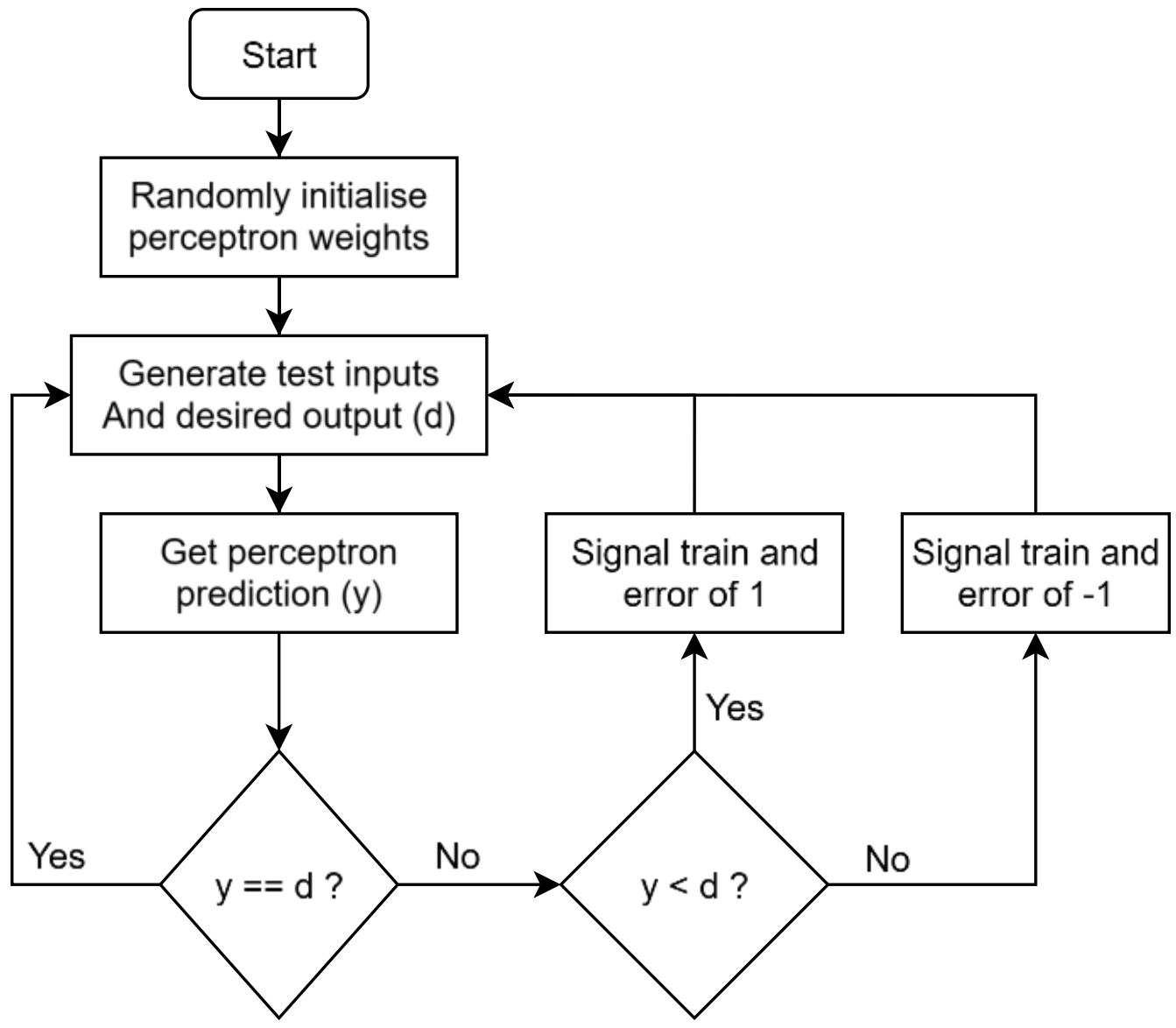

*Figure 40. testing loop for perceptron neuron verification.*

The perceptron manages to converge to a set of solution weights for all FP64 configurations, Fig.41a. Learning rate can be seen to have a significant effect on the speed of convergence, with learning rate of 0.01 leading to solution weights after 4800 observations. Conversely, other learning rates either lead to overshoot in the case of lr=0.1 or slower training in the case of lr=0.001 and lr=0.00001, but all eventually reach a set of solution weights. This demonstrates the significance of learning rate tuning to achieve an optimal trade-off between adaptability and stability of the perceptron.

Next, integer quantised perceptron configurations are tested. Two out of three INT10 configurations, with learning rates of 0.5 and 1.0, converge to a set of solution weights, Fig.41b. The third configuration with lr=0.1 never converges. This stems from bit truncation as a result of the low learning rate value. The 4-bit input truncates to a single bit when multiplied by a learning rate less than 0.125 and greater than 0.625. Effectively zeroing the weight update values except for input values > 8, leading to no training occurring for most inputs. This is an acceptable limitation, necessitated by the requirement of high frequency operation. For the two learning rates that do converge, it can be seen that lr=1.0 converges fastest. This is due to the relatively large value range that INT10 provides in comparison to the input. Which allows a high learning rate to be used without incurring significant overshoot in weight updates. Allowing lr=1.0 to converge after 10200 observations compared to 18600 for lr=0.5.

INT8 quantised configurations show a similar sensitivity to learning rate as the INT10 configurations, with only lr=0.2 converging to solution weights. This stems from the same issue of value truncation that hinders the INT10 configuration for lr=0.1 where no weight updates are made. For lr=0.5, there is significant overshoot due to the more comparable value ranges of the INT8 weights and the INT4 inputs. This overshoot leads the weights to overcorrect and never converge to a solution weights set. However, the perceptron still manages to partially fit to the problem and achieves an accuracy of 85% almost immediately. This further demonstrates the importance of learning rate parameter tuning to achieve optimal performance. The rapid convergence of lr=0.2 reinforces this, achieving a set of solution weights after just 2000 observations. Exhibiting exceptional performance ahead of even the FP64 quantised configurations. This is hypothesised to result from the learning rate of 0.2 leading to weight updates being in range of 4-0, 64x less than the range of the weights. Similar to the highest performing INT10 method that had a weight update range of 16-0 which was also 64x less than the range of INT10 weights. This means that the learning rate of 0.2 leads to an optimal trade-off between learning speed (rate) and bit truncation. Allowing the weights to step sufficiently but not overshoot while also not truncating the training inputs so much as to result in no weights stepping.

These results show that low bit depth quantised perceptron implementations are both functional and sufficiently capable of fitting to arbitrary functions. INT8 quantised weights are viable through the careful consideration of input data range and tuning of the learning rate parameter. Leading to a viable implementation capable of rapid convergence towards a set of general solution weights, satisfying the deliverables requirement $\boldsymbol{O_1}$.

### *3.4.4b Prediction Latency*

An important aspect of the perceptron neurons design is its ability to achieve high speed low latency prediction as part of the deliverable requirement $\boldsymbol{O_2}$. Achievable prediction latency is assessed by performing static timing analysis (STA) using the OpenROAD EDA platform [41], which allows the use of the open source ASAP7 predictive 7nm PDK [42]. OpenROAD flow scripts (ORFS) automates the entire RTL -> GDSII flow, fully synthesising and placing all designed logic. The perceptron logic targets the super low voltage threshold (SLVT) library of ASAP7, which is ideal for high speed digital logic where increased current leakage is an acceptable trade-off for faster switching.

The perceptron's achievable operating frequency are determined by the ability of ORFS to realise the design without timing violations. To do this, a clock period of 1000ps in addition to external input and output signal delay of 100ps is specified in a constraints.sdc file. The ORFS RTL -> GDSII flow is then ran with the flags shown in Table 13 to target the ASAP7 SLVT libraries.

*Table 13. ORFS configuration variables.*

| | |
|---|---|
| ASAP7_USE_VT | SLVT |
| SYNTH_HIERARCHICAL | false |
| SYNTH_HDL_FRONTEND | slang |
| ABC_SPEED | true |
| CLUSTER_FLOPS | true |
| CORE_UTILIZATION | 70% |

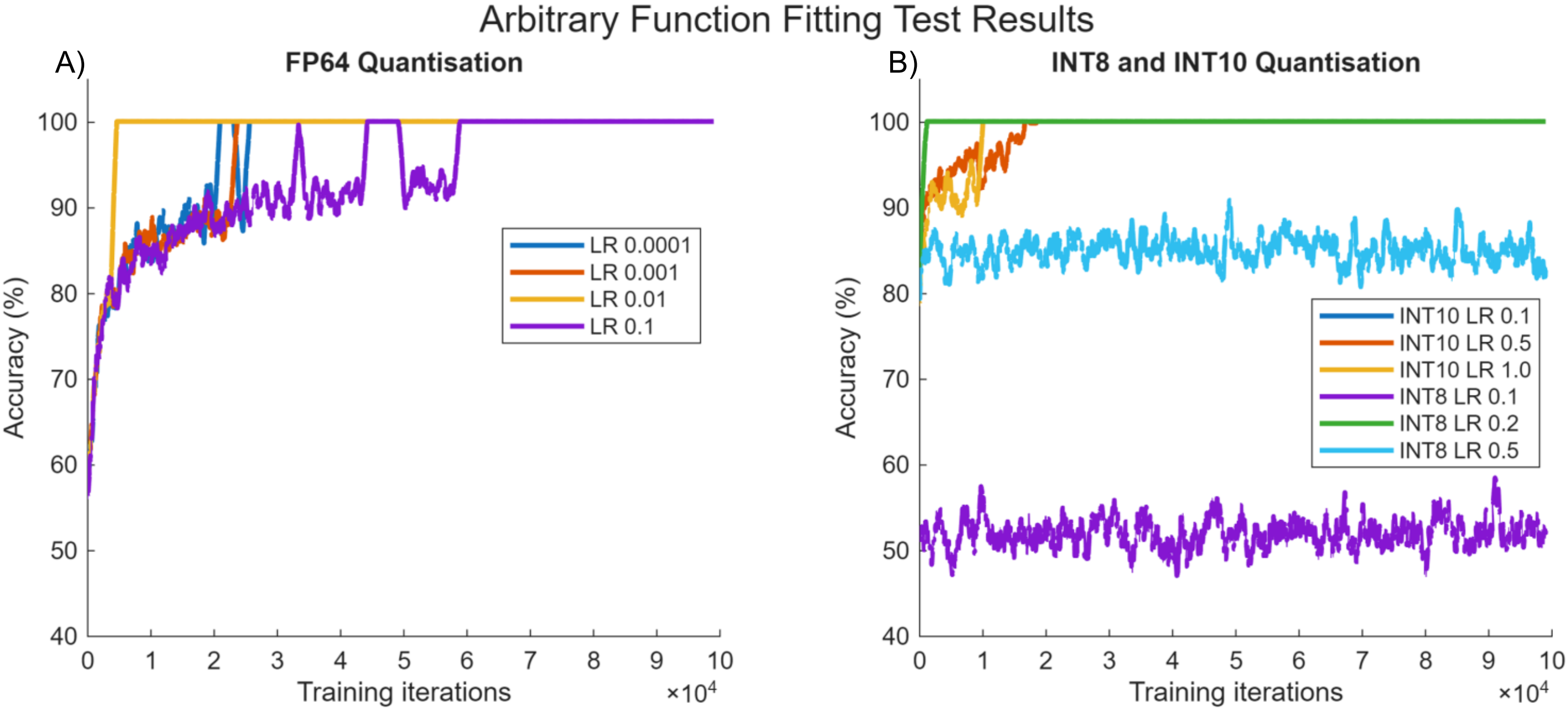

*Figure 41. arbitrary function fit test results for FP64 quantised configurations (a), and integer quantised configurations (b).*

ORFS is set to favour speed over area by the ABC_SPEED flag. The optimisation of clustering multi bit flipflops is enabled by the CLUSTER_FLOPS flag. ORFS is further instructed to target a high core utilisation of 70% to achieve shorter wires and hence reduce critical path timing by the CORE_UTILIZATION flag.

Both INT8 and INT10 weight quantisation is considered for timing analysis, the resulting physical designs are shown in Fig.42. INT 8 configurations with 8, 16, 32, and 48 inputs passed STA; achieving 339ps, 222ps, 78ps, and 13ps of slack respectively. With positive slack indicating that timing requirements are met. INT10 configurations of 16 and 32 inputs passed STA with slack of 256ps and 71ps respectively. The main limitation in achieving faster operation for both INT8 and INT10 configurations is the combinatorial delay added by the adder tree as part of the prediction logic. Resulting in INT8 configurations with 64 inputs failing due to a mix of increased wire length from larger area and the addition of an extra level to the adder tree. With INT10 weights configuration failing for 48 or more inputs for the same reason.

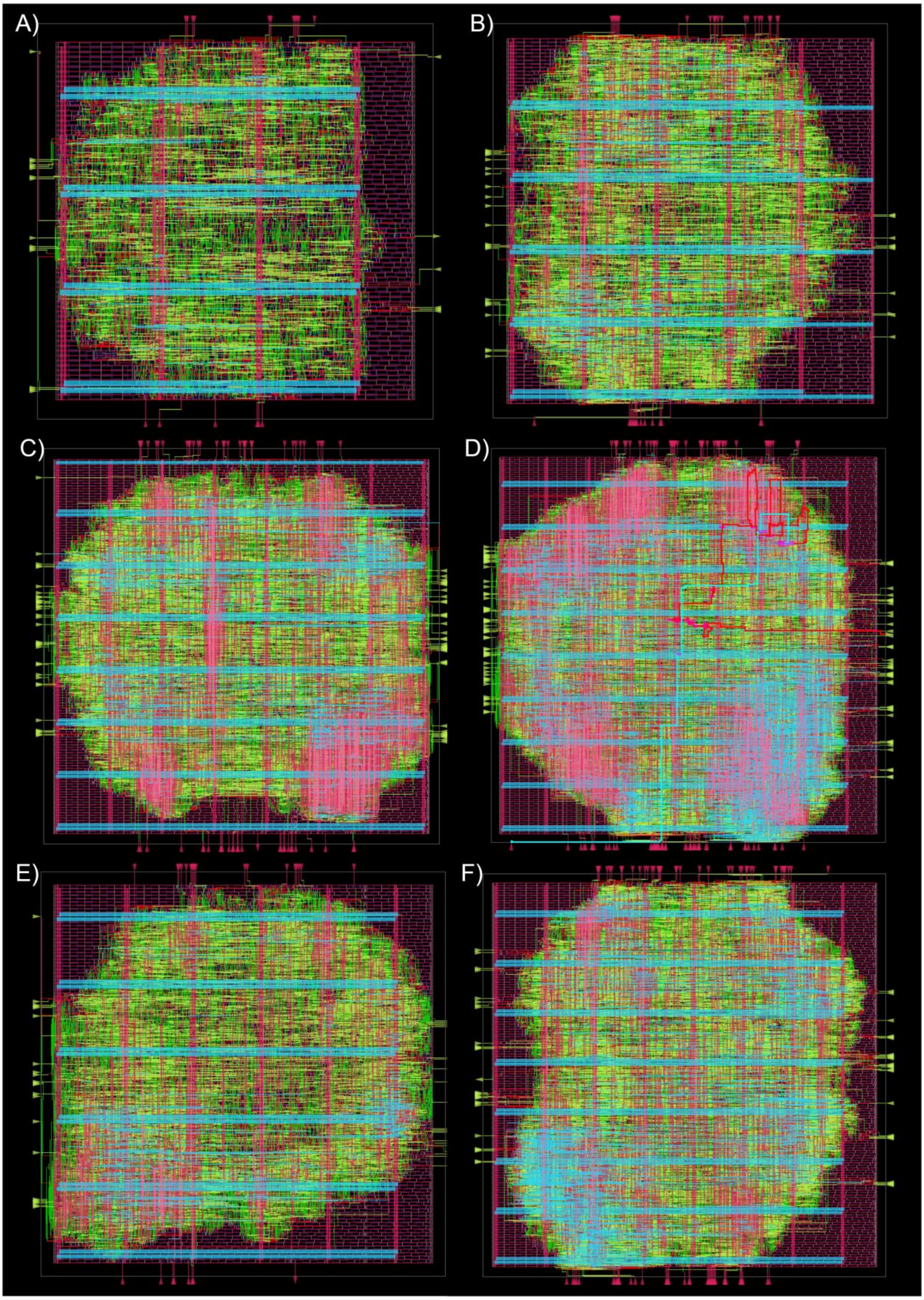

*Figure 42. GDSII results from ORFS for 8 inputs with 8-bit weights (a), 16 inputs with 8-bit weights (b), 32 inputs with 8-bit weights (c), 48 inputs with 8-bit weights (d), 16 inputs with 10-bit weights(e), and 32 inputs with 10-bit weights (f).*

## 3.5 Full System Integration

This section presents the successful full-system integration and validation of the proposed RC BPU architecture within the Siemens Symphony AMS mixed-signal simulation environment. Validation using a temporal sequence detection workload demonstrates real-time adaptability and sequence classification accuracy exceeding 70%, satisfying requirement $\boldsymbol{F_1}$. Testing further investigates the impact of reservoir configuration and memory characteristics on system performance, fulfilling requirements $\boldsymbol{F_2}$ and $\boldsymbol{F_3}$. While operation at 1GHz demonstrates compatibility with the target prediction latency of ≤2 ns, satisfying requirement $\boldsymbol{F_4}$. Additionally, the modular mixed-signal implementation and automated reservoir generation framework enable efficient design-space exploration across varying reservoir configurations and operating conditions, satisfying requirement $\boldsymbol{F_5}$.

### 3.5.1 Implementation

The implementation of the RC BPU design from Section 3.2 consists of five SV and VAMS modules, Fig.43, split across analogue and digital domains. The reservoir design and implementation framework developed in Section 3.3 is encapsulated inside the reservoir harness module alongside 1-bit input DACs and 4-bit output ADCs. The reservoir harness is then encapsulated inside the RC BPU module alongside the digital RC BPU control logic and perceptron readout layer. The reservoir and perceptron modules remain unchanged from their design in Sections 3.3 and 3.4 respectively.

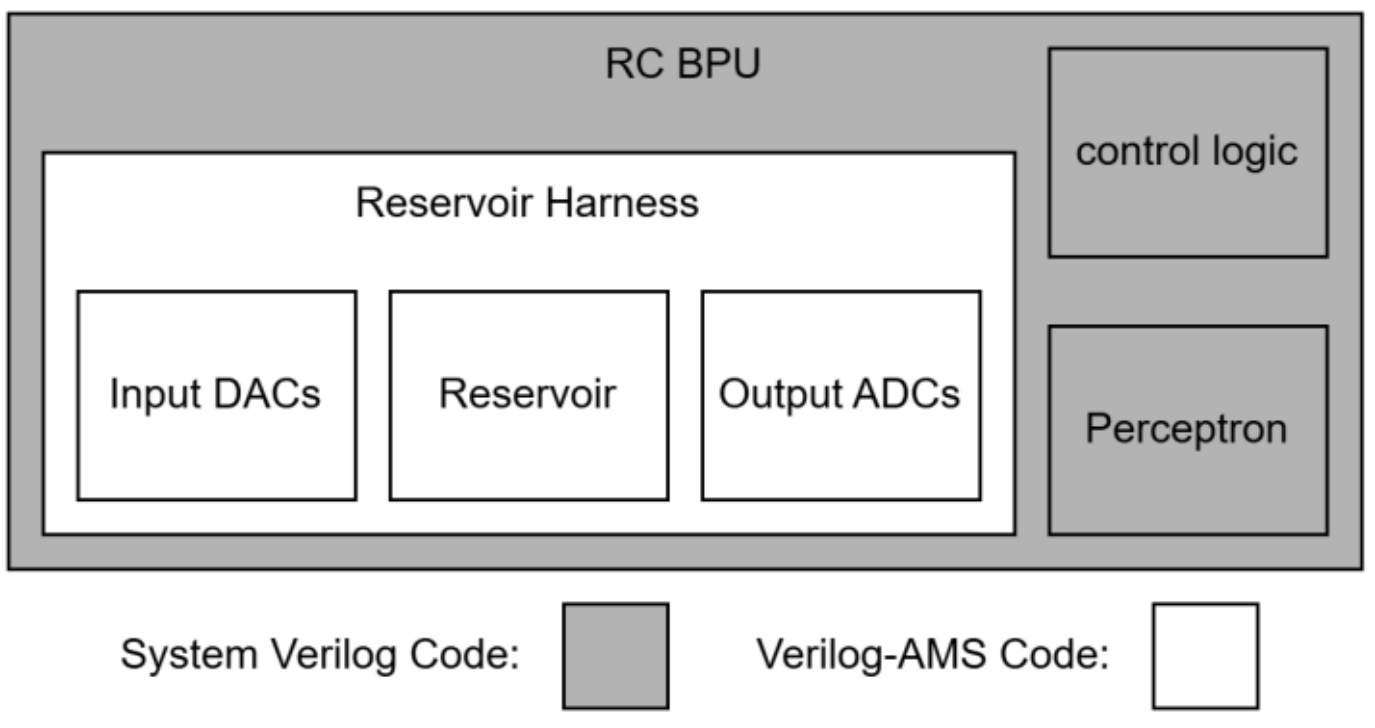

*Figure 43. RC BPU design implementation component overview.*

#### *3.5.1a RC BPU Module*

The RC BPU module is the top level module which encapsulates the entire RC BPU design, providing a single module interface for RC BPU testing. It has the instantiation parameters given in table 16 and the module port interface given in table 14.

The RC BPU module directly implements the BPUs control logic, Fig.43, which is responsible for control and coordination of each subcomponent. Control logic performs the following functions:

- Error signal to perceptron
- Flopping the RC BPU prediction output
- Unpacking the reservoir harness output data
- Flopping the predict enable signal to the perceptron

The error signal to the perceptron is given as a continuous assignment of the perceptron's prediction output. When the perceptron predicts true (1) and the correct output is false (0) the error is -1, when the perceptron predicts false (0) and the correct output is true (1) the error is 1. As the perceptron error signal is 2s complement, this allows the buffered perceptron prediction from the previous cycle to be directly fed back to it as the error signal.

The RC BPU control logic also handles the flopping of output ports to synchronise outputs to the positive clock edge. This involves buffering the perceptron prediction for 1 clock cycle as perceptron prediction logic is asynchronous. The predict enable module input signal is also buffered for one cycle to align the predict enable signal arrival to the perceptron with the relevant sampled reservoir state resulting from the predictor data input.

*Table 14. RC BPU module ports.*

| Port Name | Direction | Type | Description |
|---|---|---|---|
| clk | in | logic | Clock signal |
| rst_n | in | logic | Negative Reset |
| pred_en | in | logic | Prediction enable |
| trn_en | in | logic | Train enable |
| dat_inp | in | logic [N:0] | Input data |
| pred_oup | out | logic | Prediction result |

#### *3.5.1b Reservoir Harness Module*

The reservoir harness module encapsulates all the RC BPUs analogue behaviour components. It was developed as a result of a bug in Siemens Symphony AMS simulator which resulted in erratic and undefined behaviour when handing unpacked arrays. This manifested as incorrect DAC output values from the reservoir input layer. The reservoir harness module mitigates this by providing a simplified interface between the analogue and digital simulation domains by using a packed array type at the module interface instead of unpacked.

The module port interface is given in table 15. It consists of a clock signal for synchronising reservoir output layer ADCs and two digital data connections. *Dat_inp* is the input binary bit vector for the reservoir input layer, with N being the number of reservoir input nodes. *Dat_oup* is a flattened one dimensional binary value array of the digitised voltages from the reservoir output layer ADCs. Its size, M, is given by the number of output nodes multiplied by the bit depth of the reservoir output layer ADCs.

*Table 15. Reservoir harness module ports.*

| Port Name | Direction | Type | Description |
|---|---|---|---|
| clk | in | logic | Clock signal |
| dat_inp | in | logic [N-1:0] | Negative Reset |
| dat_oup | out | logic [M-1:0] | Prediction enable |

The reservoir input layer is implemented directly by the reservoir harness module. VAMS *generate* syntax is used to implement an array of 1-bit ADCs which transition the voltage at each input node to either 0V or $V_{DD}$ as described in Section 3.2.1.

The reservoir output layer is implemented by instantiating a generic ADC module using VAMS *generate* syntax. $V_{ref}$ and ADC bit depth are set by instantiation parameters dictated by the RC BPU module parameters VDD and RES_OUP_DEPTH, table 15. These are passed through the reservoir harness as the only instantiation parameters of the reservoir harness module.

The final aspect of the reservoir harness is the reservoir itself. The reservoir is instantiated by macros provided by the header file that *resgen* creates during reservoir generation. The use of macros allows any reservoir configuration to be generated and passed to the simulator with no code modification required for integration. The provided definitions of *`RESERVOIR_INPUTs* and *`RESERVOIR_OUTPUTS* from the reservoir header file set the port widths globally. Bypassing the need for module level reconfiguration via instantiation parameters.

*Table 16. RC BPU module instantiation parameters.*

| Parameter | Type | Default Value | Description |
|---|---|---|---|
| PN_USE_REAL_VALS | bool | 0 | Enables the use of FP64 emulation for the perceptron neuron |
| RES_OUP_DEPTH | integer | 4 | Controls the bit depth of reservoir output voltage sampling ADCs |
| PN_WEIGHTS_DEPTH | integer | 8 | Controls the bit depth of perceptron neuron weights when PN_USE_REAL_VALS is set to 0. |
| PN_LR | real | 1 | Controls the learning rate of the perceptron neuron |
| VDD | real | 3.3 | Controls the operating voltage of the reservoir input DACs. |

### 3.5.2 Sequence Detection Workload

Number sequence detection is used to validate the functionality of the completed RC BPU design. It involves monitoring a stream of data to identify the presence of predefined numerical patterns. Requiring the system to recognise and distinguish structured patterns while also processing temporal information, making it an ideal workload with which to demonstrate the RC BPU design.

The effectiveness of sequence detection depends largely on the reservoir's memory capacity and its ability to process temporal information. In particular, the system must retain relevant past inputs (fading memory) while transforming them into meaningful internal representations that support accurate classification. And as a result, sequence detection provides a meaningful representation of the RC implementations performance.

Testing is carried out by first generating a predefined sequence of numbers with a fixed length. This sequence is then embedded into a longer input stream, where random numbers are inserted between occurrences to act as noise. A transient simulation is performed in which this input stream is applied to the RC BPU, allowing it to process and predict the presence of the sequence in real time. During the simulation, the output predictions of the RC BPU are continuously monitored. Whenever an incorrect classification is detected, the training enable signal is asserted to allow the RC BPU to update the perceptron weights to better fit to the problem. This approach allows the models ability to adapt dynamically and ability to detect an embedded sequence to be assessed with a single test.

Testing is carried out in the Siemens Symphony AMS simulator environment using the python script based verification flows as part of the project's methodology. This allows the variation of sequence length for detection alongside operating voltage and frequency through simple script variable modifications.

### 3.5.3 Testing Results

*3.5.3a Initial RC System Testing*

The first step in testing is to verify the functionality of the RC BPU and its ability to fit to the sequence detection problem in real-time. The RC BPUs ability to fit to the sequence detection problem within 500 observations is tested for varying sequence lengths. This is done using the test configurations in Table 17, with results shown in Table 18 as the rate of true positive and true negative classifications and Fig.44 as the rate of true positive classifications. The rate of true positive classifications is defined as the percentage of sequences accurately detected at the final input value of the sequence. The rate of true negative classifications is defined as the percentage of inputs accurately classified as not being the end of a sequence.

*Table 17. Testing configurations for initial RC BPU sequence detection testing.*

| Parameter | Value |
|---|---|
| RC BPU Configuration | |
| Perceptron Weights | INT10 |
| Reservoir Output Bit-Depth | 4-bit |
| Perceptron Learning Rate | 1 |
| $V_{DD}$ | 3.3V |
| Reservoir Configuration | |
| Memristor Count | 200 |
| Memristor Parameters | Table 7 |
| Maximum Node Connections | 10 |
| Minimum Node Connections | 6 |
| Number of Inputs | 8 |
| Number of Outputs | 32 |
| Input Placement Algorithm | Even |
| Output Placement Algorithm | Even |
| Test Configuration | |
| Operating Frequency | 1GHz |
| Noise padding | 100 values |
| Sequence length | 5, 10, 50, 100 values |
| Sequence count | 500 |

The initial results for sequence detection, table 18, shows that the RC BPU performs excellently at rejecting inputs that do not signify the end of the sequence. With the minimum true negative rate of 99.51%. However, it significantly underperforms when it comes to detection of the sequence; achieving a maximum overall true positive detection rate of 81.8% over 500 sequence samples. This results from the finite adaptability of the RC BPU, which is exposed to upwards of 20x more inputs that do not signify the presence of a sequence. Resulting in the bias towards predicting no sequence detected. Additionally, the overall true positive rates are brought down by early inaccuracy before the RC BPU is able to sufficiently fit to the problem, Fig. 44.

*Table 18. Results of initial sequence detection testing.*

| Sequence Length | True Positive Rate | True Negative Rate |
|---|---|---|
| 5 | 55.40% | 99.51% |
| 10 | 81.80% | 99.82% |
| 50 | 44.20% | 99.56% |
| 100 | 73.40% | 99.84% |

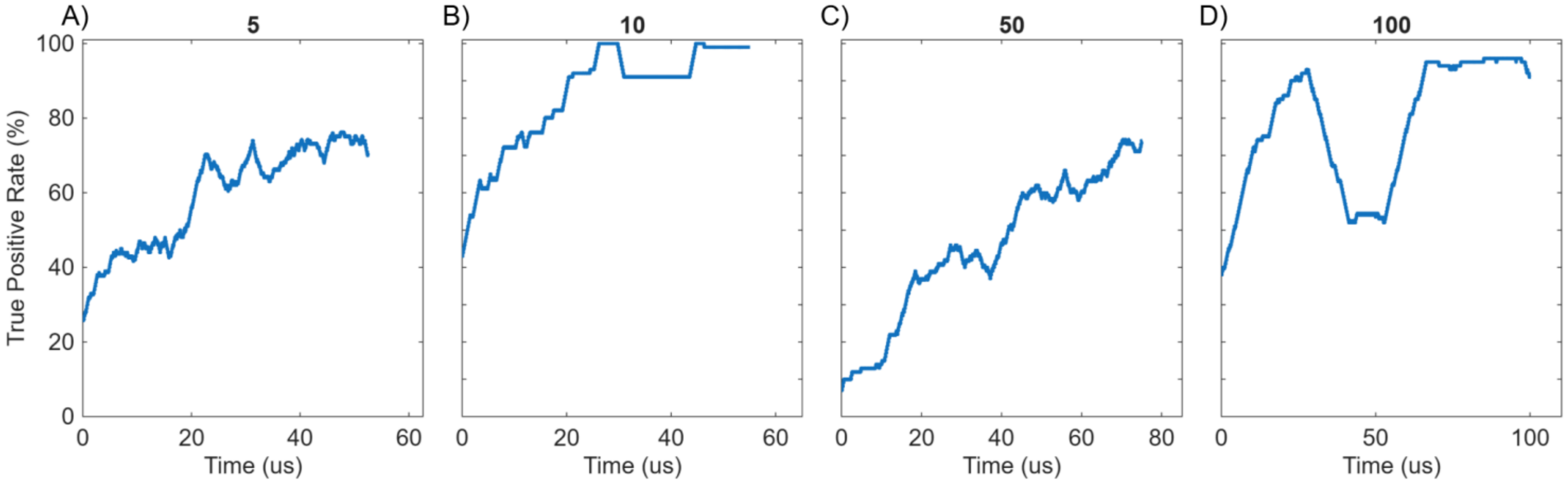

*Figure 44. True positive rate for initial sequence detection test with setup from Table 16. For sequence lengths of 5 (a), 10 (b), 50 (c), and 100 (d).*

Plots of the true positive rate over time show that the RC BPU is able to achieve a final true positive rate exceeding 70% for all sequence lengths. The improvement of true positive rate over time for all sequence lengths demonstrates the correct functioning of the perceptron's training logic to achieve real-time workload adaptability. Importantly, results show that the singular RC BPU configuration is able to not just adapt in real-time but also interpret a wide variety of sequence lengths accurately. This signifies its ability to both adapt to predict infrequent large scale patterns and more localised shorter patterns.

### *3.5.3b Output Dimensionality Testing*

The next important property to assess is the effect of the reservoir output dimensionality on the ability of the RC BPU to adapt to a given problem. This is tested due to the indicated lack of importance of output dimensions beyond the first 10-12 given by SVD analysis in Section 3.3.4d. The testing parameters, Table 19, are designed to test medium sequence length of 50 with significant noise padding of 100 samples in-between each sequence.

*Table 19. Varying reservoir dimensionality testing.*

| Parameter | Value |
|---|---|
| RC BPU Configuration | |
| Perceptron Weights | FP64 |
| Reservoir Output Bit-Depth | 4-bit |
| Perceptron Learning Rate | 0.1 |
| $V_{DD}$ | 3.3V |
| Reservoir Configuration | |
| Memristor Count | 200 |
| Memristor Parameters | Table 7 |
| Maximum Node Connections | 10 |
| Minimum Node Connections | 6 |
| Number of Inputs | 8 |
| Number of Outputs | 8, 16, 32 |
| Input Placement Algorithm | Even |
| Output Placement Algorithm | Even |
| Test Configuration | |
| Operating Frequency | 1GHz |
| Noise padding | 100 values |
| Sequence length | 50 values |
| Sequence count | 2000 |

The results, Fig. 45 and Table 20, show that the number of reservoir outputs has a significant effect on both the rate of adaptability and the achievable sequence detection performance.

*Table 20. Results of varying reservoir output dimensionality testing.*

| Reservoir Outputs | True Positive Rate | True Negative Rate |
|---|---|---|
| 8 | 12.50% | 99.35% |
| 16 | 28.95% | 99.40% |
| 32 | 71.50% | 99.78% |

Reducing the number of reservoir outputs results in a substantial degradation in both the final true positive rate and the adaptability of the RC system. Specifically, decreasing the number of outputs from 32 to 16 leads to a reduction in the overall true positive rate from 71.5% to 28.95%, as shown in Table 20. This decline is accompanied by a markedly slower convergence towards higher true positive rates, as illustrated in Fig. 45. These results run contrary to the conclusions inferred from the singular value decomposition (SVD) analysis presented in Section 3.3.4d. As the reservoir output dimensionality is instead insufficient rather than excessive for achieving improved system performance.

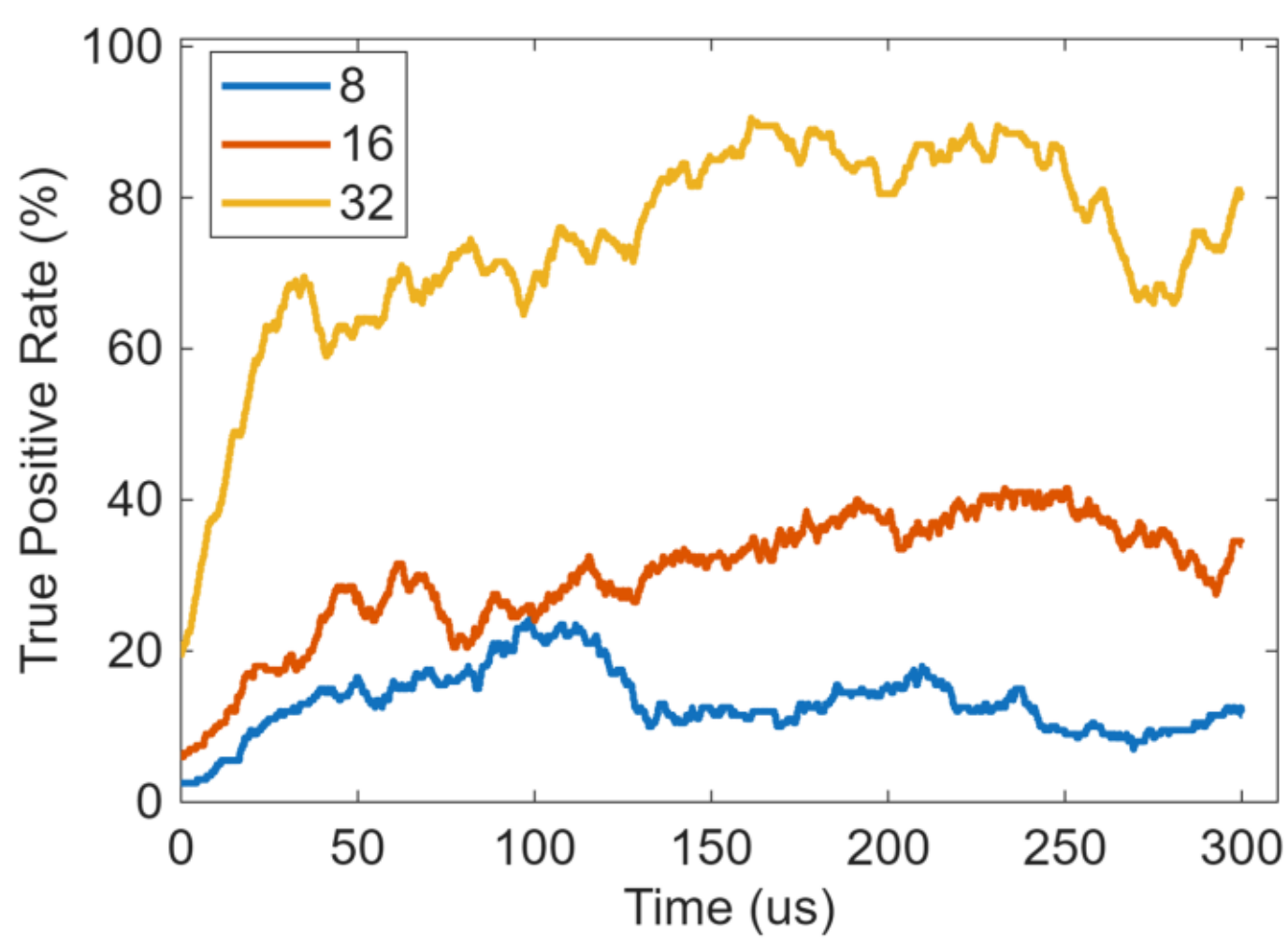

*Figure 45. Pattern recognition test results with test configuration from table 18 with 8, 16, and 32 outputs.*

A further reduction in reservoir outputs from 16 to 8 causes the RC system to become largely incapable of effectively fitting to the sequence detection task. Although the true positive rate initially increases from approximately 3% to 22% at 100 µs, performance subsequently deteriorates to approximately 14%. Considering that only 1 in 150 input samples corresponds to a valid sequence detection, this result indicates that the reservoir retains a limited capacity for information processing. However, the achieved performance remains substantially below the threshold required for reliable and practical sequence detection.

One similarity is that all reservoir configurations achieve nearly identical true negative rates. This can be attributed not to the reservoir's dynamics processing information, but to the 1:150 ratio of inputs labelled as being the end of a sequence and inputs labelled as not being the end of a sequence. Resulting in a bias towards prediction that no sequence is detected, and hence the consistently high true negative results.

#### *3.5.3c Reservoir Density Testing*

Results from Section 3.3.4a lead to the hypothesis that more densely connected reservoirs are preferable due to their more consistent fading memory. Testing is performed to investigate the validity of this hypothesis. Using the testing parameters given in Table 21, the RC system is benchmarked for its ability to learn to detect a sequence of 50 consecutive samples embedded in random noise. Three different reservoir densities are tested, mimicking those tested in Section 3.3.4a.

*Table 21. Testing parameters for varying reservoir density testing.*

| Parameter | Value |
|---|---|
| RC BPU Configuration | |
| Perceptron Weights | FP64 |
| Reservoir Output Bit-Depth | 4-bit |
| Perceptron Learning Rate | 0.1 |
| $V_{DD}$ | 3.3V |
| Reservoir Configuration | |
| Memristor Count | 200 |
| Memristor Parameters | Table 7 |
| Maximum Node Connections | 10, 7, 5 |
| Minimum Node Connections | 6, 4, 3 |
| Number of Inputs | 8 |
| Number of Outputs | 32 |
| Input Placement Algorithm | Even |
| Output Placement Algorithm | Even |
| Test Configuration | |
| Operating Frequency | 1GHz |
| Noise padding | 100 values |
| Sequence length | 50 values |
| Sequence count | 500 |

The testing results presented in Table 22 demonstrate that reservoir connection density has a significant influence on system performance. However, the observed trend contradicts the hypothesis proposed in Section 3.3.4a, as lower-density reservoir configurations consistently outperform their denser counterparts. Reducing the connection density to 5–3 connections per node increases the overall true positive rate to 64.80%, compared to 44.20% obtained during the initial testing presented in Section 3.5.3a. This improvement suggests an increase in information processing capability arising from the reduced structural homogeneity of lower-density reservoirs. The resulting increase in dynamical diversity produces more varied and complex reservoir responses, thereby enhancing the system's ability to process and discriminate temporal patterns.

*Table 22. Overall true positive and true negative rates for varying reservoir density testing.*

| Reservoir Density | True Positive Rate | True Negative Rate |
|---|---|---|
| 5 - 3 | 64.80% | 99.76% |
| 7 - 4 | 46.20% | 99.57% |
| 10 - 6 | 43.20% | 99.53% |

The hypothesis of enhanced information processing capability in sparser reservoir configurations is further supported by the rate at which the system learns to detect the target sequence. As illustrated in Fig. 46, sparser reservoirs achieve substantially higher initial true positive rates at t=1µs. In particular, the 5–3 connections-per-node configuration attains an initial true positive rate of 22%, compared to only 3% for the 10–6 configuration. This trend persists throughout training, with RC systems with lower-density reservoirs consistently adapting to the task more rapidly than their denser counterparts.

Although sparser reservoirs exhibit faster learning dynamics, their improvement in true positive rate is characterised by increased variability, including localised decreases in accuracy and temporary performance plateaus. In contrast, denser reservoirs display more stable but slower convergence behaviour. This increased instability is likely attributable to the higher degree of system noise observed in Section 3.3.4a. Nevertheless, the negative impact of this variability is outweighed by the substantial increase in computational and representational capacity provided by the sparser reservoir configurations.

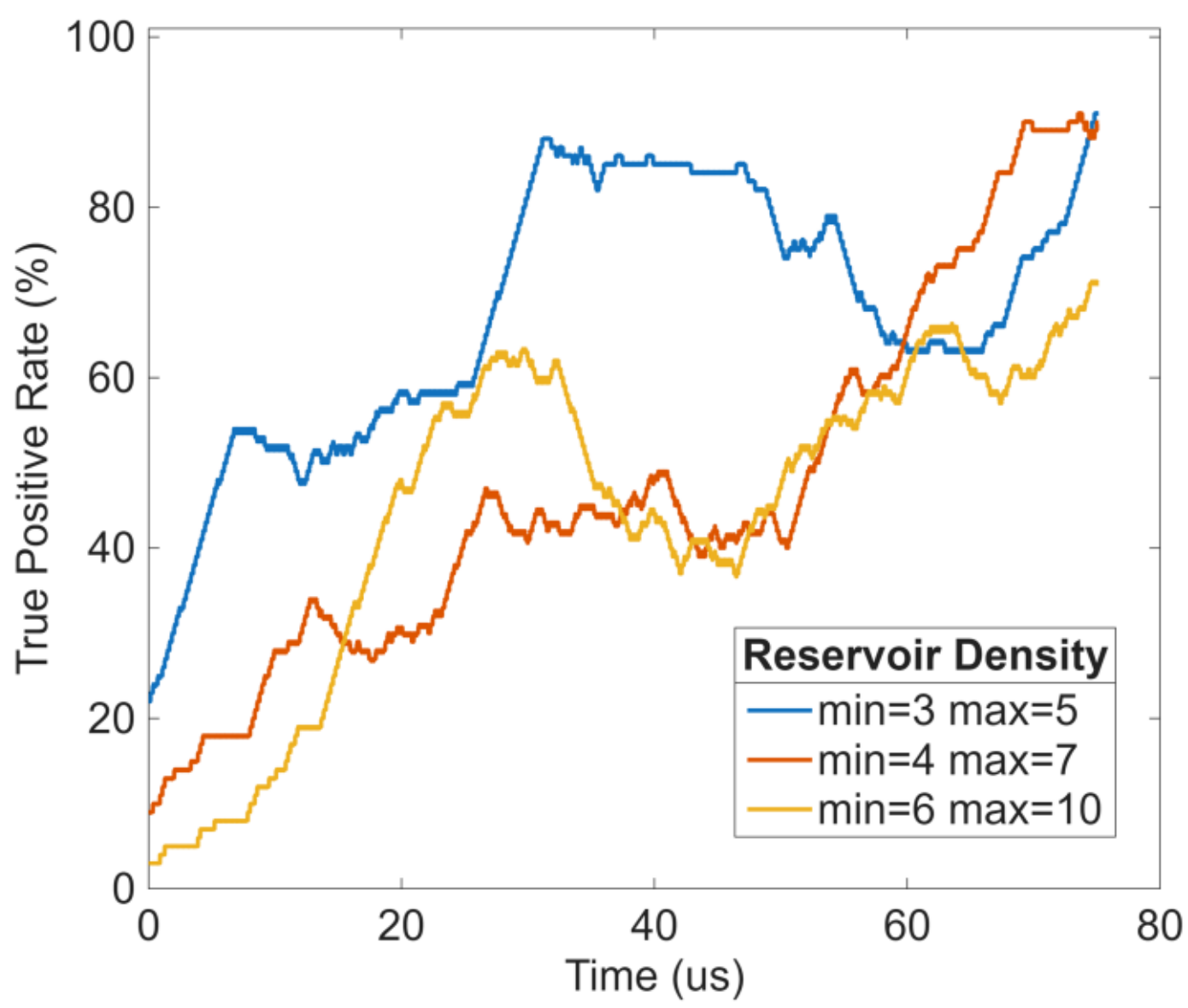

*Figure 46. Pattern recognition test true positive rate over time for reservoir density variation testing at densities from Table 21.*

Testing demonstrates that reservoir density is a significant design parameter governing the trade-off between fading memory consistency and dynamical complexity within the reservoir. Increasing reservoir sparsity enhances the diversity and richness of the nonlinear reservoir dynamics, thereby improving the computational capability of the RC system for temporal pattern classification. However, this occurs at the expense of reduced fading memory consistency and increased response variability. Reservoir density therefore represents a critical optimisation parameter in final RC BPU design. Being closely coupled to the broader architectural trade-offs between operating voltage, prediction latency, dynamical stability, and overall computational capability.

# 4 Final System Testing

This section presents the final benchmarking and evaluation of the developed RC BPU using the Dhrystone benchmark as a workload for branch prediction. Testing uses the RISC-V RV64GC ISA [43] as it is an open-source and royalty-free specification with broad industry adoption and the required toolchain support. Using the Dhrystone workload, the RC BPU is evaluated in terms of prediction accuracy, convergence behaviour, and adaptability across a range of different configurations. Multiple reservoir input encoding schemes and architectural parameters are investigated to determine the key trade-offs governing RC BPU performance.

The benchmarking methodology, workload generation flow, and instruction trace preprocessing stages are first described, after which various RC BPU design configurations are benchmarked and analysed. Through testing, all project aims are successfully verified, with the final RC BPU design operating at 1GHz with a 2.8V reservoir operating voltage while achieving 96% overall prediction accuracy on the Dhrystone workload. However, comparison against the TAGE predictor shows that the developed RC BPU implementation remains significantly less effective than current state-of-the-art branch predictors, with TAGE converging to 100% accuracy approximately 15× faster than the RC BPU. Achieving an overall accuracy of 99.93%.

## 4.1 Testing Setup

The Dhrystone benchmark, detailed in Section 4.1.1, is used to assess the performance of the developed RC BPU design for branch prediction. The benchmark is first ran through a RISC-V functional model to generate an instruction stream, detailed in Section 4.1.2. The instruction stream is then processed into usable inputs for the RC BPU, detailed in Section 4.1.3, which consist of input values (X) and desired classification labels (D). The RC BPU is then tested in simulation using Siemens Symphony AMS simulator with the test setup shown in Fig. 47. The prepared inputs and labels are fed into the RC BPU which then gives an output prediction with a 2 cycle delay. The output prediction is then compared to the classification label to signal train enable when an incorrect prediction is made, providing the feedback mechanism for training. The classification accuracy of taken (T) or not taken (NT) is recorded to a .csv file for later analysis using MATLAB.

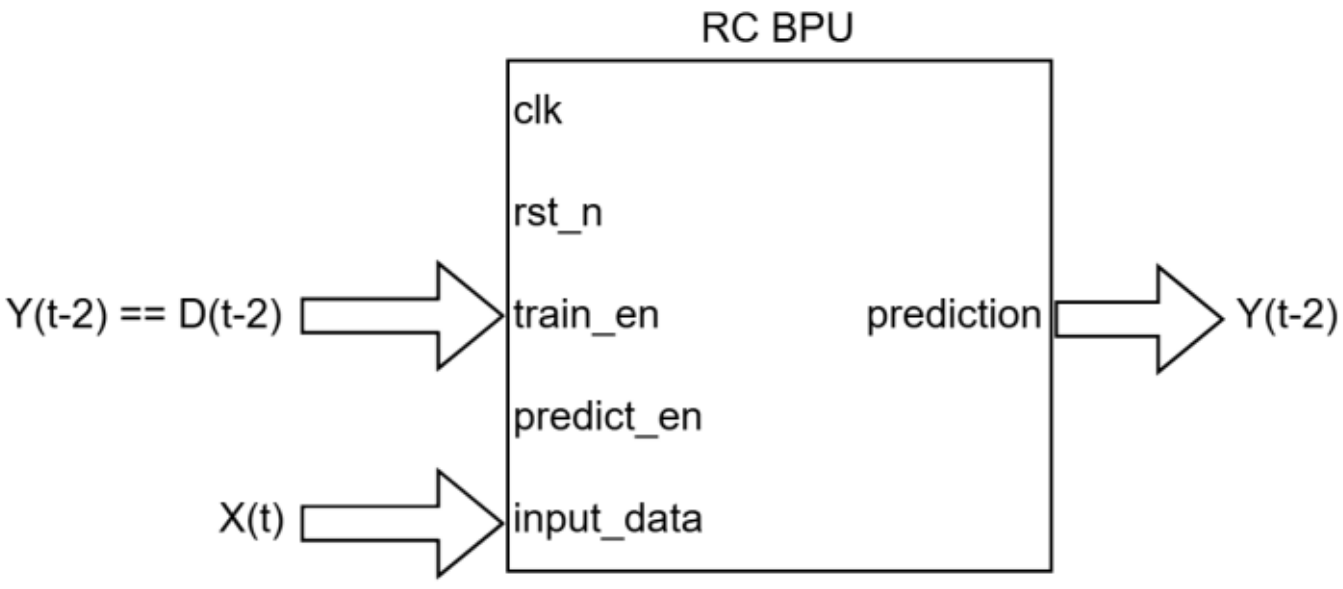

*Figure 47. RC BPU testing setup for BP workload benchmarking.*

### 4.1.1 Workload

Dhrystone [23] is used for assessing BPU performance, with code provided as part of the Condor RISC-V functional model discussed in Section 4.1.2. It was initially developed in 1984 as a CPU and compiler performance benchmark representative of contemporary workloads. Roughly 26% of Dhrystone code is control statements that represent branches [23], with a mix of infrequently taken hard to predict branches and frequently taken branches from loops. Dhrystone as a benchmark is relatively outdated and has been superseded by benchmarks such as SPEC INT and MiBench [44]. Despite this, Dhrystone's simplicity and generous use of branching instructions coupled with availability of ready to run code makes it suitable for BP workload benchmarking.

Three Dhrystone configurations are tested based on the code from Condor [22] with 1000, 2000, and 4000 benchmark iterations.

*Table 23. Dhrystone benchmark configurations.*

| Configuration | Iterations | Instruction Trace Size |
|---|---|---|
| 1 | 1,000 | 287,010 |
| 2 | 2,000 | 574,020 |
| 3 | 4,000 | 1,152,020 |

### 4.1.2 Trace Generation

The bare metal Dhrystone binaries are converted into traces using the Condor fork of the RISC-V spike functional model [22]. This emulates a RISC-V CPU and executes the compiled Dhrystone binaries and generates an output trace.

The generated trace contains a list of all the executed instructions in the order that they were ran. This provides a detailed chronological record of every instruction executed by the CPU while running the Dhrystone benchmark. It also provides additional information such as recording if branches are taken or not taken and the target addresses and program counter for each instruction.

### 4.1.3 Trace Processing

The traces generated by the spike functional model are converted into usable RC BPU inputs by a developed C++ tool. The tool produces a .csv file with the RC BPU inputs, encoded as integers, in column one and associated classification labels in column two.

There are three different RC BPU input encoding schemes that are implemented. Each encoding scheme is designed to distil the instructions down to a minimal representation that contains only necessary information to classify taken or not taken. This is necessary as inputting all 32 instruction bits would require a likely infeasibly large reservoir.

The first is to assign every instruction in the RISC-V ISA a unique integer which is its encoding for RC BPU input. The second approach is to input only branching instructions to the RC BPU using one-hot encoding. The third approach is to input only branching instructions; encoded as integers with address bits concatenated to form the full input.

#### *4.1.3a Approach One*

The first input encoding scheme assigns each instruction in the selected RISC-V ISA a unique integer label, which is then used as the reservoir input encoding. The chosen RV64GC ISA contains 208 unique instruction encodings, representing 168 distinct instructions and requiring only 8 reservoir input nodes. This scheme offers the advantage of a direct 1:1 mapping between a decoded instruction and its corresponding reservoir input, enabling implementation through a simple lookup table (LUT).

However, this approach requires every decoded instruction to be supplied to the reservoir. Making it impractical for execution pipelines that support parallel instruction decode, which would require the pipeline to stall to wait for each instruction to be input to the reservoir one at a time. Furthermore, this approach does not include instruction program counter information in the encoded reservoir input. Depriving the RC BPU of critical information that is highly correlated with branch outcomes. Nonetheless, approach one provides a useful baseline for initial RC BPU evaluation as a naïve adaptation to BP workload. Providing a minimal reservoir input node requirement, allowing the use of smaller reservoir structures, and a simple low cost input encoding scheme.

#### 4.1.3b Approach Two

The second input encoding scheme evaluated only supplies branching instructions to the RC BPU using one-hot encoding. There are eight instructions that conditionally modify the program counter in the chosen RV64GC ISA configuration, allowing the use of the encoding scheme shown in Fig. 48. This approach removes the execution pipeline bottleneck of the first approach by only presenting branching instructions to the RC BPU. Additional to this, the one-hot encoding scheme necessitates similarly few reservoir input nodes as approach one. Allowing for the use of area efficient and speed optimised reservoir designs.

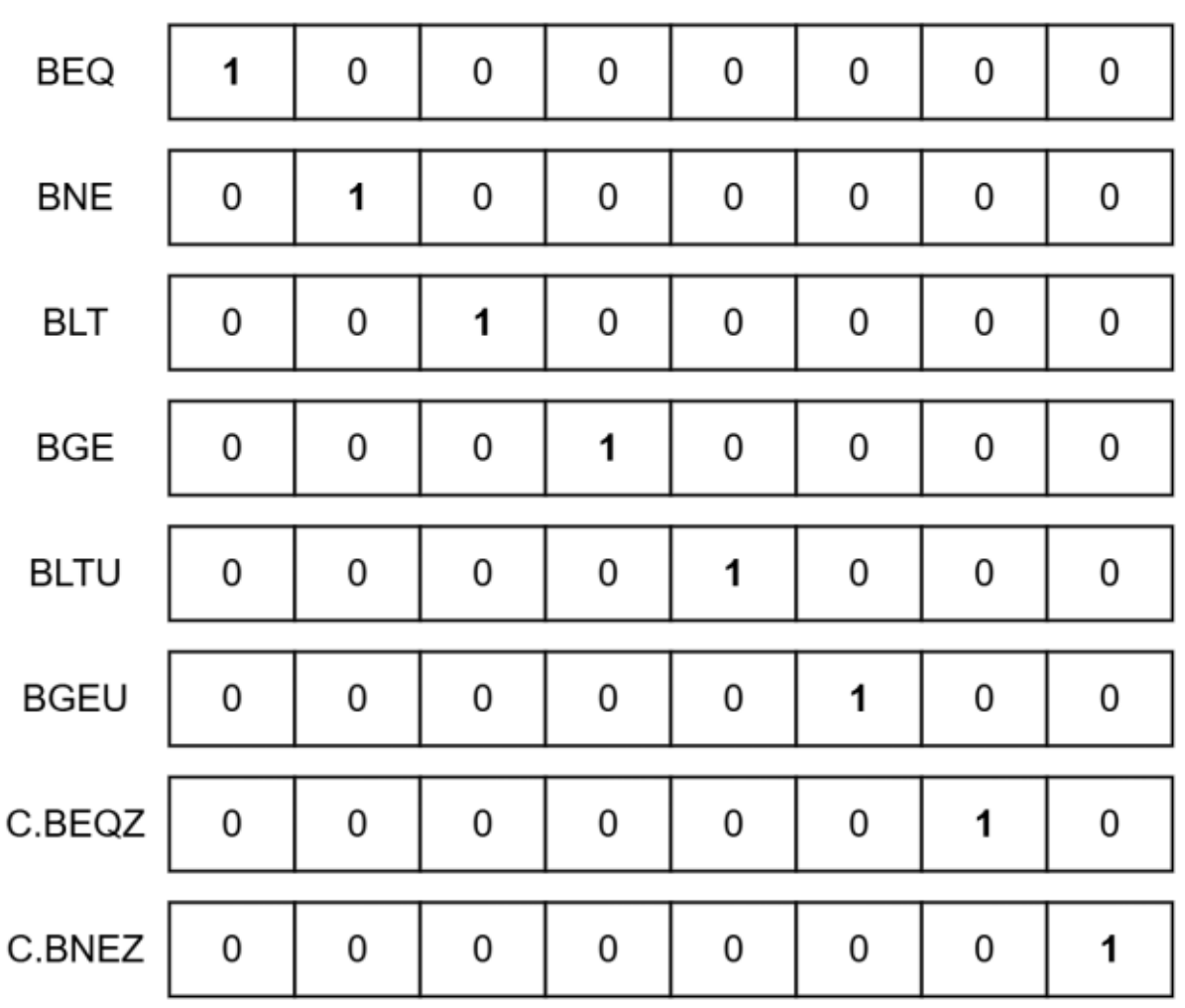

*Figure 48. Instruction encodings for RC BPU input for input encoding approach two.*

However, this approach has the same drawback as approach one in that it does not provide instruction PC information to the reservoir. The absence of which deprives the reservoir of critical information that is often highly correlated with branching outcomes.

#### 4.1.3c Approach Three

The third reservoir input encoding approach supplies only branching instructions to the reservoir while using the same integer classification scheme introduced in Approach One. This method also incorporates instruction address information into the encoded input in the form of tag bits. The resulting instruction encoding format, shown in Fig. 49, uses the first four bits to represent the branching instruction type. The remaining input bits encode the lower bits from the address in program memory that the instruction was fetched from. Using only address tag bits serves as an optimisation to keep reservoir size requirements manageable. By limiting the amount of address information provided, the number of required reservoir input nodes can remain relatively low while still providing useful program structure information. Depending on the desired level of address granularity, approach three can employ either 4-bit or 8-bit address encoding.

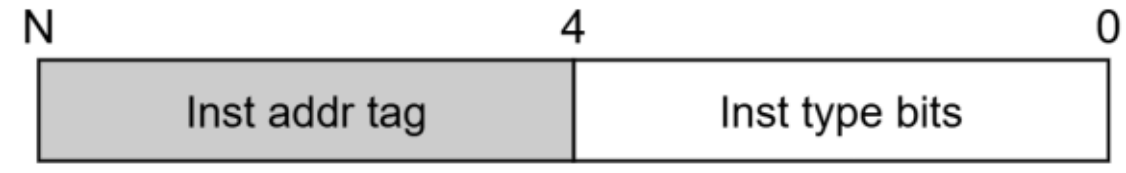

*Figure 49. Instruction encoding for RC BPU input for input encoding approach three.*

This approach improves upon the first two encoding methods by providing both the type of branching instruction encountered and information about its location within the program. As a result, the RC BPU can develop internal representations not only of the instruction sequences that lead to taken or not-taken branches, but also of the behaviour of specific branches identified through their address tag bits. This enables the RC BPU to associate particular branch locations with their likely execution outcomes, improving its ability to analyse and predict branching behaviour. Additionally, this approach is implementable as a simple LUT and concatenation with address bits in hardware. Providing a low overhead input encoding scheme that captures highly relevant information for predicting branch outcomes.

## 4.2 TAGE Predictor

The TAGE predictor [44] was implemented as a comparison baseline for the developed RC BPU architecture. As discussed in Section 2.2, TAGE is widely regarded as one of the highest-performing conventional branch prediction architectures and is therefore representative of near state-of-the-art branch prediction performance. Consequently, it provides an effective reference against which the developed RC BPU can be evaluated in terms of overall prediction accuracy, adaptability, and convergence behaviour.

TAGE, short for *TAgged GEometric history length*, is a hybrid branch predictor architecture that combines a base bimodal predictor with multiple tagged predictor tables indexed using geometric history lengths derived from the global branch history [45]. The bimodal predictor serves as the base prediction mechanism, generating an initial prediction of taken or not-taken behaviour. While the tagged predictor tables provide refined predictions based on longer and more complex branch history correlations.

The bimodal predictor is implemented as a two-bit saturating counter table indexed using program counter bits. Each counter represents one of four prediction states: strongly not-taken (SN), weakly not-taken (WN), weakly taken (WT), and strongly taken (ST). The counter state is updated according to the actual branch outcome, as shown in Fig. 50. Allowing the predictor to adapt dynamically to changing branch behaviour patterns and providing the primary training mechanism for the base predictor.

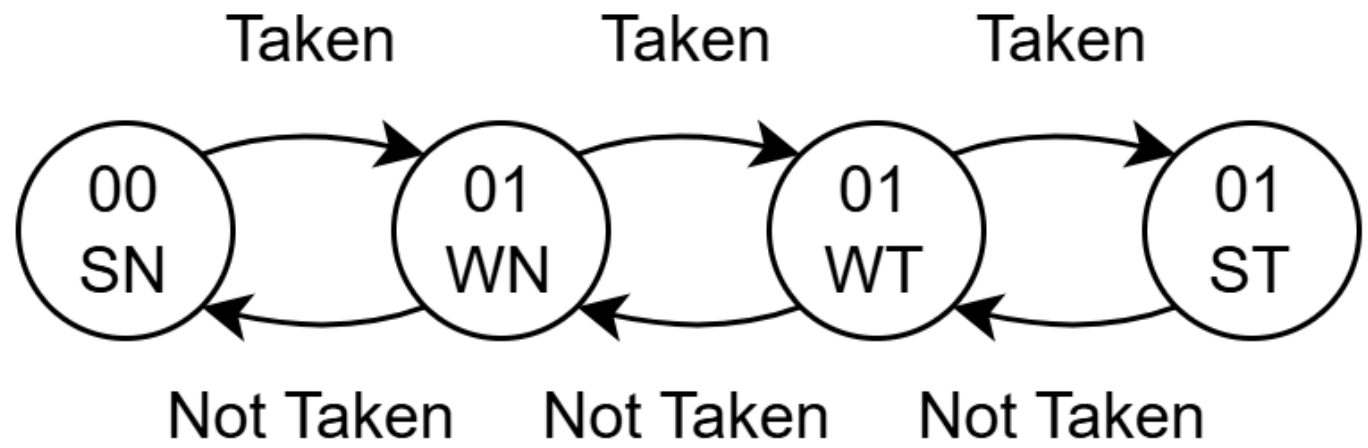

*Figure 50. Bimodal predictor 2-bit counter values of strongly taken (ST), weakly taken (WT), weakly not taken (WN), and strongly not taken (SN).*

The tagged predictor tables are implemented as a series of data structures with geometrically increasing history lengths, enabling the predictor to capture both short-term and long-term branch correlations. Each table entry contains a signed saturating prediction counter, tag bits, and a usefulness bit [45]. The signed counter functions similarly to the bimodal predictor counter and provides the branch prediction output. The tag bits are generated using a hash of the program counter and selected global history bits, allowing entries to be uniquely associated with specific branch history patterns during table lookup. The usefulness bit is used to indicate the effectiveness of a table entry and assists the replacement policy in determining which entries should be retained or overwritten during predictor updates [45].

The TAGE predictor used for comparison with RC BPU is implemented in C++ as a functional model. Where the instruction trace, Section 4.1.2, is used as input and the functional model outputs the same TP, TN, FP, FN results based on the TAGE predictors classification of each branch it encounters.

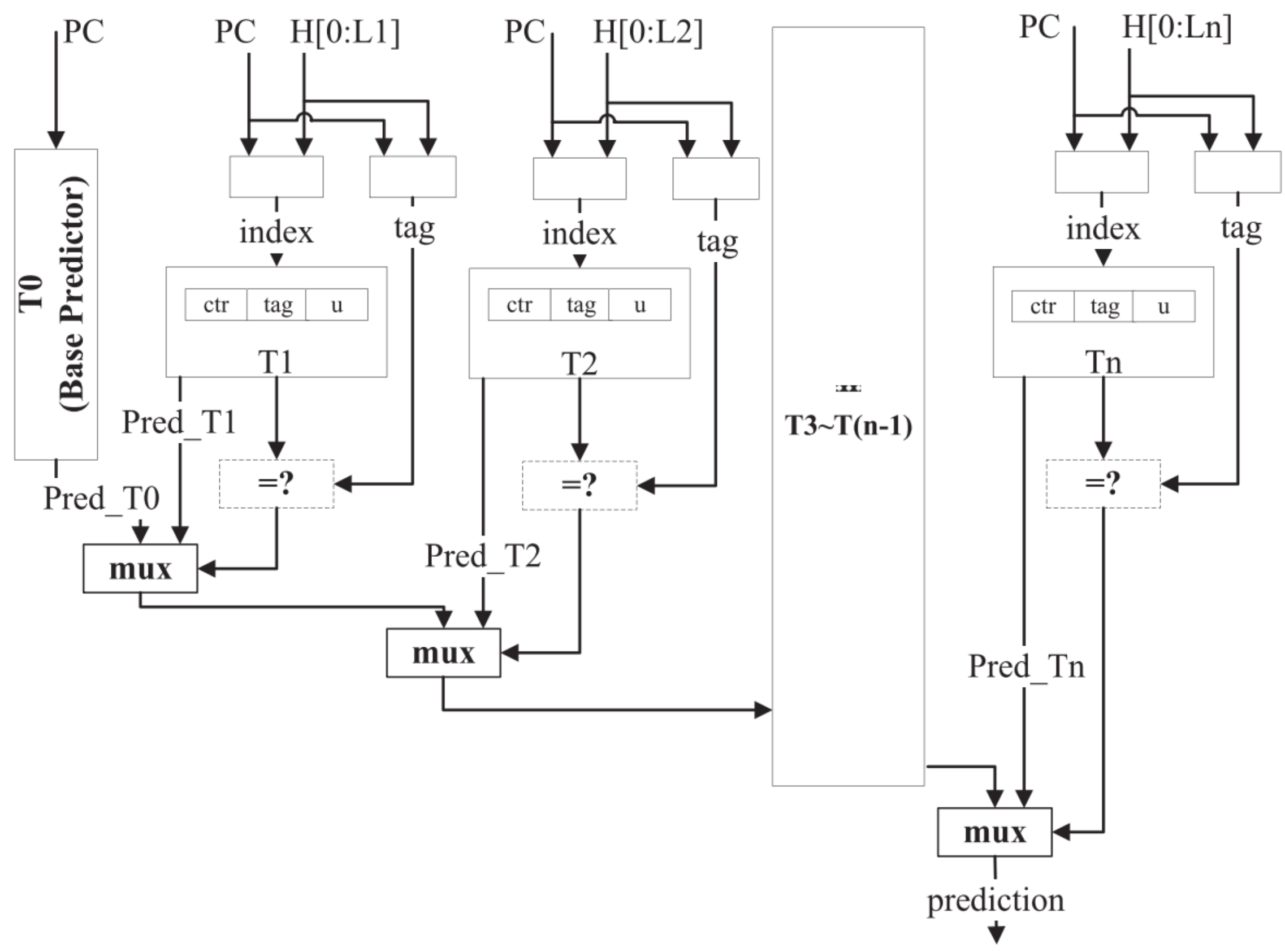

*Figure 51. Structure of TAGE predictor. [45]*

## 4.3 Test Results

### 4.3.1 Input Encoding Scheme

Each of the three input encoding schemes are assessed to determine which one provides the best input data representation to allow for meaningful internal representations. With input encoding that allows fast adaptation and high accuracy being of importance. The RC BPU configuration shown in Table 24 is used for testing. The performance of the input encoding schemes is isolated by using FP64 quantisation for the perceptron output layer. Meaning that the only significant bottleneck to RC BPU performance is the reservoir dynamics and the input data encoding scheme.

The results shown in Fig. 52 indicate that the first input encoding strategy yields the lowest overall classification accuracy. This outcome can be attributed to the increased task complexity introduced by including all instructions in the reservoir input. In this configuration, the 32-dimensional output space appears insufficient to achieve effective linear separability of the underlying decision boundary. Nevertheless, this approach exhibits the most rapid initial learning behaviour, reaching approximately 75% accuracy within the first 1,000 observed branches. This suggests that the inclusion of non-branching instructions does provide correlated contextual information that improves the adaptability of the RC BPU to changes in branching behaviour.

In contrast, the second approach, which is based solely on one-hot encoding of branch instructions, achieves the highest final accuracy among all tested encoding schemes. Reaching 100% accuracy after approximately 18 thousand branches encountered. This result indicates the importance of reservoir input data choice and representation, as the simpler feature space appears to facilitate more stable and consistent transformations within the reservoir. Thereby improving classification performance as the reservoir dynamics are able to more effectively make the problem linearly separable.

*Table 24. Input encoding Scheme test setup.*

| Parameter | Value |
|---|---|
| RC BPU Configuration | |
| Perceptron Weights | FP64 |
| Reservoir Output Bit-Depth | 4-bit |
| Perceptron Learning Rate | 0.1 |
| $V_{DD}$ | 3.3V |
| Reservoir Configuration | |
| Memristor Count | 200 |
| Memristor Parameters | Table 7 |
| Maximum Node Connections | 10 |
| Minimum Node Connections | 6 |
| Number of Inputs | 8 |
| Number of Outputs | 32 |
| Input Placement Algorithm | Even |
| Output Placement Algorithm | Even |
| Test Configuration | |
| Operating Frequency | 1GHz |
| Dhrystone Configuration | 1 |
| Input Encoding | Approach 1, 2, and 3 |

However, the absence of supplementary contextual signals (such as program counter information or preceding non-branch instructions) reduces the rate of convergence. Consequently, this approach demonstrates the slowest adaptation, with full performance only attained late in program execution at 75% completion.

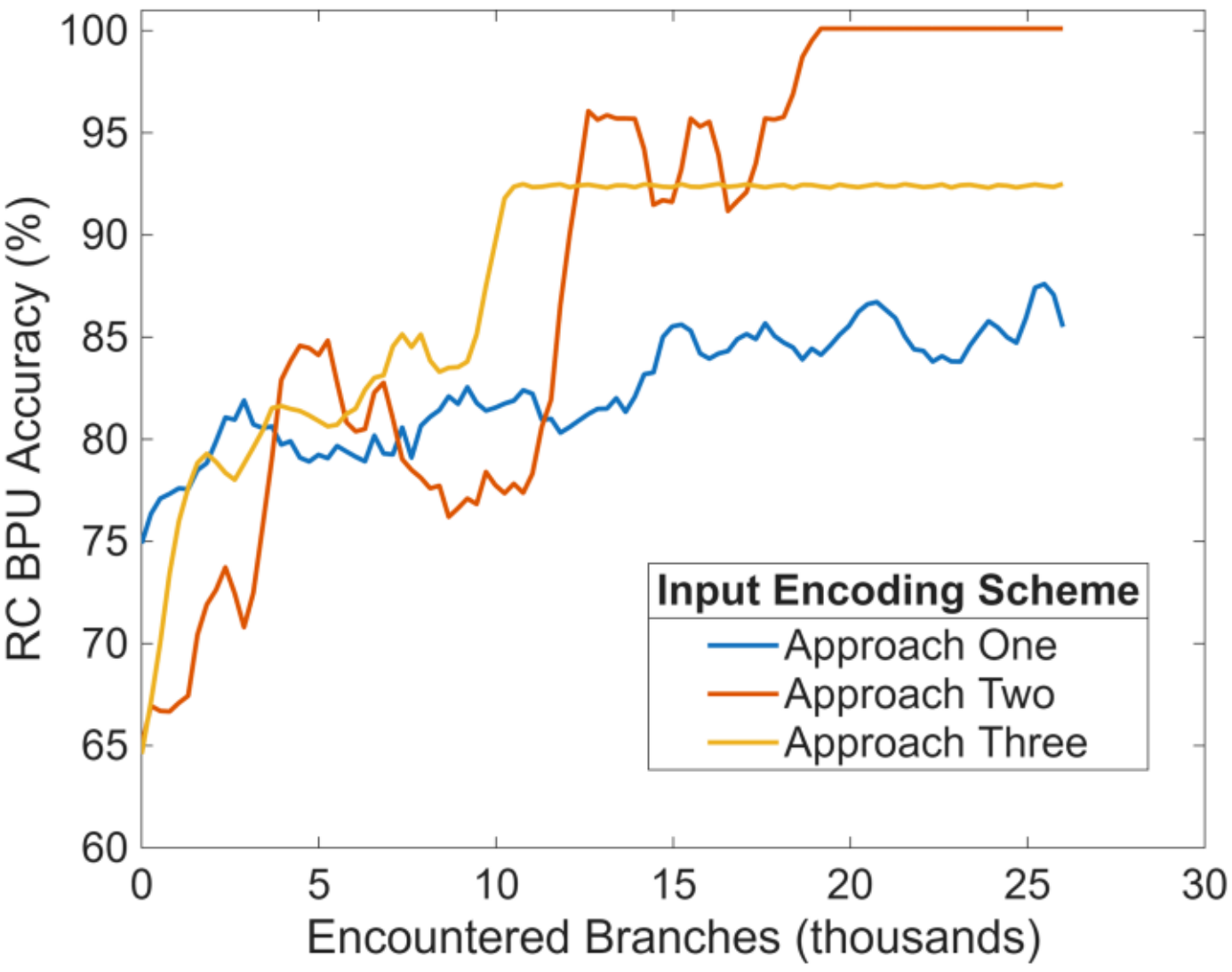

*Figure 52. RC BPU accuracy during Dhrystone benchmark for each input encoding scheme.*

The third approach yields intermediate performance in terms of final accuracy, achieving approximately 93% correct branch prediction. Notably, it demonstrates superior adaptability relative to the other methods, reaching its peak performance at around 40% program completion after encountering 10 thousand branches. Compared to approximately 60% (~15 thousand encountered branches) for the first approach and 75% (~18 thousand encountered branches) for the second. This accelerated convergence suggests that the inclusion of instruction address tag bits provides useful discriminative information that enhances early learning. However, this additional information also increases the complexity of the input space, which appears to exceed the reservoir's processing capacity. As a result, while convergence is faster, the method exhibits reduced overall accuracy relative to the second input encoding scheme.

*Table 25. Initial input encoding scheme test overall branch prediction accuracies.*

| Input Encoding Scheme | Overall accuracy (%) |
|---|---|
| Approach One | 82.3% |
| Approach Two | 87.7% |
| Approach Three | 87.4% |

Additional testing is performed to investigate the hypothesised benefit of increased output dimensionality for input encoding approaches two and three. The testing configuration is given in table 26, with the number of reservoir outputs increased from 32 to 48 and the number of memristors comprising the reservoir increased to 300 to accommodate this. It is hypothesized that the increase in the number of output nodes of the reservoir and the reservoir size will enhance its ability to process the more complex input encodings of approaches two and three. Bringing the benefit of higher overall accuracy and a more rapid increase in accuracy due to faster adaptability.

The results of additional testing, Fig.53, show a significant and meaningful increase in performance for input encoding approach three. Which manages to converge to 100% accuracy after encountering just 9 thousand branches, a significant improvement over its initial performance in Fig. 52. Similarly, approach two sees an improvement, reaching peak accuracy after just 3 thousand branches. However, its peak accuracy regresses compared to initial testing results shown in Fig. 52 of 100% accuracy. And its convergence profile becomes partially unstable as its accuracy temporarily drops between 6000 and 9000 observed branches.

*Table 26. Further input encoding scheme testing setup.*

| Parameter | Value |
|---|---|
| RC BPU Configuration | |
| Perceptron Weights | FP64 |
| Reservoir Output Bit-Depth | 4-bit |
| Perceptron Learning Rate | 0.1 |
| $V_{DD}$ | 3.3V |
| Reservoir Configuration | |
| Memristor Count | 300 |
| Memristor Parameters | Table 7 |
| Maximum Node Connections | 10 |
| Minimum Node Connections | 6 |
| Number of Inputs | 8 |
| Number of Outputs | 48 |
| Input Placement Algorithm | Even |
| Output Placement Algorithm | Even |
| Test Configuration | |
| Operating Frequency | 1GHz |
| Dhrystone Configuration | 1 |
| Input Encoding | Approach 2 and 3 |

The overall prediction accuracy of both input encoding approaches improves in conjunction with an increased rate of adaptability. With the third input encoding scheme demonstrating the most significant improvement, with overall accuracy increasing by 5.3% to 92.7%. The remaining prediction inaccuracies for the third approach are attributable primarily to early-stage performance prior to full convergence of the model to the underlying branching behaviour.

*Table 27. Further input encoding scheme testing overall branch prediction accuracies.*

| Input Encoding Scheme | Overall accuracy (%) |
|---|---|
| Approach Two | 89.7% |
| Approach Three | 92.7% |

The reduction in peak achievable accuracy observed for the second input encoding approach is likely attributable to the limitations of the one-hot encoding mechanism when applied to the larger reservoir architecture. Because only a single input is driven to logic high ($V_{DD}$) at any given time, a substantial portion of the reservoir experiences insufficient voltage across individual devices to induce state switching. Consequently, the proportion of actively switching memristors is reduced, limiting the reservoir's effective information processing capability. This reduction in active dynamical behaviour decreases the effective dimensionality of the reservoir, thereby contributing to the decline in final prediction accuracy from 100% to 93%, as shown in Fig. 53.

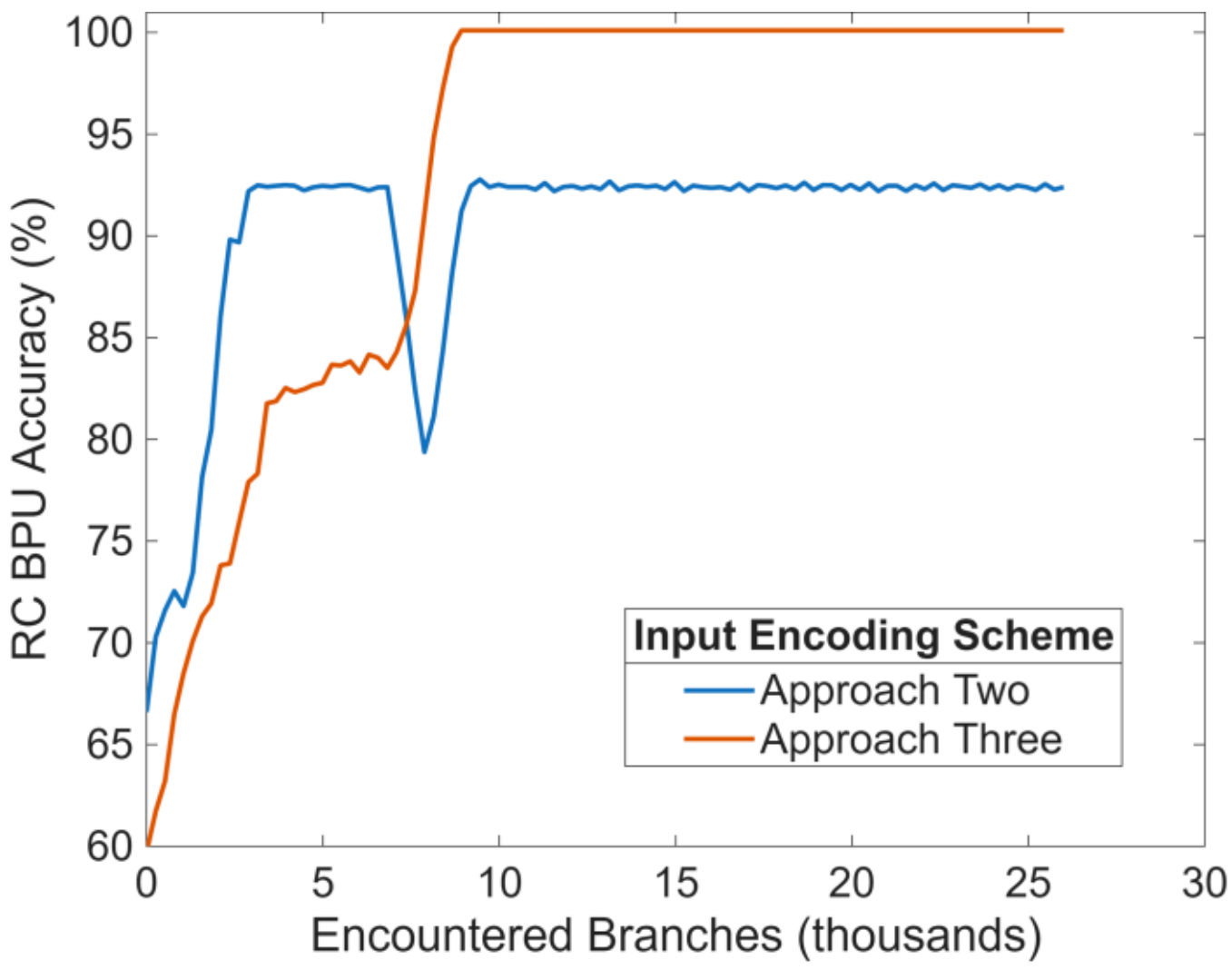

*Figure 53. RC BPU accuracy during Dhrystone benchmark for additional testing for input encoding schemes two and three.*

### 4.3.2 Reservoir Density Variation

To build upon the testing performed in Section 4.3.1 which found increased dimensionality is effective for increasing RC BPU performance. Different reservoir densities are investigated for their impact on overall RC BPU performance and effect on the rate of convergence and adaptability of the RC BPU. Testing in Section 3.5.3c demonstrated that the reservoir density is an important design parameter for achieving an adaptable and accurate RC system. With sparser reservoirs demonstrated to be more adaptable and effective for accurately performing sequence classification due to the increased complexity of reservoir dynamics providing more effective information processing.

Testing of reservoir densities effect on RC BPU performance is conducted using the testing configuration given in Table 28. With varying reservoir densities.

*Table 28. Reservoir density variation impact testing.*

| Parameter | Value |
|---|---|
| RC BPU Configuration | |
| Perceptron Weights | FP64 |
| Reservoir Output Bit-Depth | 4-bit |
| Perceptron Learning Rate | 0.1 |
| $V_{DD}$ | 3.3V |
| Reservoir Configuration | |
| Memristor Count | 300 |
| Memristor Parameters | Table 7 |
| Maximum Node Connections | 10 |
| Minimum Node Connections | 6 |
| Number of Inputs | 8 |
| Number of Outputs | 48 |
| Input Placement Algorithm | Even |
| Output Placement Algorithm | Even |
| Test Configuration | |
| Operating Frequency | 1GHz |
| Dhrystone Configuration | 1 |
| Input Encoding | Approach 3 |

The testing results presented in Fig. 54 demonstrate that reservoir connection density significantly influences branch prediction performance. However, in contrast to the behaviour observed in Section 3.5.3c, reducing the reservoir density leads to decreases in both adaptability and overall prediction accuracy. Only the two highest-density reservoir configurations successfully converge to 100% accuracy, whereas the sparsest configuration achieves a peak accuracy of approximately 92%. Representing a substantial degradation in performance relative to the results presented in Section 4.3.1. With the lowest density reservoir configuration achieving only 76.85% overall accuracy, Table 29, compared to 92.7% for the original density reservoir, Table 27.

*Table 29. Reservoir density variation impact testing overall branch prediction accuracies.*

| Reservoir Density | Overall accuracy (%) |
|---|---|
| 3-5 | 76.85% |
| 4-7 | 91.83% |
| 6-10 | 92.7% |

These findings suggest that, for the branch prediction task, the additional computational capacity provided by sparser reservoirs is outweighed by the increased inconsistency in their fading memory. This interpretation is supported by the analysis presented in Section 3.3.4a, which demonstrated that reservoirs containing 300 memristors exhibited the highest levels of fading memory inconsistency across both evaluated operating voltages. Consequently, the resulting highly irregular dynamics reduce the reservoir's ability to consistently retain and process the temporal information required for accurate branch prediction.

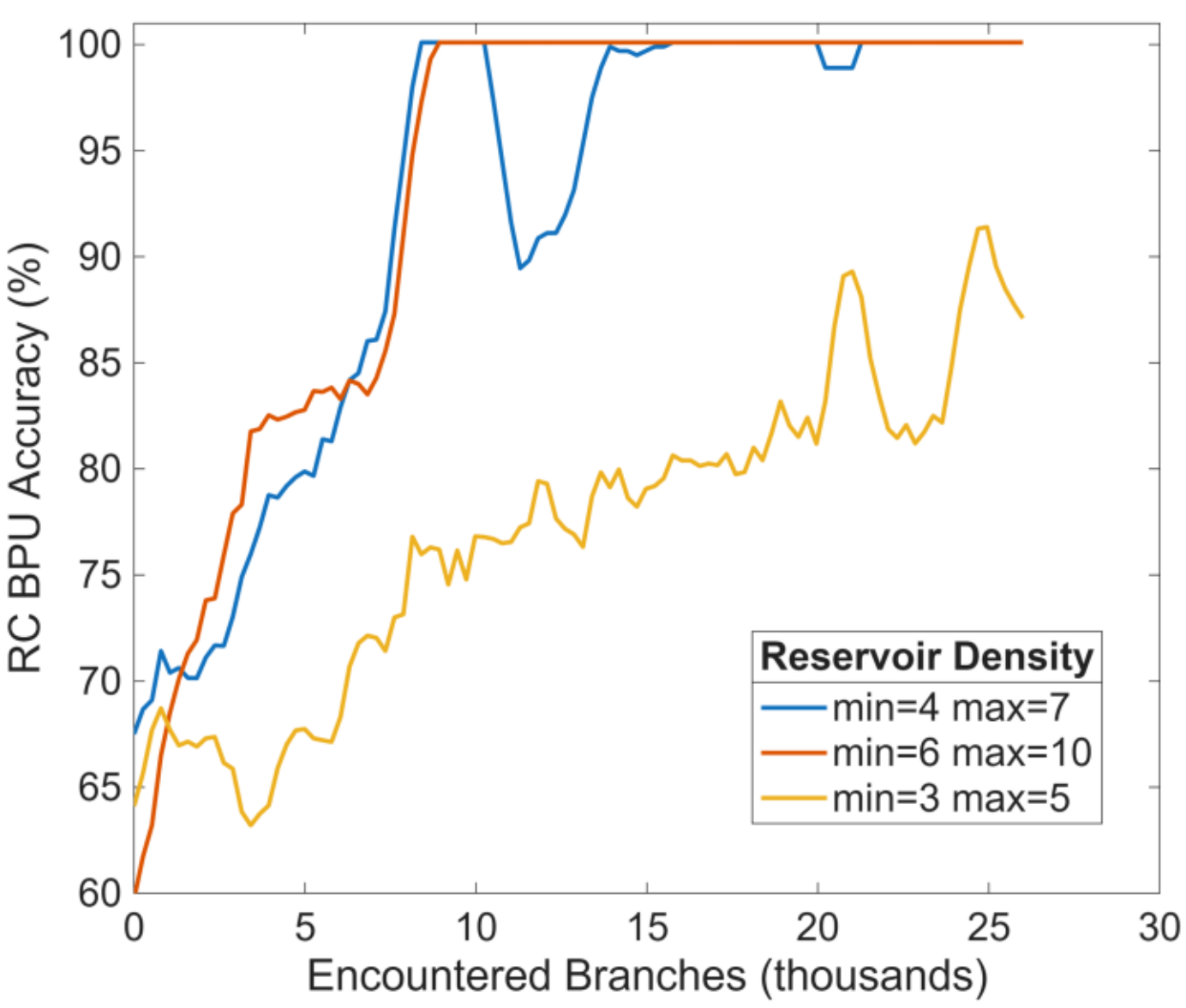

*Figure 54. RC BPU accuracy during Dhrystone benchmark for different reservoir densities.*

The highly irregular dynamics arise from the threshold switching behaviour of individual memristor devices. As the reservoir density decreases, the effective resistance of branches between the input nodes increases due to more memristors per branch. This leads to a lower voltage across individual memristors and a lower current flow. Consequently, both thermal and shot noise in addition to switching noise become more prevalent. Leading to increased instability and inconsistency in reservoir dynamics. The primary method of mitigating this would be to increase the operating voltage, thereby increasing branch currents, which would increase the signal to noise ratio of the reservoir. However, this is not practical due to the voltage limitation imposed by $\boldsymbol{P_4}$ and its derived reservoir requirement $\boldsymbol{R_2}$ of a maximum operating voltage of 3.3V.

### 4.3.3 Perceptron Weight Quantisation

To evaluate the RC BPU under realistic implementation constraints, the test configuration outlined in Table 30 is used. In this configuration, the perceptron weights were quantised to 8-bit integers (INT8). Which Section 3.4.4b demonstrated to satisfy the timing requirements for 1 GHz operation with 48 inputs when targeting the ASAP7 process node. As this quantised-weight configuration represents the intended future hardware implementation, the performance of the RC BPU under INT8 quantisation provides a key indicator of the design's practical feasibility.

*Table 30. INT8 quantised perceptron weights testing setup.*

| Parameter | Value |
|---|---|
| RC BPU Configuration | |
| Perceptron Weights | INT 8 |
| Reservoir Output Bit-Depth | 4-bit |
| Perceptron Learning Rate | 2, 1, 0.5 |
| $V_{DD}$ | 3.3V |
| Reservoir Configuration | |
| Memristor Count | 300 |
| Memristor Parameters | Table 7 |
| Maximum Node Connections | 10 |
| Minimum Node Connections | 6 |
| Number of Inputs | 8 |
| Number of Outputs | 48 |
| Input Placement Algorithm | Even |
| Output Placement Algorithm | Even |
| Test Configuration | |
| Operating Frequency | 1GHz |
| Dhrystone Configuration | 1 |
| Input Encoding | Approach 3 |

Initial testing results presented in Fig. 55 demonstrate that perceptron weight quantisation has a substantial impact on the performance of the RC BPU. INT8 quantisation results in a significant reduction in prediction accuracy compared to the equivalent reservoir configuration evaluated using FP64 precision in Section 4.3.1, Fig. 53. The highest-performing INT8 configuration achieves an overall accuracy of 77.56%, as shown in Table 31, compared to 92.7% obtained using FP64 quantisation.

The observed degradation in performance is primarily attributable to the reduced numerical precision provided by INT8 representation. This limitation becomes particularly evident at higher learning rates of 1 and 2, where the training behaviour exhibits noticeable overshoot and a lack of convergence, Fig. 55. In contrast, the lower learning rate of 0.5 produces a more stable convergence profile, albeit with a consistently lower final accuracy of approximately 77%. These results indicate a trade-off between training stability and achievable performance under low-precision weight quantisation.

*Table 31. Initial perceptron quantisation test overall branch prediction accuracies.*

| Learning Rate | Overall accuracy (%) |
|---|---|
| 2 | 77.31% |
| 1 | 77.56% |
| 0.5 | 75.31% |

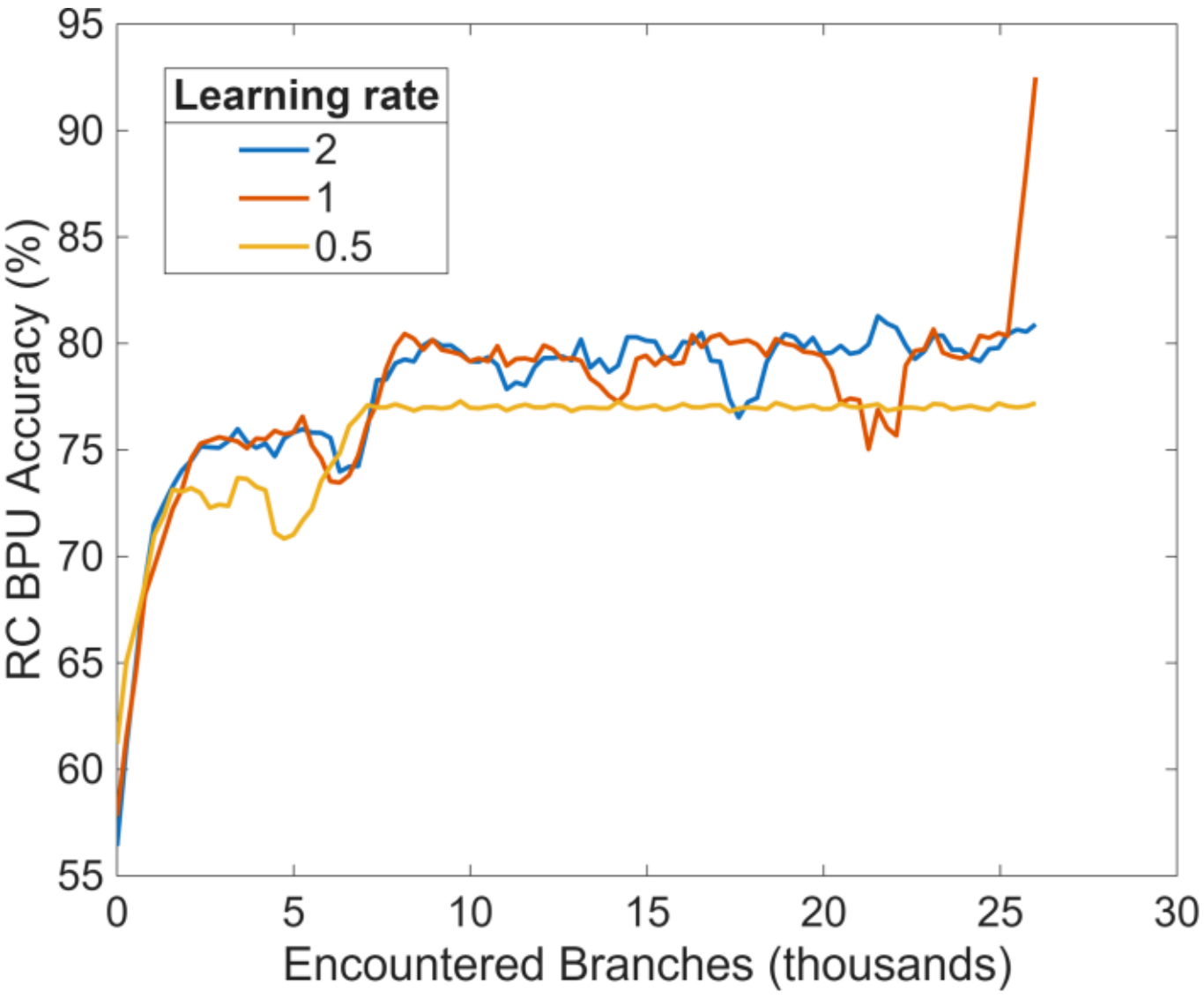

*Figure 55. RC BPU accuracy during Dhrystone benchmark for initial INT8 quantised perceptron testing.*

The significant gap in overall performance between the INT8 and FP64 quantised perceptron, from Fig. 55 and Fig. 53 respectively, indicates that INT8 quantisation is responsible for the drop in performance. INT10 exists as an alternative developed quantisation, Section 3.4.4, which would provide additional granularity to the perceptron weights to partially alleviate any issues arising from quantisation. This is investigated using the configurations from Table 32, with both INT10 and INT8 quantisation tested on identical reservoir configurations and operating speed and learning rate.

*Table 32. INT10 & INT8 quantised perceptron weights testing setup.*

| Parameter | Value |
|---|---|
| RC BPU Configuration | |
| Perceptron Weights | INT 10, INT 8 |
| Reservoir Output Bit-Depth | 4-bit |
| Perceptron Learning Rate | 2, 1, 0.5 |
| $V_{DD}$ | 3.3V |
| Reservoir Configuration | |
| Memristor Count | 200 |
| Memristor Parameters | Table 7 |
| Maximum Node Connections | 10 |
| Minimum Node Connections | 6 |
| Number of Inputs | 8 |
| Number of Outputs | 32 |
| Input Placement Algorithm | Even |
| Output Placement Algorithm | Even |
| Test Configuration | |
| Operating Frequency | 1GHz |
| Dhrystone Configuration | 1 |
| Input Encoding | Approach 3 |

Testing results, Table 33 and Fig. 56, show that for the identical learning rate and reservoir configuration, the perceptron weights quantisation has a meaningful impact on performance. Validating the hypothesis that the RC BPU accuracy would be benefitted by an increase in perceptron weight bit depth. With the INT10 LR=2 configuration outperforming INT8 LR=2 by 11.1%, Table 33.

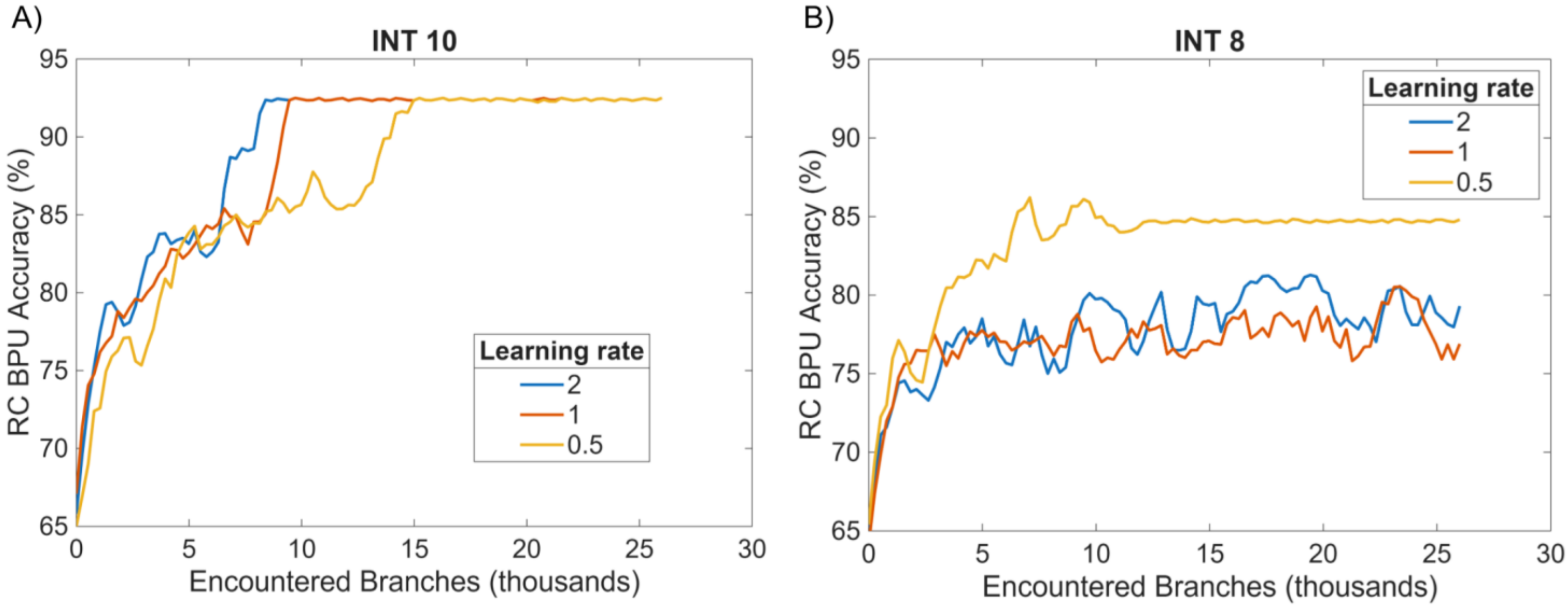

*Figure 56. RC BPU prediction accuracy during Dhrystone benchmark for INT10 (a) and INT8 (b) perceptron quantisation's.*

Additionally, the INT10 quantisation regime exhibits significantly more consistent steady-state accuracy. This is likely due to the increased weight resolution reducing update overshoot. As for an equivalent learning rate the INT10 weight updates are proportionally 4× smaller relative to the maximum representable weight value compared to INT8. This interpretation is further supported by the INT8 LR=0.5 configuration. Which exhibits increased stability in steady-state accuracy, consistent with the behaviour observed during the initial testing in Fig. 55. Showing that the lower performance of INT8 can be in part attributed to overshoot, a result of the high learning rates necessitated for practical implementation to avoid input value truncation.

The larger representable value range of the INT10 configuration also allows the use of higher learning rates without preventing convergence. Unlike INT8, INT10 does not require extremely low learning rates to maintain stable training behaviour. Allowing learning rates of 1 and 2 to preserve more of the training input information without excessive truncation. This enables more meaningful weight updates and contributes to the consistently higher steady-state accuracies of approximately 93% achieved by the INT10 configurations.

*Table 33. Further perceptron quantisation testing overall branch prediction accuracies.*

| Quantisation | Learning Rate | Overall accuracy (%) |
|---|---|---|
| INT10 | 2 | 88.57% |
| INT10 | 1 | 88.02% |
| INT10 | 0.5 | 86.43% |
| INT8 | 2 | 77.47% |
| INT8 | 1 | 76.61% |
| INT8 | 0.5 | 82.64% |

Importantly, the trade-off between affordable perceptron input count and weight quantisation is partially dependent on the underlying reservoir dynamics. This is evidenced by the differing convergence profiles observed for identical perceptron configurations in Fig. 55 and Fig. 56b when evaluated using different reservoir configurations. The results therefore suggest that perceptron and reservoir parameters cannot be optimised independently. And that a degree of system-level co-optimisation is necessary to tune the overall performance, adaptability, and stability of the RC BPU architecture.

### 4.3.4 Operating Voltage and Speed

Reservoir operating voltage is an important operating parameter that has a direct effect on the reservoir's dynamics. As explored in Sections 3.3.4a and 3.3.4c, the operating voltage of the reservoir has a direct impact on its fading memory half-life and its ability to process inputs into a higher dimensional representation. Additional to this, varying the reservoir voltage is the obvious approach to tuning the reservoirs fading memory for varying speeds of operation.

Testing is carried out, using the configuration in Table 34, to investigate the coarse variation of reservoir operating voltage and RC BPU operating frequency. Due to time limitations, this is only conducted for a small range of voltages and frequencies using the smallest Dhrystone configuration.

*Table 34. Reservoir operating voltage and frequency testing setup..*

| Parameter | Value |
|---|---|
| RC BPU Configuration | |
| Perceptron Weights | INT 10 |
| Reservoir Output Bit-Depth | 4-bit |
| Perceptron Learning Rate | 1 |
| $V_{DD}$ | 3.3V, 2.8V, 2.3V, 1.8V |
| Reservoir Configuration | |
| Memristor Count | 200 |
| Memristor Parameters | Table 7 |
| Maximum Node Connections | 10 |
| Minimum Node Connections | 6 |
| Number of Inputs | 8 |
| Number of Outputs | 32 |
| Input Placement Algorithm | Even |
| Output Placement Algorithm | Even |
| Test Configuration | |
| Operating Frequency | 1GHz, 500MHz, 100MHz |
| Dhrystone Configuration | 1 |
| Input Encoding | Approach 3 |

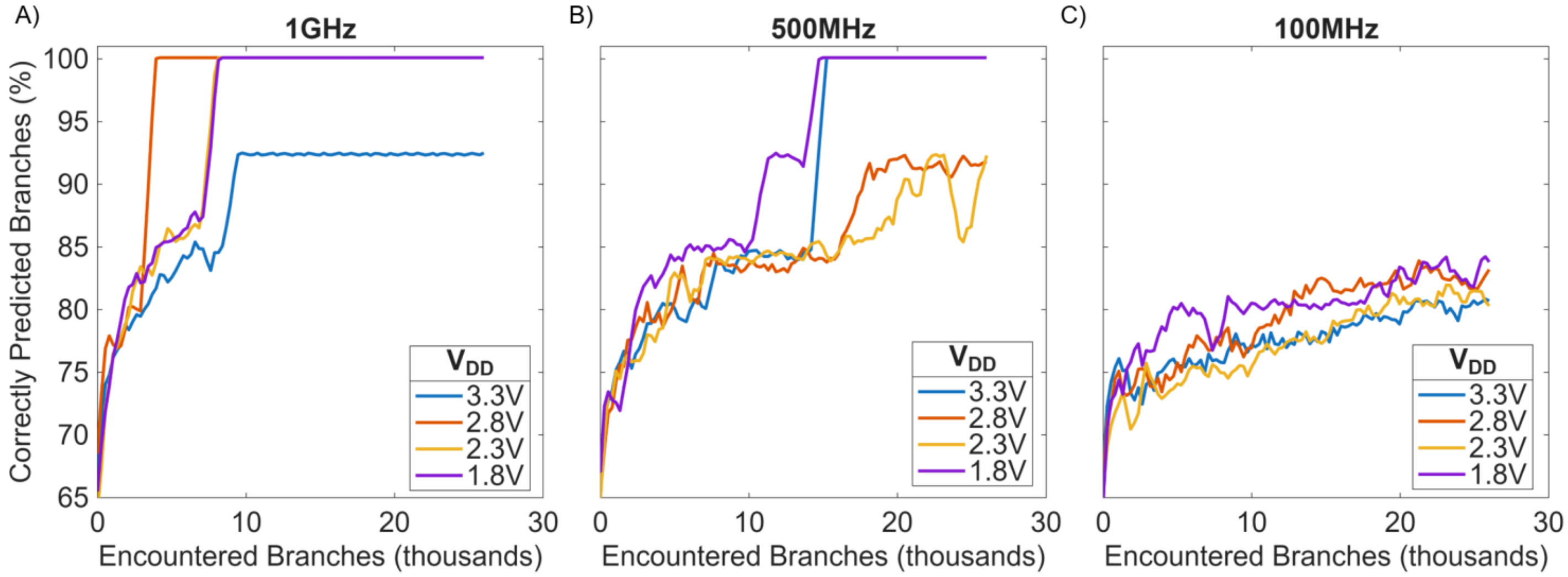

*Figure 57. RC BPU prediction accuracy during Dhrystone benchmark for INT10 weight quantisation with LR=1. For 1GHz RC BPU operating speed (a), 500MHz RC BPU operating speed (b), and 100MHz RC BPU operating speed (c).*

Testing results, shown in Fig. 57 and Table 35, demonstrate that RC BPU performance varies significantly with reservoir operating point. At 1GHz operation, convergence rate is strongly influenced by operating voltage. Unexpectedly, the 3.3V operating point performs worst among all tested voltages and fails to achieve 100% final accuracy. In contrast, operation at 2.8V produces the most favourable reservoir dynamics for high accuracy and rapid learning; achieving 100% accuracy after observing only 4000 branches. Lower operating voltages of 1.8V and 2.3V also outperform 3.3V by eventually converging to 100% accuracy, although with slower convergence rates than 2.8V. These results suggest that operation at 2.8V provides the most effective balance between the reservoirs computational capability, dynamical stability, and fading memory. This is further supported by the achieved overall prediction accuracy of 96.50%, shown in Table 35, exceeding that of all tested FP64 perceptron quantisation configurations. Further adding to the perceived importance of co-optimisation established in Section 4.3.3.

*Table 35. Reservoir operating point testing overall branch prediction accuracies.*

| RC BPU Accuracy (%) | | Operating Frequency | | |
|---|---|---|---|---|
| | | 1GHz | 500MHz | 100MHz |
| Operating Voltage | 3.3V | 88.02 | 88.93 | 77.27 |
| | 2.8V | 96.50 | 84.53 | 78.98 |
| | 2.3V | 94.22 | 83.75 | 76.90 |
| | 1.8V | 94.31 | 91.02 | 79.73 |

Operation at 500MHz, shown in Fig. 57b, exhibits a similar preference for lower operating voltages to that observed at 1GHz. However, unlike the 1GHz configuration, the highest performance at 500MHz is achieved at $V_{DD}$ of 1.8V. Initially, this behaviour may be attributed to the increased fading memory complementing the lower operating frequency. However, strong performance is also observed at 3.3V. Indicating that system performance at 500MHz is not determined solely by fading memory characteristics. The resulting behaviour is therefore more complex than initially anticipated and cannot be fully explained within the limitations of the current testing methodology.

Based on the observations from Section 3.3.4c, it is hypothesised that this behaviour arises from the threshold-switching characteristics of the memristor devices. At 1.8V, the applied voltage may be sufficient to consistently activate only a subset of devices within the reservoir, producing stable and computationally useful dynamics. As the operating voltage increases to 2.3V and 2.8V, a larger proportion of devices begin to switch, but with slower and potentially less stable dynamics due to switching variability arising from the stochastic nature of device switching. At 3.3V, the entire reservoir is likely activated with sufficiently fast switching behaviour that the relative impact of switching noise is reduced. However, due to limitations in the simulation and testing setup, insufficient evidence exists to conclusively validate this hypothesis.

Operation at 100MHz, shown in Fig. 57c, exhibits a substantial reduction in prediction accuracy compared to the higher operating frequencies tested. None of the evaluated operating voltages achieved convergence to 100% accuracy, and all configurations displayed unstable convergence behaviour throughout training. This behaviour suggests that the reservoir dynamics at 100MHz become either highly inconsistent or significantly more linear and thereby reduce the computational capacity of the reservoir. Preventing the perceptron readout layer from reliably learning to distinguish between taken and not-taken branches. The noisy convergence profiles, Fig.57c, supports both these hypotheses; as the perceptron is not able to linearly dichotomise the reservoir outputs. Indicating that the problem is not easily linearly separable.

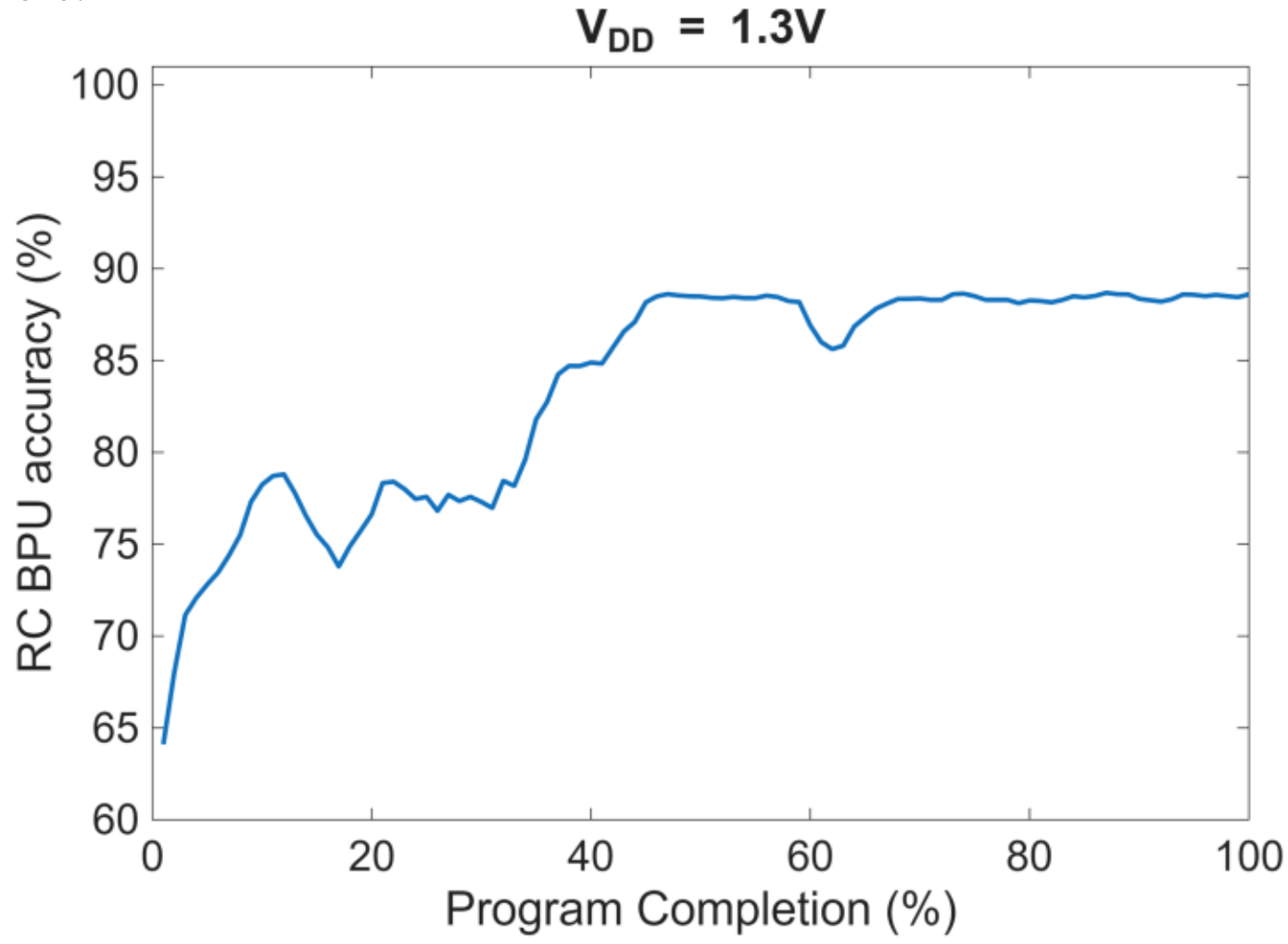

*Figure 58. RC BPU prediction accuracy for Dhrystone benchmark for operation at 100MHz and 1.3V.*

To further investigate this behaviour, an additional benchmark run was performed at $V_{DD}$ of 1.3V. Unlike the higher-voltage configurations, the 1.3V operating point achieved convergence to approximately 88% accuracy while exhibiting a significantly more stable convergence profile, as shown in Fig. 58. These results support the hypothesis that the initially tested operating voltages produced inconsistent and chaotic reservoir dynamics. Which limited the ability of the perceptron readout layer to reliably classify reservoir states. Despite the improved stability and increased prediction accuracy achieved at 1.3V, considerable variation in accuracy remains after convergence. This suggests that the reservoir dynamics remain partially deficient in either computational capacity or fading memory performance. Based on the observations from Section 3.3.4a and Section 4.3.2, the latter is considered the more significant limitation.

Furthermore, the inability of the RC BPU to converge to 100% accuracy is likely an inherent consequence of operating the reservoir at reduced voltages and frequencies. Both of which reduce the overall nonlinearity of the reservoir dynamics, as observed in Section 3.3.4c, limiting the reservoir's computational capability and its effectiveness at separating temporal patterns. In addition, lower operating voltages further exacerbate inconsistencies in the reservoir's fading memory behaviour, as demonstrated in Section 3.3.4a. The combined reduction in nonlinear computational capacity and fading memory consistency is therefore hypothesised to jointly contribute to the reduced prediction performance observed during 100MHz operation. Giving the reservoir an effective range of operation from 500MHz to 1GHz based on conducted testing.

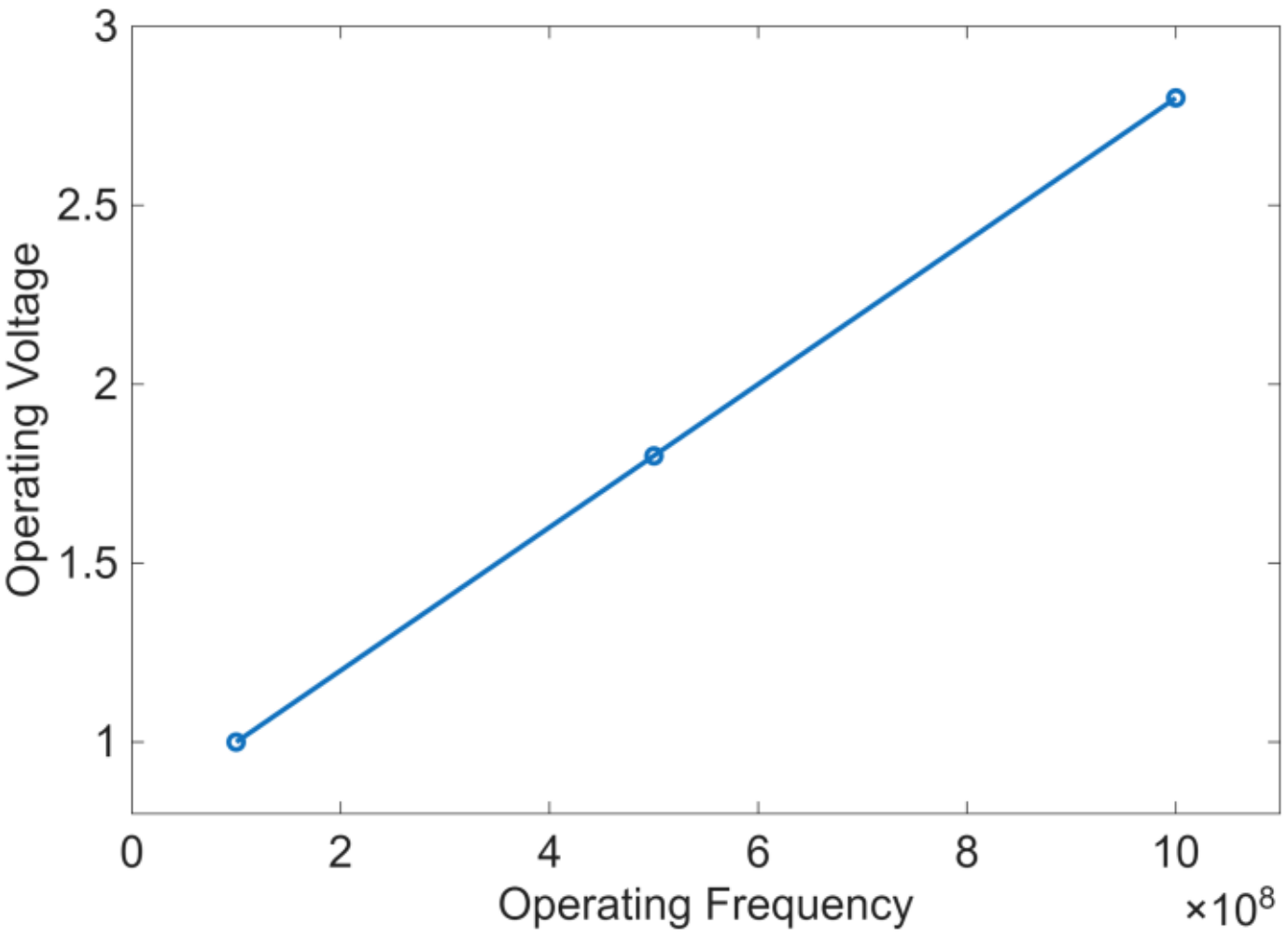

*Figure 59. Plot of operating frequency vs voltage at which the highest overall accuracy is achieved.*

Figure 59 was generated by plotting operating frequency against the operating voltage corresponding to the highest overall prediction performance. The resulting trend demonstrates a linear relationship between operating frequency and optimal reservoir operating voltage. This observation is significant, as it suggests that conditioning the reservoir for operation at different frequencies may require only a linear adjustment of the operating voltage. From a hardware implementation perspective, such behaviour is considerably simpler to realise than a nonlinear relationship, reducing the complexity of dynamic operating point control circuitry. Consequently, these results suggest that an effective RC implementation may be capable of supporting variable-frequency operation through adaptive voltage scaling. However, investigation of dynamic frequency and voltage scaling behaviour is beyond the scope of this project and was therefore not explored further.

### 4.3.5 Final Proposed System

With the key trade-offs between reservoir operating point, adaptability, and prediction accuracy established. A final RC BPU configuration, Table 36, is proposed and assessed against the project specification and against the TAGE predictor for the Dhrystone workload.

*Table 36. Final proposed RC BPU configuration for specification validation.*

| Parameter | Value |
|---|---|
| RC BPU Configuration | |
| Perceptron Weights | INT 10 |
| Reservoir Output Bit-Depth | 4-bit |
| Perceptron Learning Rate | 1 |
| Input Encoding | Approach 3 |
| Reservoir Configuration | |
| Memristor Count | 200 |
| Memristor Parameters | Table 7 |
| Maximum Node Connections | 10 |
| Minimum Node Connections | 6 |
| Number of Inputs | 8 |
| Number of Outputs | 32 |
| Input Placement Algorithm | Even |
| Output Placement Algorithm | Even |
| Operating Point | |
| Operating Frequency | 1GHz |
| $V_{DD}$ | 2.8V |

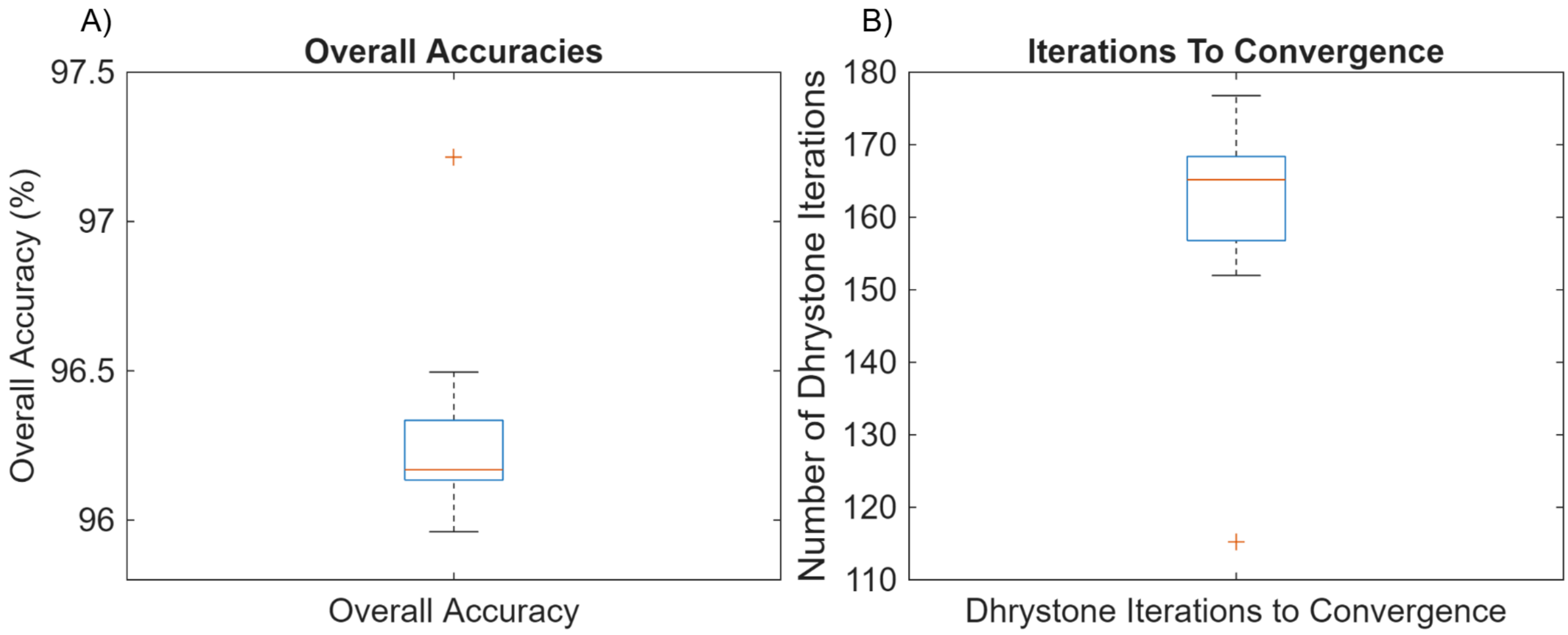

*Figure 60. RC BPU overall accuracy (a) and iterations to convergence (b) for 40 different initial reservoir states.*

The design configuration presented in Table 36 was selected based on the results from Section 4.3.4, where it achieved the highest overall prediction accuracy of 96.50% and exhibited the fastest convergence behaviour among all tested RC BPU configurations. To evaluate the reproducibility and consistency of this result, the Dhrystone benchmark test was repeated 40 times using significantly different initial reservoir states for each run. The resulting performance distributions, shown in Fig. 60, demonstrate that the RC BPU maintains both high accuracy and stable convergence behaviour across a wide range of initial reservoir conditions. Indicating strong robustness and generalisability of the proposed architecture as it is consistently able to reach the general solution in a mostly regular amount of time.

Across all benchmark repetitions, the RC BPU achieved a median overall prediction accuracy of 96.17% and a median convergence at 166 Dhrystone iterations. With upper and lower adjacent values of 96.5% and 95.96% respectively for overall accuracy, and 176 and 152 respectively for Dhrystone iterations to convergence. Only a single statistical outlier was observed, corresponding to superior performance of 97.2% overall accuracy with convergence achieved after 115 Dhrystone iterations. The limited variation across repeated experiments suggests that the developed RC BPU architecture is not highly sensitive to initial reservoir state and is capable of producing consistent performance despite the stochastic nature of the reservoir dynamics. This also signifies that the bottleneck of performance is likely to lie more with the perceptron than the reservoir dynamics due to the significant variety of initial reservoir states all converging similarly to each other.

*Table 37. TAGE predictor configuration for testing.*

| Parameter | Value |
|---|---|
| TAGE Configuration | |
| Table History Lengths | 4, 8, 16, 32 |
| Tagged Table Size | 1024 Entries |
| Tag bits | 10 |
| Global History Register Length | 64 Entries |
| Bimodal Size | 4096 Entries |

The TAGE predictor described in Section 4.2 was evaluated using the same 1000-iteration Dhrystone benchmark employed for RC BPU testing to enable direct performance comparison between the two branch predictors. The resulting performance, shown in Fig. 61 and Table 38, demonstrates that the TAGE predictor significantly outperforms the developed RC BPU implementation in both convergence speed and overall prediction accuracy. The TAGE predictor converges to 100% prediction accuracy after only 10 benchmark iterations, approximately 15× faster than the RC BPU, allowing it to achieve an overall prediction accuracy of 99.93%.

*Table 38. RC BPU and TAGE overall accuracies.*

| Predictor | Overall accuracy (%) |
|---|---|
| RC BPU | 96.50% |
| TAGE | 99.93% |

These results indicate that, although the developed RC BPU demonstrates functional branch prediction capability and strong temporal learning behaviour. Substantial further development is required before memristor-based reservoir compute can be considered competitive with state-of-the-art conventional branch prediction architectures. In particular, improving adaptability and reducing convergence time represent critical challenges for future RC-based branch prediction research. Which, as identified above, are likely to be achievable by further refinement of the perceptron neuron design.

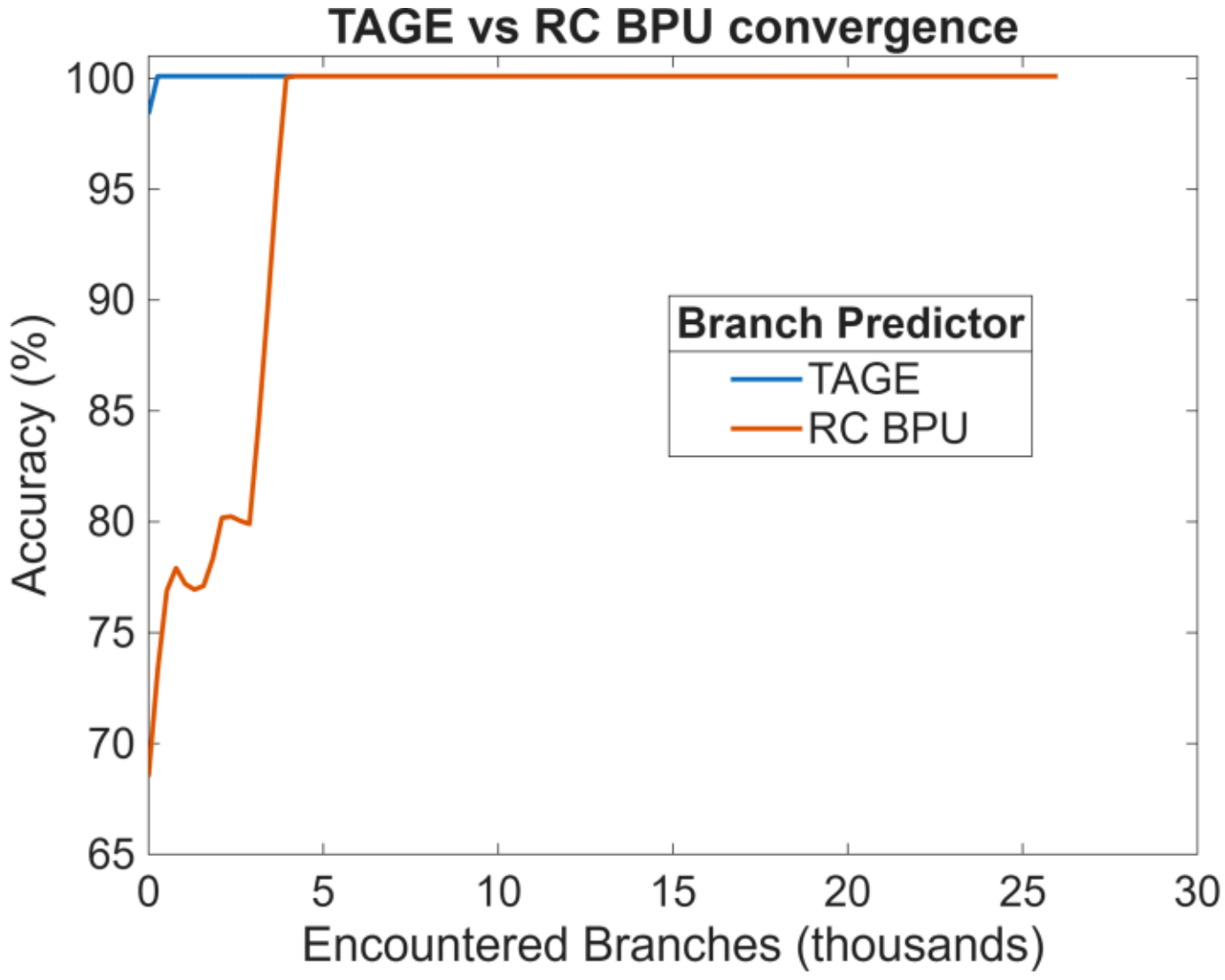

*Figure 61. TAGE and RC BPU convergence profiles for 1000 iteration Dhrystone benchmark.*

Despite these limitations, the RC BPU architecture remains a promising approach for future branch prediction research. The high-dimensional reservoir state space provides a significant theoretical advantage by enabling the perceptron readout layer to represent and classify a large number of unique temporal branching patterns simultaneously. Which is mostly limited by the neuron capacity, Cover's theorem. As such, if the current limitations in convergence speed and adaptability can be sufficiently addressed; the high-dimensional nature of the reservoir dynamics may allow the RC BPU to maintain more stable prediction performance across varying workloads and execution contexts. This could reduce the frequency of retraining required following workload changes or context switches, potentially improving long-term prediction stability in comparison to more conventional branch prediction approaches. However, testing was not able to confirm this due to compute and time limitations making further benchmarking infeasible.

The developed RC BPU implementation successfully satisfies all the project aims and specification requirements. The system consistently achieves an overall prediction accuracy of approximately 96%, substantially exceeding the target accuracy requirement of 70%, $\boldsymbol{P_3}$. With its only main deficiency in comparison to TAGE being the rate of convergence to solution weights.

The developed implementation further satisfies project aims $\boldsymbol{P_1}$ and $\boldsymbol{P_2}$ by achieving the performance target defined by $\boldsymbol{P_3}$ while operating at a frequency of 1GHz. This corresponds to an overall prediction latency of 2ns and a prediction throughput of 1 billion predictions per second. These results are significant, as they demonstrate that reservoir computing architectures may be capable of operating at conventional processor frequencies. While also maintaining sufficiently low prediction latency to support speculative execution without introducing excessive pipeline stalls.

Finally, the developed RC BPU satisfies project aim $\boldsymbol{P_4}$ through compliance with requirement $\boldsymbol{R_2}$. Achieving optimal operation at an operating voltage of 2.8V, which is below the maximum target voltage of 3.3V. This result is particularly significant as it suggests that with further memristor device engineering and reservoir optimisation, future RC implementations may be capable of operation at voltages closer to those used in modern CMOS digital logic. Which is typically within the range of 0.8V to 1.4V. Such operation would improve both energy efficiency and practical integration compatibility with conventional processor architectures.

# 5 Conclusion

## 5.1 Project Specification Completion

All primary specification requirements outlined in Section 1.3 were successfully achieved, in addition to several stretch-goal objectives. The only specification points not fully completed were $\boldsymbol{R_4}$ and $\boldsymbol{O_4}$, primarily due to limited remaining project time following a refocus towards the collection and analysis of quantitative experimental results during the later stages of the project.

The final project is believed to represent the first proposed application of memristor-based reservoir computing to the branch prediction workload to be developed using industry standard modelling tools. A novel RC implementation framework was developed and evaluated with consideration for the practical constraints associated with physical realisation. The architecture and design methodology draw upon existing reservoir computing and branch prediction literature while introducing a series of original design decisions tailored specifically towards the requirements of high-speed real-time branch prediction. The developed design framework is also believed to be the first memristor-based RC architecture targeting gigahertz operating frequency with real-time inference and online training for branch prediction within a CMOS-compatible implementation framework using ReRAM devices. In addition, the project contributes a set of target ReRAM device characteristics and operating requirements that may inform future device-engineering efforts aimed at the physical realisation of the proposed reservoir architecture.

*Table 39. Project specification point fulfilment status.*

| Specification Point | Pass / Fail | Description |
|---|---|---|
| | | Overall Project Goals |
| $\boldsymbol{P_1}$ | PASS | Demonstrated in Section 4.3.4, prediction latency of 2ns is achieved. |
| $\boldsymbol{P_2}$ | PASS | Demonstrated in Section 4.3.4, through operation at 1GHz. |
| $\boldsymbol{P_3}$ | PASS | Demonstrated in Section 4.3.5, median accuracy of 96.16% is achieved. |
| $\boldsymbol{P_4}$ | PASS | Demonstrated in Section 4.3.4, reservoir operation at 2.8V, below 3.3V limit. |
| | | Memristor Behavioural Model |
| $\boldsymbol{M_1}$ | PASS | Demonstrated in Section 3.1.4, I-V sweep and analysis performed. |
| $\boldsymbol{M_2}$ | PASS | Demonstrated in Section 3.1.2 and 3.1.3, VTEAM used for computational simplicity. |
| $\boldsymbol{M_3}$ | PASS | Demonstrated in Section 3.1.4a, consistent response over 1 million I-V sweeps. |
| $\boldsymbol{M_4}$ | PASS | Achieved in Section 3.1.3, table 1 gives 13 different parameters to control behaviour. |
| $\boldsymbol{M_5}$ | PASS | Demonstrated in Section 3.1.4b, behaviour variations with change in oxide thickness. |
| $\boldsymbol{M_6}$ | PASS | Demonstrated in Section 3.1.4c, behaviour changes with operating temperature. |
| | | Reservoir Architecture and Design |
| $\boldsymbol{R_1}$ | PASS | Demonstrated in Section 3.3.4a, fading memory showed state change in <1ns. |
| $\boldsymbol{R_2}$ | PASS | Demonstrated in Section 3.3.4, fading memory and non-linearity demonstrated at 3.3V. |
| $\boldsymbol{R_3}$ | PASS | Demonstrated in Section 3.3.4a, fading memory characterised at 3.3V and 1.8V. |
| $\boldsymbol{R_4}$ | FAIL | Perceptron neuron timing results resulted in this requirement being superfluous. |
| $\boldsymbol{R_5}$ | PASS | Demonstrated in Section 3.3.4, reservoir size, density, and operating voltage investigated. |
| | | Trainable Output Layer |
| $\boldsymbol{O_1}$ | PASS | Demonstrated in Section 3.4.4a, in-situ learning demonstrated for arbitrary function. |
| $\boldsymbol{O_2}$ | PASS | Demonstrated in Section 3.4.4b, STA gives no negative slack for tested configurations. |
| $\boldsymbol{O_3}$ | PASS | Achieved in Section 3.4.3, 7 configurable instantiation parameters given in Table 12. |
| $\boldsymbol{O_4}$ | FAIL | Critical path target not feasible with target process node at a reasonable perceptron size. |
| | | Full System Integration |
| $\boldsymbol{F_1}$ | PASS | Demonstrated in Section 3.5.3, true positive and true negative exceeded 70%, Table 18. |
| $\boldsymbol{F_2}$ | PASS | Demonstrated in Section 3.5.3b and c, reservoir density and output count tested. |
| $\boldsymbol{F_3}$ | PASS | Demonstrated in Section 3.5.3b and c, reservoir density and output count tested. |
| $\boldsymbol{F_4}$ | PASS | Demonstrated in Section 3.5.3, as $\boldsymbol{F_1}$ achieved at 1GHz operating frequency. |
| $\boldsymbol{F_5}$ | PASS | Demonstrated in Section 3.5.1, modular and configurable implementation detailed. |
| | | BP Workload Benchmarking |
| $\boldsymbol{W_1}$ | PASS | Demonstrated in Section 4.3.1, $\boldsymbol{W_3}$ achieved for 1GHz operation, Table 25. |
| $\boldsymbol{W_2}$ | PASS | Demonstrated in Section 4.3.4, optimal operation found to be at 2.8V at 1GHz. |
| $\boldsymbol{W_3}$ | PASS | Demonstrated in Section 4.3.5, final configuration achieved 96.50% overall accuracy. |
| $\boldsymbol{W_4}$ | PASS | |

## 5.2 Reflection on Project Challenges

Various challenged, both expected and unexpected, where faced during the project. The primary challenge faced was with the Siemens Symphony AMS simulation platform. While other challenges, such as memristor device modelling and physical design, were also significant during project development.

Initially, the project was planned around the use of the Questa-ADMS simulation package. However, due to a miscommunication regarding software licensing availability, the package was not included within the university's Siemens license. Consequently, the project transitioned to the alternative Siemens Symphony AMS simulation platform. This transition introduced substantial delays, as Symphony AMS required extensive manual configuration and depended heavily on legacy operating system support and outdated software dependencies. As a result, the development of the simulation and modelling framework did not begin until the 17th of November, compared to the originally planned start date of the 27th of October. This approximately three-week delay had a significant downstream impact on the project schedule, with modelling and verification work not recovering to the planned timeline until early February.

Despite these setbacks, the adoption of the agile development methodology proved highly beneficial in mitigating the impact of the delays. During the three week delay owing to difficulty setting up Siemens Symphony for modelling work, alternative project tasks were reprioritised. This primarily involved expanding the literature review and conducting additional background research related to memristor device technologies, reservoir computing architectures, and contemporary branch prediction methods. While unplanned, this additional research ultimately strengthened the theoretical foundation of the project and contributed to a broader understanding of the application space and relevant implementation challenges.

Further difficulties were encountered throughout the use of the Siemens Symphony AMS platform due to inconsistent compliance with the Verilog-AMS language specification [46]. One notable example occurred during subsystem integration, discussed in Section 3.5.1b, where the simulator incorrectly handled unpacked array types. This issue necessitated module redesign work, resulting in the reservoir harness module to mitigate the issue. Similar simulator errata occurred repeatedly during development and frequently resulted in unexpected debugging overhead and increased implementation complexity. Although many of these issues were too minor or numerous to document individually within the main body of the thesis, several of the most significant examples are summarised in Table 40.

A further significant challenge encountered during the project was the modelling of memristor device behaviour. The VTEAM model was identified early in the project as a suitable foundation for behavioural modelling. However, additional literature review conducted during the Siemens Symphony AMS simulator setup delay showed that the complexity of physical devices was vastly higher than the VTEAM model could effectively capture. As real-world memristor behaviour was considerably more sophisticated than initially anticipated, particularly regarding device conduction mechanisms. As discussed in Section 3.1, practical SiOx memristor devices may exhibit multiple simultaneous conduction mechanisms. Including Ohmic conduction, Schottky emission, and Pool–Frenkel emission. The coexistence and complexity of these varying conduction mechanisms would have introduced significant numerical model complexity and additional modelling effort. Which there was less time for due to simulator setup delays.

These findings presented a major challenge as the chosen VTEAM model and the assumption of ohmic conduction were key to achieving fast model implementation and computational simplicity. Accurately modelling the full range of observed conduction mechanisms would have required substantially more complex physical models together with extensive parameter extraction and validation work. In addition to significantly increasing development effort, such models would likely have introduced substantial simulation performance overheads and reduced numerical stability. Directly conflicting with the project requirements for a highly configurable and computationally efficient behavioural model.

*Table 40. Encountered Symphony AMS simulator issues.*

| Problem | Fix |
|---|---|
| Incorrect *$rdist_poisson* behaviour in VAMS | The Poisson process for shot noise is approximated to a normal distribution and *$rdist_normal* is used instead. |
| Incorrect array indexing for unpacked arrays in VAMS | The reservoir was encapsulated in an intermediary module, the reservoir harness, to mitigate this. |
| Incorrect array indexing for integer indexed arrays in VAMS | All non-analogue modelling was moved to SV, which handled array indexing with genvars and integers properly. |
| Simulator crash when using array of type electrical with little endian size declaration. | Redefined all electrical type arrays to packed big endian. |
| Incorrect D2A domain boundary cross resulting in incorrect assignment of analogue voltages for DAC | Related to incorrect array indexing, reservoir harness and explicit code generation (not using genvar, using C++) to prevent incorrect value conversion from D2A. |
| Generate syntax resulting in inconsistent and incorrect instantiation of analogue procedural assignments | *Resgen* developed to address this, all reservoir code is generated in C++, with all arrays given via immediate value sizing and indexing. Eliminating the need for generate statements. |
| Incorrect handling of signed logic arithmetic in VAMS | All non-analogue logic was moved to SV, which handled the logic arithmetic properly. |

Following detailed review of the literature [10, 15, 11, 8, 19], it was ultimately concluded that attempting to comprehensively model all observed device behaviours within a single unified framework was impractical within the scope of the project. Furthermore, the variability between reported device implementations suggested that no single physical model would adequately capture the behaviour of all relevant memristor technologies. Consequently, the adopted approach focused on cautiously extending the VTEAM model with only the minimum additional functionality necessary to reproduce the key behavioural characteristics relevant to reservoir computing operation. Although this approach sacrifices some physical accuracy, the resulting model proved to be computationally efficient, numerically stable, and sufficiently flexible to support the development and evaluation of the RC BPU architecture.

The final major challenge encountered during the project related to the physical design and timing analysis of the developed RC BPU architecture and its constituent subcomponents. The original project aims included investigation of physical implementation and floor planning considerations, as outlined in Section 1.3. However, this aspect of the work was ultimately deprioritised due to the limited practical value of estimating physical implementation metrics solely from behavioural simulation models. In the absence of available physical memristor devices for experimental validation, any estimation of silicon area or physical reservoir density would have been highly speculative and insufficiently grounded in practical device behaviour. Consequently, it was determined that meaningful conclusions regarding physical implementation size and layout could not be reliably drawn within the scope of the project.

Despite this, several aspects of the original physical implementation objectives remained highly relevant, particularly those concerning system timing performance. As a result, the project focus shifted towards evaluating whether the developed RC BPU architecture could theoretically satisfy the timing requirements necessary for practical high-speed processor integration. These objectives were retained through the timing-related deliverable requirements for both the reservoir and perceptron output layer, specifically $\boldsymbol{R_1}$, $\boldsymbol{R_4}$, $\boldsymbol{O_2}$, and $\boldsymbol{O_4}$. This approach proved substantially more meaningful, as it allowed the project to identify the memristor switching characteristics and system-level operating conditions required to achieve processor-compatible operating frequencies. Thus, instead of providing a potentially poorly founded conclusion on the actual performance, area, and speed; the project instead delivered comprehensive analysis of possible implementation requirements and performance.

Testing simulations and STA demonstrated that the baseline timing requirements of 1ns critical path operation were successfully achieved for both the memristor reservoir and the perceptron readout layer. In particular, the reservoir exhibited greater performance potential than initially anticipated due to the lower-than-expected operating voltages required to sustain optimal operation at 1GHz. However, the perceptron output layer emerged as the primary limiting factor for higher-speed operation. Although the perceptron was capable of satisfying the 1GHz timing target, this was only achievable under carefully selected configurations that balanced weight quantisation bit depth and input dimensionality. With STA results for the perceptron, Section 3.4.4b, showing that faster than 1GHz operation was not feasible as insufficient slack remained. As a consequence, the stretch goal timing requirements of 0.5ns operation for both the reservoir, $\boldsymbol{R_4}$, and perceptron neuron, $\boldsymbol{O_4}$, were ultimately unachievable.

## 5.3 Reflection on Project Wider Background

This project delivers novel contributions in three principal areas. The first is the development of a design framework for high-speed memristor-based reservoir computing systems through behavioural modelling. The second is the iterative optimisation and evaluation of this framework through extensive simulation and workload-driven benchmarking. The third is the identification of memristor device performance targets required to achieve effective reservoir computing operation under realistic high-speed workload conditions. Collectively, these contributions provide an original and practically focused investigation into the feasibility of applying reservoir computing to branch prediction workloads. While maintaining consideration of implementation constraints, operating requirements, and architectural practicality.

Although the developed RC BPU architecture was ultimately shown to be significantly less adaptable than the state-of-the-art TAGE branch predictor; the broader outcomes of the research remain highly applicable beyond the specific scope of branch prediction. The work done demonstrates that memristor-based RC systems are theoretically capable of performing complex temporal computation at high operating speeds and reasonable operating voltages. As RC is fundamentally a form of general-purpose temporal information processing, the developed framework is transferable to a wide range of workloads. Most of which have more favourable operation requirements than branch prediction. Consequently, the research has significant potential relevance for emerging edge AI and embedded AI applications. Where RC can provide a real-time and intelligent form of information processing. Potential application areas include speech recognition, audio denoising, sensor data processing, time-series forecasting, and more computationally intensive workloads such as image classification and object detection. In these applications, memristor based RC has the potential to provide higher throughput, lower latency, and more substantial computation compared to existing systolic array style machine learning inference accelerators. This is particularly important given the rapidly increasing computational and environmental cost associated with modern AI systems [13]. As the global artificial intelligence market is projected to exceed 4.8 trillion dollars by 2033 [47], there exists substantial commercial incentive for the development of alternative computing architectures to achieve higher performance and energy efficiency.

The development of memristor RC as a powerful form of universal computation also has the potential for a positive societal impact. Such as for healthcare, where RC could form a natural extension of information processing from sensors for patient diagnostics. Or for personal and recreational use, where embedded memristor based RC could enable (where useful) smarter appliances. Such as enabling more natural human device interaction through accurate ultra-low latency speech and gesture recognition. In this context, the developed RC framework could be further refined into a configurable hardware IP block suitable for deployment across a range of embedded intelligent sensing applications.

This project also provides a target memristor device performance that would enable practical and realisable workloads, such as BP. Having a wider impact on device engineering efforts. With the main identified requirements for potential divider reservoir architecture of low threshold voltage and slow switching behaviour. In addition to the relaxed requirement for ON/OFF ratio, which runs contrary to current development efforts. Additionally, this project demonstrates the necessity of a more physically accurate and comprehensive memristor behavioural model. Reinforcing the necessity of such accurate model's development to allow widespread investigation of potential memristor applications while device manufacturing remains niche and limited.

## 5.4 Project Further Steps

The primary focus of future work should be addressing the substantially slower convergence behaviour of the developed RC BPU relative to the TAGE predictor. Although the RC BPU was capable of achieving high overall prediction accuracy, the TAGE predictor converged to 100% accuracy approximately 15× faster. Representing the RC BPU designs most significant performance deficiency identified by testing. The results obtained throughout this project suggest that the primary limitation is not the reservoir itself, but rather the perceptron readout layer and its ability to rapidly adapt to changing temporal patterns.

The first area for future research to focus on is the improvement of the perceptron readout neuron. One promising avenue for improvement is the investigation of continuous activation functions rather than the current threshold-based activation function. Continuous activation functions would enable scaled weight updates proportional to prediction error and confidence, potentially resulting in more stable and significantly faster convergence behaviour. This approach could draw inspiration from conventional predictors such as TAGE, which utilise multi-bit saturating counters to encode both prediction direction and confidence. Another potential direction is the implementation of the perceptron readout layer using analogue or mixed-signal circuitry. Existing research has demonstrated that memristors can be used to implement a perceptron neuron for BP. By encoding weights as memristor resistance states and exploiting Kirchhoff's current law to inherently perform multiply-and-accumulate operations through current summation [48]. Such approaches may significantly reduce prediction latency and hardware complexity within the reservoir readout stage. However, implementing effective in-situ training within analogue perceptron architectures would likely introduce substantial additional design complexity.

Longer-term research should focus on the development of a physical test chip together with investigation of variable-frequency and voltage operation. A fabricated test chip would provide substantially more definitive insight into the practicality of memristor-based reservoir computing for branch prediction workloads than behavioural simulation alone. In particular, physical implementation would enable direct investigation of reservoir topologies designed to improve noise immunity. Such as architectures incorporating multiple parallel memristors or alternative device configurations. These characteristics are difficult to accurately evaluate through simulation due to the stochastic and highly device-dependent nature of memristor switching behaviour.

A physical test platform would additionally enable investigation of long-term reliability and device degradation effects, which were beyond the scope of this thesis. As all physical electronic systems degrade over time, practical reservoir implementations would need to be verified for stable long-term operation to be considered viable for commercial processor integration. A physical test chip would therefore enable future work to investigate the impact of device ageing and degradation on reservoir dynamics, prediction stability, and computational consistency. Engineering efforts should target a mean time between failure exceeding approximately 140 thousand operational hours, corresponding to roughly 15 years of continuous operation. This target is increasingly relevant as modern computing devices commonly continue operating in low-power or sleep states rather than fully powering down when not actively in use. Thus, the RC implementation must be at least as reliable as the CPU that it is to be integrated into, with 15 years being a typical target.

## 5.5 Reflection on Project Management

Project management throughout execution was broadly characterised by the opportunistic completion of work. Per the agile methodology, weekly targets were set and worked towards. This was effective as the flexible schedule of project completion allowed the partially effective mitigation of unexpected roadblocks of simulator access and bugs. However, the unorganised nature of project execution also led to significant redundant work.

The first and primary success of the project management and execution was the adoption of the agile methodology and decomposition of deliverables into multiple semi-independent stages. This was critical in allowing progress to continue while waiting for the resolution of external factors to the project such as the licensing issues encountered with Siemens software. Where the agile approach allowed focus to be shifted temporarily to other work while awaiting a resolution. This represented a large success in project management as the cumulative simulator downtime exceeded one month, during which other work was effectively re-prioritised and worked on. Allowing effective schedule recovery with minimal directly attributable downstream effects, see progress review one. The primary example of this would be the work done in Section 2.0, which was conducted while waiting for licensing issues with Questa-ADMS and Siemens Symphony to be resolved at the start of the project.

The weekly supervisor meetings also provided a good structure of weekly goals, which were reviewed upon completion. This structure allowed the tracking of project progress informally through weekly meetings, and feedback on progress and methods to inform further development. This was particularly impactful during reservoir development, where supervisor feedback proved valuable in assessing the application of different tests and metrics to evaluate the reservoir design. Alongside this, the weekly supervisor meetings also provided a forum for evolving the project aims towards more meaningful targets as the project progressed. This involved decisions to scrap the physical design specification points and fully detailed modelling of memristor devices, which were vital decisions to keep the project on track and in scope.

The project also faced some major setbacks due to deficiencies in project management and a lack of clarity in the initial stated plan of execution. The primary deficiency was the lack of formal progress tracking and management; this resulted in time being spent aligning deliverable outcomes with work which had not yet been completed. Which wasted significant time as this led to substantially more retroactive refactoring than would be considered optimal, see in progress review five. One such example would be the memristor behavioural model, which was refactored three times. With it eventually being simplified to more closely resemble the VTEAM model as a result of the significant complexity of modifications to better model physical effects such as different conduction mechanisms.

Another major setback to the project management was the lack of clarity on the project's goals. The initial project proposal was over ambitious. Goals such as physical design and verification against physical memristor devices were at odds with the speculative and forwards looking nature of the research. Which was compounded by the lack of a physical memristor device for testing to verify the memristor behavioural model against. This led to significant time wasted pursuing irrelevant project development such as the temperature variation modelling for the memristor. Distracting from the projects true goal of developing a forward's looking implementation and design framework in simulation to assess the general application of RC to BP workload.

In conclusion, the project management was largely sub-par. Although the agile methodology and week to week work prioritisation was pivotal in recovering from encountered externalities. The lack of a clear vision in planning and formal progress tracking lead to significant time spent on redundant work which was not relevant to the final thesis outcome. For future projects, an approach of rushing to minimum viable product, with further refinement from there would be more appropriate. Alongside more formal tracking of progress to allow the reduction in redundant work. All five project review proformas can be found in the appendix in Section 7.

# 6 References

[1] Intel corp, “Intel® 64 and IA-32 Architectures Optimization Reference Manual,” NA 01 2024. [Online]. Available: https://cdrdv2-public.intel.com/821613/355308-Optimization-Reference-Manual-050-Changes-Doc.pdf. [Accessed 17 03 2026].

[2] I. Healy, P. Giordano and W. Elmannai, “Branch Prediction in CPU Pipelining,” in *IEEE Annual Ubiquitous Computing, Electronics & Mobile Communication Conference (UEMCON)*, New York, 2023.

[3] F. Vikas, P. Gratz and D. Jiménez, “Workload Characterization for Branch Predictability,” 17 12 2025. [Online]. Available: https://arxiv.org/abs/2512.15827. [Accessed 17 04 2026].

[4] A. Seznec, “A new case for the TAGE branch predictor,” in *2011 44th Annual IEEE/ACM International Symposium on Microarchitecture (MICRO)*, Porto Alegre, 2011.

[5] Y. Lu, Y. Liu and H. Wang, “A study of perceptron based branch prediction on Simplescalar platform,” in *IEEE International Conference on Computer Science and Automation Engineering (CSAE)*, Shanghai, 2011.

[6] G. H. Loh, “Revisiting the performance impact of branch predictor latencies,” in *2006 IEEE International Symposium on Performance Analysis of Systems and Software*, Austin, 2006.

[7] W. Sun, W. Zhang, Y. Li, Z. Guo, J. Lai, D. Dong, X. Zheng, F. Wang, S. Fan, X. Xu, D. Shang and M. Liu, “3D Reservoir Computing with High Area Efficiency (5.12 TOPS/mm2) Implemented by 3D Dynamic Memristor Array for Temporal Signal Processing,” in *2022 IEEE Symposium on VLSI Technology and Circuits (VLSI Technology and Circuits)*, Honolulu, 2022.

[8] W. Sun, J. Yu, D. Dong, X. Zheng, J. Lai, S. Fan, H. Wang, J. Gao, J. Liu and X. Xu, “Investigation on the 3D Memristor Array Architecture for 3D Reservoir Computing System Implementation,” *IEEE Electron Device Letters, vol. 45, no. 8,* pp. 1445-1448, 21 06 2024.

[9] Y. Chen, “ReRAM: History, Status, and Future,” *IEEE Transactions on Electron Devices,* vol. 67, no. 4, pp. 1420-1433, 2020.

[10] A. Bricalli, E. Ambrosi, M. Laudato, M. Maestro, R. Rodriguez and D. Ielmini, “Resistive Switching Device Technology Based on Silicon Oxide for Improved ON–OFF Ratio—Part I: Memory Devices,” *IEEE Transactions on Electron Devices,* vol. 65, no. 1, pp. 115-121, 2018.

[11] S. Pi, P. Lin, H. Jiang, C. Li and Q. Xia, “Device engineering and CMOS integration of nanoscale memristors,” in *IEEE International Symposium on Circuits and Systems (ISCAS)*, Melbourne, 2014.

[12] M. t. Vrugt, “[2412.13212] An introduction to reservoir computing,” 12 12 2024. [Online]. Available: https://arxiv.org/abs/2412.13212. [Accessed 19 03 2026].

[13] BBC, “Data centre power use 'to surge six-fold in 10 years',” BBC, 26 03 2024. [Online]. Available: https://www.bbc.co.uk/news/technology-68664182. [Accessed 10 10 2025].

[14] P. Valerio, “TSMC Price Hikes End the Era of Cheap Transistors,” EE Times, 01 10 2025. [Online]. Available: https://www.eetimes.com/tsmc-price-hikes-end-the-era-of-cheap-transistors/. [Accessed 1 10 2025].

[15] S. Ibrahim, “Probing the resistance switching mechanisms in SiOx Ag RRAM devices_ edited_version.pdf,” N/A 09 2019. [Online]. Available: https://discovery.ucl.ac.uk/id/eprint/10092556/1/Probing%20the%20resistance%20switching%20mechanisms%20in%20SiOx%20Ag%20RRAM%20devices_%20edited_version.pdf. [Accessed 17 03 2026].

[16] S. Kvatinsky, M. Ramadan, E. G. Friedman and A. Kolodny, “VTEAM: A General Model for Voltage-Controlled Memristors,” *IEEE Transactions on Circuits and Systems II: Express Briefs,* vol. 62, no. 8, pp. 786-790, 2015.

[17] H. Padberg, A. Regev, G. Piccolboni, A. Bricalli, G. Molas, J. F. Nodin and S. Kvatinsky, “Experimental Demonstration of Non-Stateful In-Memory Logic With 1T1R OxRAM Valence Change Mechanism Memristors,” *IEEE Transactions on Circuits and Systems II: Express Briefs,* vol. 71, no. 1, pp. 395-399, 2024.

[18] C. Wanjun, C. Yiping, G. Jun, M. ZeLin, C. XuCheng, D. Shanqing, L. Zhiyu and P. Shusheng, “Intrinsic resistive switching in ultrathin SiOx memristors for neuromorphic inference accelerators,” *Applied Surface Science,* vol. 625, pp. 157-191, 2023.

[19] H.-X. G. Yang, G. Xi-Xi, G. J. Zhou and Y. Tang, “Bipolar Resistive Switching Characteristics and Mechanismin the TiN/SiOx/Pt Structure,” in *2016 13th IEEE International Conference on Solid-State and Integrated Circuit Technology (ICSICT)*, Hangzhou, 2016.

[20] L. Chua, “Memristor-The missing circuit element,” *IEEE Transactions on Circuit Theory,* vol. 18, no. 5, pp. 507-519, 1971.

[21] A. Mehonic, A. Shluger, D. Gao, I. Valov, E. Miranda, D. Ielmini, A. Bricalli, E. Ambrosi, C. Li, J. Yang, Q. Xia and A. Kenyon, “Silicon Oxide (SiOx) - A Promising Material for Resistance Switching?,” 29 06 2018. [Online]. Available: https://discovery.ucl.ac.uk/id/eprint/10051514/1/Kenyon_Silicon_Oxide_SiOx.pdf. [Accessed 28 03 2026].

[22] Condor Computing, “GitHub - condorcomputing/condor.riscv-isa-sim,” Condor Computing, 04 06 2025. [Online]. Available: https://github.com/condorcomputing/condor.riscv-isa-sim. [Accessed 21 02 2026].

[23] R. P. Weicker, “Dhrystone: a synthetic systems programming benchmark,” *Commun. ACM,* vol. 27, no. 10, pp. 1013-1030, 1984.

[24] Wikimedia Foundation, “Reservoir computing - Wikipedia,” Wikimedia Foundation, 05 11 2025. [Online]. Available: https://en.wikipedia.org/wiki/Reservoir_computing. [Accessed 25 11 2025].

[25] C. Fernando and S. Sojakka, “Pattern Recognition in a Bucket,” in *Advances in Artificial Life*, Dortmund, Springer Berlin Heidelberg, 2003, pp. 588-597.

[26] C. Du, F. Cai, M. Zidan, W. Ma, S. H. Lee and W. D. Lu, “Reservoir computing using dynamic memristors for temporal information processing,” *Nature Communications,* vol. 8, no. 1, p. 2204, 2017.

[27] X. Yang, M. You, L. Pang and B. Du, “Reservoir Computing Based on Memristor Arrays in Random States,” *IEEE Transactions on Circuits and Systems I: Regular Papers,* vol. 71, no. 7, pp. 3256-3268, 2024.

[28] Intel Corp, Intel 64 and IA-32 Architectures Software Developer's Manual, Hillsboro: Intel Corp, 2024.

[29] L. C.-K. Tarsa and S. J, “Branch Prediction Is Not a Solved Problem: Measurements, Opportunities, and Future Directions,” 13 06 2019. [Online]. Available: https://arxiv.org/abs/1906.08170. [Accessed 18 04 2026].

[30] W. Jin, J. Dong, K. Lu and Y. Li, “The Study of Hierarchical Branch Prediction Architecture,” in *2011 14th IEEE International Conference on Computational Science and Engineering*, Dalian, 2011.

[31] A.-F. Sburlan, “Discovering Predictive Patterns: A Study of Contextual Factors for Next Generation Branch Predictors,” 19 06 2023. [Online]. Available: https://www.imperial.ac.uk/media/imperial-college/faculty-of-engineering/computing/public/distinguished-projects/2223-ug-projects/Discovering-Predictive-Patterns-A-Study-of-Contextual-Factors-for-Next-Generation-Branch-Predictors.pdf. [Accessed 18 04 2026].

[32] T. Sadi, L. Wang, L. Gerrer, V. Georgiev and A. Asenov, “Self-consistent physical modeling of SiOx-based RRAM structures,” in *2015 International Workshop on Computational Electronics (IWCE)*, West Lafayette, 2015.

[33] W. H. Ng, A. Mehonic, M. Buckwell, L. Montesi and A. J. Kenyon, “High-Performance Resistance Switching Memory Devices Using Spin-On Silicon Oxide,” *IEEE Transactions on Nanotechnology,* vol. 17, no. 5, pp. 884-888, 2018.

[34] A. J. Kenyon, A. Mehonic, W. Ng, L. Zhao, H. Cox, M. Buckwell, K. Patel, A. P. Knights, D. J. Mannion and A. L. Shluger, “Defining the performance of SiOx ReRAM by engineering oxide microstructure,” in *2022 11th International Conference on Modern Circuits and Systems Technologies (MOCAST)*, Bremen, 2022.

[35] A. M. Kenyon, M. M.S, B. N, M. L, B. M, S. A.L and J. A, “Intrinsic resistance switching in amorphous silicon oxide for high performance SiOx ReRAM devices,” *Microelectronic Engineering,* vol. 103, no. 1, p. 98, 2017.

[36] S. Kvatinsky, E. G. Friedman, A. Kolodny and U. C. Weiser, “TEAM: ThrEshold Adaptive Memristor Model,” *IEEE Transactions on Circuits and Systems I: Regular Papers,* vol. 60, no. 1, pp. 211-221, 2013.

[37] Wikimedia Foundation, “Johnson–Nyquist noise - Wikipedia,” Wikimedia Foundation, 28 11 2025. [Online]. Available: https://en.wikipedia.org/wiki/Johnson%E2%80%93Nyquist_noise. [Accessed 10 12 2025].

[38] Wikimedia Foundation, “Erlang distribution - Wikipedia,” Wikimedia Foundation, 15 09 2025. [Online]. Available: https://en.wikipedia.org/wiki/Erlang_distribution. [Accessed 21 04 2026].

[39] D. Pilati, A. Ceni, F. Michieletti, C. Gallicchio, C. Ricciardi and G. Milano, “RCbench: a unified framework for benchmarking reservoir computing systems,” *Neuromorphic Computing and Engineering,* vol. 6, no. 1, p. 014012, 2026.

[40] Wikimedia foundation, “Learning rule - Wikipedia,” Wikimedia foundation, 15 09 2025. [Online]. Available: https://en.wikipedia.org/wiki/Learning_rule. [Accessed 02 04 2026].

[41] T. Ajayi, V. A. Chhabria, M. Fogaca, S. Hashemi, A. Hosny, A. B. Kahng, M. Kim, J. Lee, U. Mallappa, M. Neseem, G. Pradipta, S. Reda and M. Saligane, “INVITED: Toward an Open-Source Digital Flow: First Learnings from the OpenROAD Project,” in *2019 56th ACM/IEEE Design Automation Conference (DAC)*, Las Vegas, 2019.

[42] L. T. C. Yeric, V. Vinay, S. Lucian, G. Aditya, S. Saurabh, C. Brian and R. Chandarasekaran, “ASAP7: A 7-nm finFET predictive process design kit,” *Microelectronics Journal,* vol. 53, no. 1, pp. 105-115, 2016.

[43] RISC-V International, “1.1. Introduction :: RISC-V Ratified Specification Library,” RISC-V International, 20 01 2026. [Online]. Available: https://docs.riscv.org/reference/isa/unpriv/intro.html. [Accessed 15 04 2026].

[44] M. R. Guthaus, J. S. Ringenberg, D. Ernst, T. M. Austin, T. Mudge and R. B. Brown, "MiBench: A free, commercially representative embedded benchmark suite," in *Proceedings of the Fourth Annual IEEE International Workshop on Workload Characterization. WWC-4*, Austin, 2001.

[45] "Efficient architectural exploration of TAGE branch predictor for embedded processors," *Microelectronics Journal,* vol. 88, pp. 88-98, 2019.

[46] Accellera Systems Initiative, Verilog-AMS Language Reference Manual, Elk Grove: Accellera Systems Initiative Inc, 2024.

[47] UN trade & development, "AI market projected to hit $4.8 trillion by 2033, emerging as dominant frontier technology," UN trade & development, 07 04 2025. [Online]. Available: https://unctad.org/news/ai-market-projected-hit-48-trillion-2033-emerging-dominant-frontier-technology. [Accessed 12 04 2026].

[48] J. Wang, Y. Tim, W.-F. Wong and H. Helen Li, "A practical low-power memristor-based analog neural branch predictor," in *International Symposium on Low Power Electronics and Design (ISLPED)*, Beijing, 2013.

[49] V. Mladenov and S. Kirilov, "A Modified Metal Oxide Memristor Model," in *International Conference on Modern Circuits and Systems Technologies (MOCAST)*, Bremen, 2022.

[50] Intel Corp, "Intel Core TM Ultra 9 Processor 285K," Intel, 24 10 2024. [Online]. Available: https://www.intel.com/content/www/us/en/products/sku/241060/intel-core-ultra-9-processor-285k-36m-cache-up-to-5-70-ghz/specifications.html. [Accessed 09 10 2025].

[51] "Tech megacaps plan to spend more than $300 billion in 2025 as AI race intensifies," CNBC, 08 02 2025. [Online]. Available: https://www.cnbc.com/2025/02/08/tech-megacaps-to-spend-more-than-300-billion-in-2025-to-win-in-ai.html. [Accessed 19 10 2025].

[52] Siemens AG, "QuestaADMS analogue and mixed-signal simulation | Siemens Software," Siemens AG, N/A N/A 2025. [Online]. Available: https://eda.sw.siemens.com/en-US/ic/questa-one/adms/. [Accessed 19 10 2025].

[53] NA, "Tcl Developer Site," NA, NA NA NA. [Online]. Available: https://www.tcl-lang.org/about/language.html. [Accessed 22 10 2025].

# 7 Appendix

## 7.1 Progress Review One

| **Summarised Planned State of Project:** | **Actual Progress Since Last Review** |
|---|---|
| - *Proposal submitted*<br>- *Initial software setup and configured*<br>- *Initial test module complete to demonstrate synthesis and simulation of simple AMS module* | - *Proposal submitted.*<br>- *Issues with Questa ADMS, do not have license.* |
| **Next Steps and Supervisor Feedback**<br>- *Look for alternative software*<br>- *Work on other tasks such as research and lit review in the meantime*<br>- *Reach out to Yiqun for help with AMS licensing* | |

## 7.2 Progress Review Two

<table>
<tr><th>Summarised Planned State of Project:</th><th>Actual Progress Since Last Review</th></tr>
<tr><td>- Proposal submitted.<br>- Memristor model fully implemented.<br>- Benchmark research underway.<br>- Research on reservoir configuration underway.<br>- Research on memristor models underway.<br>- Research on manufacturability and device performance complete.</td><td>- Research on reservoir configuration underway.<br>- Research on memristor models underway.<br>- Siemens Symphony as an alternative to Questa-ADMS is up and running. Initial test of memristor linear ion diffusion modelling complete.</td></tr>
<tr><td colspan="2">Next Steps and Supervisor Feedback<br>- Work on memristor device modelling.<br>- Start work on reservoir design and architecture<br>- Progress is ok, need to get back on track.<br>- Re focus project on a simulation framework due to the infeasibility of verification against real devices.</td></tr>
</table>

## 7.3 Progress Review Three

<table>
<tr><th>Summarised Planned State of Project:</th><th>Actual Progress Since Last Review</th></tr>
<tr><td>- Reservoir design complete<br>- Reservoir design characterisation underway<br>- Final system integration underway<br>- Output layer design underway<br>- Basic workload benchmarking framework developed</td><td>Memristor design:<br>- Memristor VAMS model complete<br>- Noise source modelling complete<br>- Behaviour characterisation complete<br>- Groundwork for proposed operating requirements complete<br>Reservoir design:<br>- Design complete<br>- Design space exploration underway<br>- Resgen tool implemented to allow design space exploration<br>Output layer:<br>- Int 8 quantised perceptron neuron designed<br>- Int 8 quantised perceptron neuron demonstrated on test workload<br>Full system integration:<br>- Underway and almost complete<br>Workload benchmarking:<br>- Investigated branch prediction workload benchmarking approaches<br>- Assessed MiBench and use of functional RISC-V models and ran provisional test workloads on functional model</td></tr>
<tr><td colspan="2">Next Steps and Supervisor Feedback<br>- Progress is good<br>- Need to focus on thesis structure and data generation for presentation<br>- Need to formalise performance evaluation metrics for BP workload and reservoir design<br>- Finalise full system integration to demonstrate non BP test workload<br>- Write thesis sections on memristor device modelling and reservoir design</td></tr>
</table>

## 7.4 Progress Review Four

<table>
<tr><td>Summarised Planned State of Project:<br><br>- Final system design complete<br>- Final system testing underway<br>- Thesis 50% complete<br>- BP workload benchmarking underway<br>- Perceptron & Gshare implementation underway</td><td>Actual Progress Since Last Review<br><br>Reservoir design:<br>- Design space exploration underway<br>- Identified Lyapunov exponent as metric<br>- Identified histogram of quantised states as metric<br>- Modifications made to resgen to enhance support for greater variety of reservoir configs<br>- Modifications made to resgen to produce more efficient output code<br>Full system integration:<br>- Complete<br>Workload benchmarking:<br>- Identified RISC-V functional model and Dhrystone workload for BP workload testing<br>- Working on test data generation<br>- Comparison to Perceptron & Gshare dropped due to time limitations.<br>Thesis:<br>- Thesis introduction almost complete<br>- Thesis memristor behavioural model section almost complete<br>- Thesis reservoir design section underway</td></tr>
<tr><td colspan="2">Next Steps and Supervisor Feedback<br><br>- Progress is not ideal<br>- Need to shift focus from design to testing and generation of meaningful metrics for design success<br>- Thesis should be the main focus moving forwards<br>- Focus on example workload testing to demonstrate RC and then BP workload testing</td></tr>
</table>

## 7.5 Progress Review Five

<table>
<tr><td>Summarised Planned State of Project:<br><br>- Final system design complete<br>- Final system testing complete<br>- Thesis 90% complete<br>- BP workload benchmarking almost done</td><td>Actual Progress Since Last Review<br><br>Reservoir design:<br>- Lyapunov testing complete.<br>Workload benchmarking:<br>- Initial test data generated<br>- Chosen Dhrystone benchmark<br>- Test data processing and initial benchmarks complete.<br>Memristor:<br>- memristor behavioural model re-work underway to fix modelling inaccuracy of oxide thickness modelling.<br>- Added initial verification flows to better characterise memristor device.<br>Thesis:<br>- Thesis introduction re written to increase writing quality<br>- Thesis memristor behavioural model re-work underway to fix modelling inaccuracy of oxide thickness modelling.<br>- Thesis reservoir design section underway<br>- Took brief thesis feedback and worked on more graphs and figures to illustrate design and testing setup.<br><br>AFS license is warning of upcoming expiration.</td></tr>
<tr><td colspan="2">Next Steps and Supervisor Feedback<br><br>- Progress is not on track for thesis, need to shift focus towards thesis.<br>- Reach out to Yiqun for AFS license issues.<br>- Continue the shift towards gathering performance data and plots.<br>- Current thesis structure is not ideal; wording is subpar and should be improved moving forwards.</td></tr>
</table>

**Statement of originality**

- I have read and understood the University of Nottingham Academic Misconduct Policy.
- I am aware that failure to act in accordance with the Regulations Governing Academic Integrity may lead to the imposition of penalties which, for the most serious cases, may include termination of programme.
- I consent to the University copying and distributing any or all of my work in any form and using third parties (who may be based outside the EU/EEA) to verify whether my work contains plagiarised material, and for quality assurance purposes.

***You must change the statements in the boxes to reflect your individual submission.***

You must acknowledge all sources of information (e.g. ideas, algorithms, data) using citations to auditable, authoritative sources.

You must also put quotation marks around any sections of text that you have copied without paraphrasing and cite the original source.

If any figures or tables have been taken or modified from another source, you must explain this in the caption, and cite the original source.

I have acknowledged all sources, and identified any content taken from elsewhere.

If you have used any code (e,g. open-source code), reference designs, or similar resources that have been produced by anyone else, you must list them in the box below. In the report, you must explain what was used and how it relates to the work you have done.

I have used the following resources that have been produced by someone else:

- *Condor fork of RISC-V Spike ISA sim*
- *STFLib*
- *Siemens Symphony Simulator*
- *OpenROAD EDA Flow Scripts (ORFS)*

I have acknowledged all input (hardware, software, intellectual) taken from elsewhere and explained how this was used.

If you have used generative AI tools in the preparation of your assessed work, or the underlying code/designs/underpinning work, its use must be declared in a signed statement in the box below. You must declare the tools employed and the method and extent of use. You can do this You can do this using the table below:

Generative AI tools have been used in the preparation of this work as follows:

| *Tools* | *Method* | *Extent of use* |
|---|---|---|
| *Chat GPT* | *Grading thesis* | *Used ChatGPT to grade my thesis so I could have an idea of what grade I would get.* |

I have acknowledged, in context, the use of all generative AI tools in preparation of this assessment, including textual, referencing, hardware, software, and intellectual inputs to my submitted work. All outputs were critically reviewed, edited, and verified by me.

To add to this, AI did not write anything or do anything other than give me an estimated % grade.

You can consult with teaching staff/demonstrators, but you should not show anyone else your work (this includes uploading your work to publicly accessible repositories e,g. Github, unless expressly permitted by your supervisor). Nor should you help other students with their work.

You must get permission in writing before you seek outside assistance, e.g. a proofreading service, and declare it here.

I did all the work myself, and have not helped anyone else.

You must not have fabricated, modified or distorted any data, evidence, references, experimental results, or other material used or presented in the report. You must clearly describe your experiments and how the results were obtained, and include all data, source code and/or designs (submitted as a separate file) so that your results could be reproduced.

The material in the report is genuine, and I have included all my data/code/designs.

If your work involved research studies (including surveys) on human participants, their cells or data, or on animals, you must have been granted ethical approval before the work was carried out, and any experiments must have followed these requirements. You must give details of this in the report, and list the ethical approval reference number(s) in the box below.

<table>
<tr><td colspan="3">not applicable</td></tr>
<tr><td rowspan="2">My responses to these questions are full and complete, and I have discussed them with my supervisor.</td><td>Student signature</td><td>H.J</td></tr>
<tr><td>Date</td><td>19.05.2026</td></tr>
</table>